\documentclass[review]{elsarticle}
\graphicspath{ {./Figure/} }
\usepackage{hyperref}
\usepackage{float}
\usepackage{verbatim} %comments
\usepackage{apalike}
\restylefloat{figure}
\restylefloat{table}

\journal{Expert Systems with Applications}

\biboptions{authoryear}

\usepackage{graphicx}
\usepackage{epsfig}
\usepackage{subcaption}
\usepackage{calc}
\usepackage{amssymb}
\usepackage{amstext}
\usepackage{amsmath}
\usepackage[acronym]{glossaries}
\newacronym{eeg}{EEG}{electroencephalography}
\newacronym{mi}{MI}{motor imagery}
\newacronym{bci}{BCI}{brain–computer interface}
\newacronym{mrcp}{MRCP}{motor related cortical potential}

\newacronym{auc}{AUC}{area under the curve}

\newacronym{ica}{ICA}{independent component analysis}
\newacronym{pca}{PCA}{principal component analysis}
\newacronym{rpca}{RPCA}{robust PCA}

\newacronym{fbcsp}{FBCSP}{filter-bank common spatial pattern}
\newacronym{csp}{CSP}{common spatial pattern}
\newacronym{vmd}{VMD}{visual mode decomposition}
\newacronym{stft}{STFT}{short-time Fourier transform}
\newacronym{psd}{PSD}{power spectral density}
\newacronym{dtw}{DTW}{dynamic time warping}

\newacronym{ml}{ML}{machine learning}
\newacronym{dl}{DL}{deep learning}
\newacronym{cnn}{CNN}{convolutional neural networks}
\newacronym{vae}{VAE}{variational autoencoder}
\newacronym{ffnn}{FFNN}{feed-forward neural network}
\newacronym{elbo}{ELBO}{evidence lower bound}
\newacronym{gan}{GAN}{generative adversarial network}
\newacronym{lda}{LDA}{linear discriminant analysis}
\newacronym{svm}{SVM}{support vector machine}
\newacronym{mse}{MSE}{mean square error}
\newacronym{lstm}{LSTM}{long-short term memory}

\newacronym{knn}{kNN}{k-nearest neighbors}

\newacronym{dffn}{DFFN}{densely feature fusion CNN}
\newacronym{cwcnn}{CW-CNN}{channel-wise CNN}
\newacronym{tsgl}{TSGL-EEGNet}{temporary-constrained sparse group Lasso-enhanced EEGNet}
\newacronym{mbshallow}{MBShallow ConvNet}{multi-branch shallow CNN}
\newacronym{qeegnet}{Q-EEGNet}{quantized EEGNet}

\newacronym{elu}{ELU}{exponential linear unit}

\newacronym{rh}{RH}{right hand}
\newacronym{lh}{LH}{left hand}
\newacronym{to}{TO}{tongue}
\newacronym{fe}{FE}{feet}

\newacronym{vmse1}{vEEGNet-ver1}{vEEGNet-ver1}                                      % vEEGNet-ver1 = Variational-EEGNet with MSE e CLF (Praga, ICT4AWE'23)

\newacronym{vmse2}{vEEGNet-ver2}{vEEGNet-ver2}                                      % vEEGNet-ver2 = Variational-EEGNet modificata (nei layer conv) with MSE e CLF (estensione su Journal di ICT4AWE'23)

\newacronym{vmse}{vEEGNet-MSE-NOCLF}{Variational-EEGNet with MSE}                   % vEEGNet-variante di vEEGNet-ver1 SENZA CLF [NON USATA]

\newacronym{vdtw}{vEEGNet-ver3}{vEEGNet \emph{version 3}}                                       % vEEGNet-ver3 = Variational-EEGNet with DTW

\newacronym{hvmse}{hvEEGNet-MSE-NOCLF}{Hiearchical-variational-EEGNet with MSE}     % hvEEGNet-variante [NON USATA]

\newacronym{hvdtw}{hvEEGNet}{\emph{hierarchical} vEEGNet}                                              % hvEEGNet = Hiearchical-variational-EEGNet with DTW
\usepackage{blindtext}
\usepackage{makecell} %celle e tabelle
\usepackage{siunitx} 
\usepackage{soul}
\usepackage{tabularray}
\usepackage{apalike}
\usepackage{multirow}
\usepackage[table,xcdraw]{xcolor}

\DeclareMathOperator{\EX}{\mathbb{E}}% expected value

\usepackage[absolute]{textpos}
\usepackage{everyshi}

\begin{document}
\begin{frontmatter}

%% TITLE
\title{hvEEGNet: exploiting hierarchical VAEs on EEG data for neuroscience applications}

%% AUTHORS
\author[label1,label2]{Giulia Cisotto\corref{cor1}}
\ead{giulia.cisotto@unimib.it}

\author[label2]{Alberto Zancanaro}
\ead{alberto.zancanaro.1@phd.unipd.it}

\author[label1]{Italo F. Zoppis}
\ead{italo.zoppis@unimib.it}

\author[label1]{Sara L. Manzoni}
\ead{sara.manzoni@unimib.it}

\cortext[cor1]{Corresponding author.}
\address[label1]{Department of Informatics, Systems, and Communication, University of Milan-Bicocca, viale Sarca 336, 20126, Milan, Italy}
\address[label2]{Department of Information Engineering, University of Padova, via Gradenigo 6b, 35131, Padova, Italy}

%% ABSTRACT

\begin{abstract}

% CONTEXT
With the recent success of artificial intelligence in neuroscience, a number of deep learning (DL) models were proposed for classification, anomaly detection, and pattern recognition tasks in electroencephalography (EEG). EEG is a multi-channel time-series that provides information about the individual brain activity for diagnostics, neuro-rehabilitation, and other applications (including emotions recognition).
%
% RESEARCH QUESTIONS
Two main issues challenge the existing DL-based modeling methods for EEG: the high variability between subjects and the low signal-to-noise ratio making it difficult to ensure a good quality in the EEG data.
%
% METHODS
In this paper, we propose two variational autoencoder models, namely \emph{vEEGNet-ver3} and \emph{hvEEGNet}, to target the problem of high-fidelity EEG reconstruction.
We properly designed their architectures using the blocks of the well-known EEGNet as the encoder, and proposed a loss function based on dynamic time warping.
%
% RESULTS OVER THE SOTA
We tested the models on the public \emph{Dataset 2a - BCI Competition IV}, where EEG was collected from $9$ subjects and $22$ channels.
\emph{hvEEGNet} was found to reconstruct the EEG data with very high-fidelity, outperforming most previous solutions (including our \emph{vEEGNet-ver3}). Furthermore, this was consistent across all subjects.
Interestingly, \emph{hvEEGNet} made it possible to discover that this popular dataset includes a number of corrupted EEG recordings that might have influenced previous literature results.
We also investigated the training behaviour of our models and related it with the quality and the size of the input EEG dataset, aiming at opening a new research debate on this relationship.
%
% FUTURE PERSPECTIVES
In the future, \emph{hvEEGNet} could be used as anomaly (e.g., artefact) detector in large EEG datasets to support the domain experts, but also the latent representations it provides could be used in other classification problems and EEG data generation.

\end{abstract}

% \begin{comment}

%% PER PREPRINT SU ARXIV
\begin{textblock*}{17cm}(1.7cm, 0.5cm)
\noindent\scriptsize This work has been submitted to Expert Systems with Applications for possible publication. Copyright may be transferred without notice, after which this version may no longer be accessible.\\
\end{textblock*}

% \end{comment}

%% KEYWORDS: a maximum of 6 keywords

% \sep EEGNet
% \sep computational neuroscience
% \sep emotion recognition
% \sep classification

\end{frontmatter}

%% HIGHLIGHTS (al max 5, anche in file Word a parte)
\section*{Highlights}

\begin{itemize}
    \item dynamic time warping helps variational autoencoders learn time-series
    \item hierarchical VAE allows for high-fidelity reconstruction of EEG data
    \item hvEEGNet effectively learns latent representations of multi-channel EEG data
    \item input data, including quality and individual variability, affects model’s training
    \item dataset 2a is corrupted by acquisition problems, causing models’ failures
\end{itemize}

\section{Introduction}
\label{sec:intro}

The first quantitative analysis of an \gls{eeg} signal dates back to the pioneering work of Hans Berger that, in the late Twenties (1929), took a Fourier transform of an \gls{eeg} signal to quantify the spectral distribution of the brain activity under different physiological and stimulation conditions~[\cite{berger1929eeg}].
% https://www.sciencedirect.com/topics/neuroscience/electroencephalogram-quantitative-analysis#:~:text=The%20first%20QEEG%20study%20was,of%20the%20electroencephalogram%20(EEG).
% https://www.sciencedirect.com/science/article/abs/pii/B9780123745347000022
%
% The first QEEG study was by Hans Berger (1932, 1934) when he used the Fourier transform to spectrally analyze the EEG, as he recognized the importance of quantification and objectivity in the evaluation of the electroencephalogram (EEG).
%
% The earliest quantitative EEG (QEEG) reference normative database was developed in the 1950s at UCLA as part of the NASA study and selection of astronaughts for purposes of space travel (Adey et al, 1961; 1964a; 1964b).
% https://www.appliedneuroscience.com/PDFs/History_of_QEEG_Databases.pdf
Since then, a vast literature flourished and obtained very successful achievements in the modeling and classification of \gls{eeg} data for different clinical and research applications~[\cite{teplan2002fundamentals}]. From the very first quantitative Berger's analyses, a number of different methods were proposed, with \gls{ml}-based models reaching the highest popularity for, e.g., pattern recognition, classification, and compression tasks~[\cite{hosseini2020review}]. Among many others, the classification of \gls{mi}, i.e., the brain activity corresponding to the imagination of moving one specific body segment, has been largely used in basic neuroscience to understand brain mechanisms~[\cite{kaiser2012relationship}], as well as to drive \gls{bci} systems~[\cite{kodama2023thirty}] and robots~[\cite{Beraldo2022}] to support neuro-rehabilitation.

Despite the large body of literature already produced, \gls{eeg} modelling still suffers from three major issues: (1) this particular time-series has a very fast dynamics (in the range of milliseconds) making it prone to interferences from many possible sources of noise, (2) it displays a high inter-subject as well as an inherent within-subject variability, and (3) when used in more ecological environments, poor reliability is often a problem.
Standard \gls{ml} models proved to be relatively good in several tasks~[\cite{hosseini2020review}], but they still lack the flexibility to generalize over different subjects or sessions, due to the rigid feature extraction step which is typically performed based on a-priori knowledge of the domain experts, or some simple (first or second-order) statistical description of the data. Also, there is still no gold-standard pre-processing to be applied. Finally, when models are embedded on portable and lower-quality \gls{eeg} devices for usage in more ecological settings, e.g., in new Internet of things scenarios~[\cite{munari2023local}] for continuous monitoring, their performance rapidly degrade~[\cite{anders2022wearable}]. Nonetheless, the state-of-the-art (SOTA) solutions for several processing tasks in \gls{eeg} are still based on standard \gls{ml}.

More recently, \gls{dl}-based models have been increasingly employed and could often outperform the SOTA methods. As an example, in the case of \gls{mi}, \gls{fbcsp} has been employed as a reference method for years~[\cite{ang2008filter}]. However, the so-called EEGNet \gls{dl}-based architecture, proposed in 2016~[\cite{EEGNet_paper}] and its later variants were able to achieve higher performance, with less pre-processing effort and no a-priori knowledge needed~[\cite{zancanaro_CIBCB_article}].
Nevertheless, the investigation on the potentialities of new architectures is still open~[\cite{congedo2018review}].
A critical issue in these methods is the dependency on the training set, as~\cite{Gyori_2022_data_distribution_impacts_ml} recently pointed out in the domain of magnetic resonance imaging data: a training dataset of poor quality, as well as a training set distributed in a non-representative way might induce biases in the final model and, consequently, to poor results in the task the model is expected to perform e.g., classification or anomaly detection. At the same time, in the neuroscience domain it is fairly difficult to certainly exclude the above-mentioned conditions~[\cite{pion2019iclabel}].

Then, to enhance models' capability in classifying, recognising anomalies as well as automatically denoising large \gls{eeg} datasets, training a \gls{dl} model to optimally reconstruct \gls{eeg} data and to provide an effective latent representation has been recently recognized by the literature as an effective pre-processing step.

% This poor performance can be attributed to several factors. The \gls{eeg} signals are often full of noise and artifacts, there is a lot of variability between data from different subjects (cross-subject variability), and in many cases even within data from the same subject (intra-subject variability). In addition, collecting large amounts of \gls{eeg} data can be a time-consuming process and publicly available datasets are often small and in different formats. Despite all these challenges many studies have been conducted on the application of \gls{dl} techniques to \gls{eeg} data to exploit their unique capabilities.

% One of the most studied topics for the application of \gls{dl} techniques is \gls{mi} classification, i.e. the classification of the imagination of movement.

% Thus, some fundamental challenges still emerge from the SOTA review to be solved, including (1) high reconstruction error or generation of traces that are not faithful to the original signal, (2) computational/architectural complexity of the proposed DL models, and (3) lack of distinctiveness of the architectural choices made for the EEG signal (e.g., GAN best for imaging).

In this paper, we propose \gls{vdtw} and \gls{hvdtw}, two \gls{dl}-based models consisting of a \gls{vae} architecture that aims at reconstructing the \gls{eeg} data with high-fidelity. In this process, a latent representation is obtained, which could then be used, e.g., to train a classifier or to generate new data samples. More specifically, \gls{vdtw} is an improvement over our previous architectures~[\cite{Zancanaro_2023_vEEGNet, Zancanaro_CCIS}] with a significant modification of the loss function, while \gls{hvdtw} represents our best model consisting in a hierarchical version of the \gls{vae} which allowed us to achieve almost perfect reconstruction of the \gls{eeg} data.

\section{State of the art} \label{sec:state_of_the_art}

To contextualize our study, we report here that related work that addressed both the reconstruction of \gls{eeg} signals via \gls{dl}, with the extraction of a latent representation for this kind of data, and those which proposed autoencoder models to detect anomalies in \gls{eeg} data.

% \textcolor{gray}{\emph{
% IPOTESI DI FONDO: alcuni meccanismi/pattern sono comuni tra gli individui, ma poi esiste una variability INDIVIDUALE che is importante e che can portare INFO significativa=KNOWLEDGE, sia in termini di patologia che in termini di specificity individuale. However, bisogna distinguerla da variability dovuta ad artefatti.
% The two most important ones are the high temporal resolution (many high-frequency components), as well as the significant variability, both between different individuals and within the same one.
%
% The ability to model and extract the most important elements of the data requires an architecture that can accurately recreate an \gls{eeg} signal. An architecture like this may thus serve as the foundation for a wide range of operations, including classification, pattern recognition, denoising, generation, and compression~\cite{globecom2018}.
% }}
% % The use of autoencoder to compress \gls{eeg} signals to facilitate their transmission is studied in~\cite{Marridi2018_AE_reconstruction_d2a}

% \hl{Reconstruction of the EEG signal}

% There is a vast body of literature suggesting \gls{ml} and \gls{dl} models and architectures to reconstruct \gls{eeg} signals. 
% 
One of the most interesting works on the topic is~\cite{Bethge_2022_EEG2VEC}, where the authors proposed EEG2VEC, i.e., a \gls{vae}-based architecture developed to encode emotions-related \gls{eeg} signals in the latent space of the \gls{vae}.
The authors succeeded in reconstructing low-frequency components of the original \gls{eeg} signals by using the \gls{vae} latent representations. Unfortunately, the higher frequency components could not be reconstructed. Moreover, the reconstructed signals appeared to be largely attenuated (with amplitudes often in the order of half the original one). According to the authors' explanation,  this can be due to the particular design of the decoder which might have introduced aliasing and artefacts.
%\AZ{According to the authors, this can be partially explained by the upsampling components of the decoder which introduce aliasing and noise artefacts} \hl{PER AZ: COME SPIEGANO GLI AUTORI QUESTA COSA? Nel pape riportano una singola frase: This can be partly explained by the up-sampling components of the generator network which introduce aliasing frequency artifacts [32] as well as noise artifacts in the inputs are not captured}.
%
These results are in line with those we found in our previous work~[\cite{Zancanaro_2023_vEEGNet}]: we proposed a new \gls{dl}-based architecture named \emph{vEEGNet-ver1}, where we used EEGNet (a popular architecture that has tailored a \gls{cnn} to specifically process \gls{eeg} data~[\cite{EEGNet_paper}]), as an encoder and its mirrored architecture as decoder in a \gls{vae} model. We evaluated its classification and reconstruction performance on a public dataset (containing \gls{mi}-related \gls{eeg} data) and found that only low-frequency components could be recovered, while achieving state-of-the-art performance in classification. In contrast with \cite{Bethge_2022_EEG2VEC}, we were able to explain this sub-optimal behaviour as the clear effect of the filters applied at the first block of the architecture: they simply have the effect of smoothing the signal, so that the information related to higher frequency components is not further propagated along the \gls{dl} network, thus making them not available anymore for the reconstruction. Nevertheless, we recognized the reconstructed low-frequency component as the \gls{mrcp}, a well-known neurophysiological behaviour associated with movements initiation~[\cite{Bressan2021, MullerPutz2019}].
In~\cite{Marridi2018_AE_reconstruction_and_compression_d2a}, the authors implemented a convolutional autoencoder to compress and reconstruct \gls{mi}-\gls{eeg} signals (from two public datasets, including the one used in our own work). They evaluated the trade-off between the compression ratio, computed as the ratio between the size of the raw signal and the size of the autoencoder latent representation, and the reconstruction quality, measured in terms of percent root mean square distortion (PRD). The authors proved the good ability of the convolutional autoencoder to reconstruct single channel \gls{eeg} signals with a relatively high compression ratio, e.g., compression ratio up to $98\%$ with PRD of $1.33\%$. However, the authors did not discuss about the quality of the dataset and reported representative performance, only. Thus, no further insights on the relationship between model training, reconstruction performance and input data quality can be retrieved.
%
% Fig.4. represents the relation between the compression ratio and distortion rate (PRD) using CAE model on the BCIIV-2b dataset. The model was able to compress up to 93% of the data and reconstruct with a distortion of 12%. However, the designed CAE model for BCI-IV-2a was able to compress 98% of the data with 1.33% distortion; 
%
%
%
% \AZ{
\cite{Dasan_2022_ECG_EMG_EEG_compression_and_reconstruction} proposed a multi-branch denoising autoencoder to jointly compress \gls{eeg}-ECG-EMG signals, assuming them acquired in a mobile-health scenario with the aim to ensure continual learning, i.e., continuous fine tuning using incoming data during real-time health monitoring. Each signal modality (\gls{eeg}, ECG, EMG) was independently pre-processed, and then a joint latent representation was obtained to compress the signals. The authors showed the trade-off between compression ratio and reconstruction quality using three (independent) public datasets. They provided an example of reconstructed \gls{eeg}, EMG, and ECG signal, where the reconstruction appeared to be very reliable. However, the targeted \gls{eeg} signal was acquired from one only sensor using a portable device, i.e., the signal was of low quality and poorly variable, thus most probably making the reconstruction easier. Finally, they used different metrics to quantify the quality of the reconstructed signal, e.g., the reconstruction quality index. However, it was defined w.r.t. the compression ratio, thus not applicable to other reconstruction-targeting scenarios. 
%
% } % https://link.springer.com/article/10.1007/s00034-022-02071-x
% proposed a multi-branch denoising autoencoder to jointly compress \gls{eeg}-ECG-EMG signals, assuming them acquired in a mobile-health scenario, i.e., from wearabless attache on the head, the chest, and one lower limb muscle of an individual, respectively. The data were supposed to be acquired with low quality and that one sensor (or channel), only, was available for each modality. The authors aimed to show the good trade-off between a high compression ratio and a high reconstruction quality, to ensure continual learning, i.e., continuous fine tuning using incoming data during real-time health monitoring. In the proposed model, each signal modality (\gls{eeg}, ECG, EMG) is independently pre-processed, and then a joint latent representation is obtained.
% They work with single channel data and each branch is responsible for a specific type of signal. The focus was to develop a model that can work in a Si, so che non ci sono le performance ma il punto è che non riportano una metrica unica. Per ogni tipo di dato riportano multipli risultati per diversi valori di SNR. 
%
%
% \AZ{
\cite{khan_2023_autoenceder_seizures} used a shallow autoencoder to obtain an encoded representation with low dimensionality ($8$ to $64$) of a single-channel \gls{eeg} data to be used in the classification of epileptic versus healthy \gls{eeg} data (using a \gls{knn} and a \gls{svm} classifier and a public dataset~[\cite{Tran_2022_chb_mit_dataset}]). They achieved very high values for the accuracy (over $97\%$), with very high sensitivity (mostly over $96\%$) as well as specificity (over $96\%$). Also, the reconstruction quality was showed in two representative \gls{eeg} signals, with very high fidelity. Unfortunately, the authors did not report the power spectrum of the original \gls{eeg} signals, thus making it difficult to fully ensure a reproducibility of these good performance on other, more complex (i.e., with larger bandwidth), \gls{eeg} data. Also, the proposed architecture was proved to be very efficient in a channel-wise reconstruction: nevertheless, in many applications, multiple channels should be processed altogether. Then, further investigations should be needed to explore a more compact solution to obtain an encoded representation from all available \gls{eeg} channels.
% } %https://www.mdpi.com/1424-8220/23/8/4112
%
% Questo studio si potrebbe citare anche dopo perchè è molto simile ad anomaly detection come concetto
% Similar to other works [~\cite{Bethge_2022_EEG2VEC,Zancanaro_2023_vEEGNet}] they manage to reconstruct the low frequency parts of the signal but they fail to rebuild the details associated with high frequency components. Furthermore, they used the latent representation from the autoencoder as input for a \gls{svm} classifier.

% \hl{AE as anomaly detector.}

Moreover, autoencoders offer the inherent possibility to be used as anomaly detectors~[\cite{Guansong_2021_anomaly_detection_review}]: in fact, they are trained, in an unsupervised way, to learn the distribution of the \emph{normal} data that corresponds to the optimal reconstruction performance. After training, the model is used to reconstruct any new sample: when the latter is sufficiently out of the expected distribution, the model shows a large reconstruction error, signalling an anomaly.
Autoencoders-based anomaly detection has been proved effective also for \gls{eeg} data, when the latter are affected by different kinds of pathologies. In~\cite{emami_2019_autencoder_epileptic_seizure}, an autoencoder was used to detect epileptic seizures on a private dataset with 24 subjects~[\cite{emamai_2019_dataset}]. To identify anomalies they set a threshold on the reconstruction error and the \gls{eeg} samples exceeding it were labelled as anomalous. A $100\%$ accuracy in seizure detection could be obtained in 22 subjects out of 24. In~\cite{ortiz_2020_dyslexia_autoencoder_anomaly}, the authors used an autoencoder-based architecture to detect dyslexia on a public dataset~[\cite{vos_2017_dataset_used_for_ortiz_2020}]. They first extracted a number of \gls{eeg} features (in time and frequency domain), and then trained an autoencoder to reconstruct the time-series of such features. The difference, i.e., the residual, between the input and the reconstructed time-series was used as to feed an \gls{svm} classifier, aimed at distinguishing between healthy and dyslexic individuals. This solution achieved an accuracy of $96\%$, sensitivity of $86\%$, specificity of $100\%$, \gls{auc} of $92\%$.
%\AZ{They extract a list of temporal, spatial and frequency features from the signal for each channel and for five different bands (Delta, Theta, Alpha, Beta, Gamma)}
%
% \AZ{In \cite{ortiz_2020_dyslexia_autoencoder_anomaly} c'erano soggetti patologico e soggetti di controllo. In \cite{emami_2019_autencoder_epileptic_seizure} da quanto ho capito hanno classificato segmenti dello stesso soggetto: alcuni con seizure e altri pezzi clean/healthy.}
In both cases, the model training relied only on data from healthy and clean \gls{eeg} data. After training, the \gls{eeg} samples corresponding to the largest reconstruction error values were labelled as anomalous. % Besides, this automatic labelling was validated by experts, and the authors reported a match between the two different labelling procedure of \hl{su Emami YES, su Ortiz in base a diagnosi dei diversi soggetti}.

However, in the case of \gls{eeg} data, the definition of \emph{normality} can be very challenging: it is fairly difficult to have a certified clean dataset, even though the subjects are healthy. In fact, an \gls{eeg} sample could be considered as anomalous both for the presence of a pathology, but also because of any noise and interference that might occur during recordings (e.g., the so-called \emph{artefacts})~[\cite{gabardi2023multi}].
Therefore, it would be more realistic to train an autoencoder model on a mixture of clean and noisy data, in line with some other literature (not necessarily addressing biological data). For instance, in~\cite{zhou_2017_robust_autoencoder}, the authors proposed a \emph{robust autoencoder}, i.e., a combination of a \gls{rpca} and an autoencoder, where the autoencoder was used for data projection in the (reduced) principal components space (in place of the usual linear projection). Unfortunately, there is a limited literature on this kind of autoencoders, as confirmed by a recent survey~[\cite{al2021review}].
%
% \hl{PER AZ: CITA QUI I 3 ARTICOLI SCELTI ASSIEME: (1) spiking NN, (2) OFAT, (3) AE DALLA REVIEW https://www.mdpi.com/2076-3417/11/12/5320}
In~\cite{xing_2020_spiking_nn_anomalies}, the authors proposed a combination of an evolving spiking neural network and a Boltzmann machine to identify anomalies in a multimedia data stream. The proposed training algorithm was able to localize and ignore any random noise that could corrupt the training data. % according to the authors 
%Another important advantage of the proposed algorithm is the ability to isolate and reject the random noise that may be contained in the training set.
%
\cite{wambura_2020_OFAT_NN_timeseries} suggested to jointly use a \gls{cnn} and a \gls{lstm} to forecast future trends and reconstruct past trends of different types of data stream. They were able to accurately predict time-series related to three real-world scenarios, i.e., web traffic in Wikipedia, price trends of the avocado fruit, and temperature series in a city. %but they did not handle noisy samples.
In~\cite{dong_2018_threaded_autoencoder_streaming}, a model composed by an ensemble of autoencoders was employed to identify anomalies in data streams. The authors claimed that the training algorithm made the presence of noisy samples in the training data not statistically significant, thus ensuring model's robustness to noise. % IPOTESI = buffer con 1 sample noisy NON ARTEFATTUALE
In~\cite{Qiu_2019_time_series_anomaly_detection_classifier}, an architecture made by the sequence of a \gls{cnn}, an \gls{lstm}, a \gls{ffnn}, and a softmax layer was proposed to identify anomalies. Interestingly, a \gls{vae} was preliminarily used to over-sample the dataset, before training the classifier (i.e., the \gls{ffnn} with the softmax layer). The model was tested on the AIOps-KPI public dataset~[\cite{li_2022_dataset_kpi}], achieving an accuracy of $77\%$ (KP1), $75\%$ (KP2), $83\%$ (KP3), and $75\%$ (KP4). %Penso siano suddivisioni del dataset, non ho cercato oltre.
%
%
%
% Ho trovato una review (https://www.mdpi.com/2076-3417/11/12/5320) dove fanno riferimento anche ad anomaly detection su noisy data. Nello specifico c'è la tabella 4 della review che riporta un po' di paper dove propongono modelli per lavorare con noisy data.
%
% \AZ{Otherwise a possible alternative is the use of autoencoder variants designed to work directly with non-clean data, e.g. the robust autoencoder~\cite{zhou_2017_robust_autoencoder}.
%

Nevertheless, to the best of our knowledge, this kind of approaches has never been applied to \gls{eeg} data, yet.

% However, limitations remain with this approach. The main one is the assumption that the dataset used to carry out the training does not contain outliers, an assumption that is difficult to verify in very large datasets or with complex data such as \gls{eeg}.

% \hl{AE=FEATURES LEARNER FOR CLASSIFICATION. AZ: CHECK THE RESULTS FROM THE WORKS BELOW}

Another body of literature presents \gls{dl} architectures that are not specifically targeted to reconstruct \gls{eeg} data, but to extract, via autoencoders, the most relevant features to improve classification.
% These works are not interested in the reconstruction capability itself but in the features that the model can learn through the reconstruction training. 
%
In~\cite{yang_2018_DSAE_EEG}, the authors proposed a denoising sparse autoencoder, i.e., an autoencoder that imposes a sparsity condition on the latent space, as a feature learner and then used the learnt representation as the input of a linear regression model to detect different types of epileptic seizures. They successfully tested their solution on a public dataset~[\cite{yang_2018_DSAE_EEG_dataset}] to classify either two \gls{eeg} classes, i.e., healthy vs epileptic, or three classes, i.e., healthy, ictal and inter-ictal \gls{eeg} data (in both cases, the reported accuracy was $100\%$). 
% According to the authors this would lead the autoencoder to learn a more efficient representation of the \gls{eeg} signal in latent space.
% The learned features are then classified using a linear regression. They managed to achieve 100\% accuracy on a dataset collected by the Epileptology department of the Bonn University~[\cite{yang_2018_DSAE_EEG_dataset}], both in binary and 3-class classification. \AZ{Per essere precisi e' ictal classification. Le 3 classi sono normal, interictal and ictal. In binary: normal vs (inter-ictal+ictal)}
%
In~\cite{Wang_2020_EEGNet_autoencoder_1}, an EEGNet-based autoencoder was employed as a feature extractor from a high-density \gls{eeg} device aimed at acquiring evoked potentials~[\cite{lascano2017clinical}] during a repeated pain stimulation with a laser at different energy levels. Here, the authors opted for an intense pre-processing which included a filtering step in the frequency range $1$ to $30$~Hz, segmentation in $1.5$~s-epochs, each one taken from $0.5$~s before the stimuls until $1$~s after it, baseline correction (using the pre-stimulus period as the baseline), \gls{ica} decomposition to remove eye-movement related artifacts, and down-sampling from $1000$ to $250$~Hz. The features extracted from the autoencoder fed four different \gls{ml} classifiers, i.e., a \gls{knn}, a \gls{svm}, a \gls{lda}, and a logistic regression model, with the aim of detecting $10$ levels of pain. Interestingly, several latent space sizes were tested, and the size of $64$, used with a logistic regression classifier, resulted as the best solution. An accuracy of $74.6\pm11.2\%$ was obtained (with a chance level of $10\%$), thus largely outperforming the alternative solution using \gls{pca} as feature extractor (accuracy equal to $59.9\pm19\%$).
% AZ: was obtained with a latent space of 64 elements combined with a logistic regression classifier (the authors tested also different latent space sizes and other classifiers always reporting performance lowering).
% \AZ{On the dataset they collected, they achieved a $74.6\pm11.2\%$ accuracy, with $10$ different level of pain. For comparison, the authors used the same pipeline with PCA as a feature extractor, obtaining an accuracy $59.9\pm19\%$.}
%
In~\cite{liu_2020_eeg_emotion_SAE_CNN}, a \gls{cnn} was used as feature extractor from \gls{eeg}, i.e., the output of the network was used to feed an autoencoder with the goal of compressing and reconstructing the original \gls{eeg} data. The \gls{cnn} and the autoencoder were trained together, with the autoencoder forcing the \gls{cnn} to extract the most significant features. After training, the obtained features were fed to a \gls{ffnn} that acted as a classifier and it was separately trained. They tested the model on the DEAP~[\cite{koelstra_2012_DEAP_dataset}] and the SEED~[\cite{zheng_2015_SEED_dataset}] public emotions-related \gls{eeg} datasets. In the DEAP dataset, they achieved an accuracy of 89.49\% in the valence category and an accuracy of 92.86\% in the arousal one (each category has 2 labels, high and low). For the SEED dataset, they achieved an accuracy 96.77\%, on the 3 classes of the dataset (positive, neutral, negative).

% \hl{CONCLUSIONI SOTA E PREPARAZIONE A NOSTRO CONTRIBUTO:}

Thus, some fundamental challenges still emerge from the SOTA review to be solved including (1) high reconstruction error or generation of traces that are not faithful to the original signal, (2) lack of focus on the reconstruction even if the architecture have the capacity to favor classification taks, and (3) no dedicated investigations on the impact of the input \gls{eeg} data quality on the training of \gls{dl} models. %computational/architectural complexity of the proposed \gls{dl} models.

\section{Materials and Methods} \label{sec:methods}

In this section, we present the basic modules as well as the overall architecture of our proposed models, i.e., vEEGNet-ver3 and hvEEGNet. Furthermore, we describe the metrics and the methodologies we employed to evaluate our models.

\subsection*{ Variational autoencoder }
The common overall architecture of our both models is the \gls{vae}.

Unlike traditional autoencoders, i.e., producing a deterministic encoding for each input, \gls{vae} is able to learn a probabilistic mapping between the input data and a latent space, which is additionally learned as a structured latent representation~[\cite{VAE_ORIGINAL_PAPER_kingma,kingma2019introduction}].
Given the observed data $\mathbf{x}$ and assuming $\mathbf{z}$ to be the latent variables, with a proper training, a \gls{vae} learns the variational distribution $q_\phi(\mathbf{z} | \mathbf{x})$ as well as the generative distribution $p_\theta(\mathbf{x} | \mathbf{z})$, using a pair of (deep) neural networks (acting as the encoder and the decoder), parametrized by $\mathbf{\phi}$ and $\mathbf{\theta}$, respectively~[\cite{blei2017variational}].
The training loss function, denoted as $\mathcal{L}_{VAE}$, accounts for the sum of two different contributions: the Kullback-Leibler divergence between the variational distribution $q_\phi(\mathbf{z} | \mathbf{x})$ and the posterior distribution $p_\theta(\mathbf{x}|\mathbf{z})$, denoted as $\mathcal{L}_\mathcal{KL}$, and the reconstruction error, denoted as $\mathcal{L}_{R}$, which forces the decoded samples to approximate the initial inputs.
Thus, the loss function adopted for the \gls{vae} is
\begin{equation}
    \mathcal{L}_{VAE} = \mathcal{L}_{KL} + \mathcal{L}_{R} = - \mathcal{KL} [q_{\phi}(\mathbf{z}|\mathbf{x}) || p(\mathbf{z})] + \EX_q(p_\theta(\mathbf{x}|\mathbf{z}))
     \label{eq:vae}
\end{equation}
where $\mu_i$ and $\sigma^2_i$ are the predicted mean and variance values of the corresponding $i$-th latent component of $\mathbf{z}$.
Assuming normal distribution as a prior for the sample distribution in the latent space, it is possible to rewrite eq.~\ref{eq:vae} as follows
% $\mathcal{L}_{KL}$ in closed form as
\begin{equation}
    \mathcal{L}_{VAE} = - \frac{1}{2}\sum_{i = 1}^{d}(\sigma^2_i+ \mu^2_i - 1 - log(\sigma^2_i)) + \EX_q(p_\theta(\mathbf{x}|\mathbf{z}))
\end{equation} \label{e:vae_extension}
where $\mu_i$ and $\sigma^2_i$ are the predicted mean and variance values of the corresponding $i$-th latent component of $\mathbf{z}$.

In this work, we adopted this basic architecture to propose \gls{vdtw}. The details characterizing our specific \gls{vae} are reported in Section~\ref{subsec:veegnet3}.

\subsection*{ Hierarchical VAE}

A hierarchical \gls{vae}~[\cite{vahdat_2021_nvae}] is the evolution of a standard \gls{vae} enriched by a hierarchical latent space, i.e., multiple layers implementing a latent space each. In fact, standard \gls{vae}s suffer from the lack of accuracy in details reconstruction, given by the trade-off between the reconstruction loss and the Kullback-Leibler divergence contributions, thus generating the tendency to generate slightly approximated data (e.g., blurred images), only. 
Hierarchical \gls{vae}s attempt to solve this problem by using multiple latent spaces, where each of them is trained to encode different levels of detail in the input data.
Assuming a model with $L$ latent spaces, its loss function can be written as
\begin{equation}
    \mathcal{L}_{HVAE} = \mathcal{L}_{KL} + \mathcal{L}_{R} 
    \label{eq:hvae}
\end{equation} 
where
\begin{equation}
    \mathcal{L}_{KL} =  - \mathcal{KL} [q_{\phi}(\mathbf{z_1}|\mathbf{x}) || p(\mathbf{z_1})] - \sum_{l= 2}^{L} \mathcal{KL} [q_{\phi}(\mathbf{z_l}|\mathbf{x},\mathbf{z_{<l}} ) || p(\mathbf{z_l}|\mathbf{z_{<l}})],
\end{equation}
with $q_{\phi}(\mathbf{z_l}|,\mathbf{z_{<l}} ) = \prod_{i = 1}^{l - 1}q_{\phi}(\mathbf{z_l}|\mathbf{x},\mathbf{z_{<i}})$ as the approximate posterior up to the $(l-1)$ level and the conditional in each prior $p(z_l|z_{<l})$ and approximate posterior $q_{\phi}(\mathbf{z_l}|\mathbf{x},\mathbf{z_{<i}})$  is represented as a factorial normal distribution. The symbol $\mathbf{z}_{<i}$ is taken from the original paper~[\cite{vahdat_2021_nvae}] and it means that the random variable is conditioned by the output of all latent spaces from $1$ to $i$.

In this work, we adopted this basic architecture to propose \gls{hvdtw}. The details characterizing our specific hierarchical \gls{vae} are reported in Section~\ref{subsec:hveegnet}.

\subsection{ vEEGNet - ver3} \label{subsec:veegnet3}

\begin{figure}[htbp!]
    \centering
    \includegraphics[width = 0.7\textwidth]{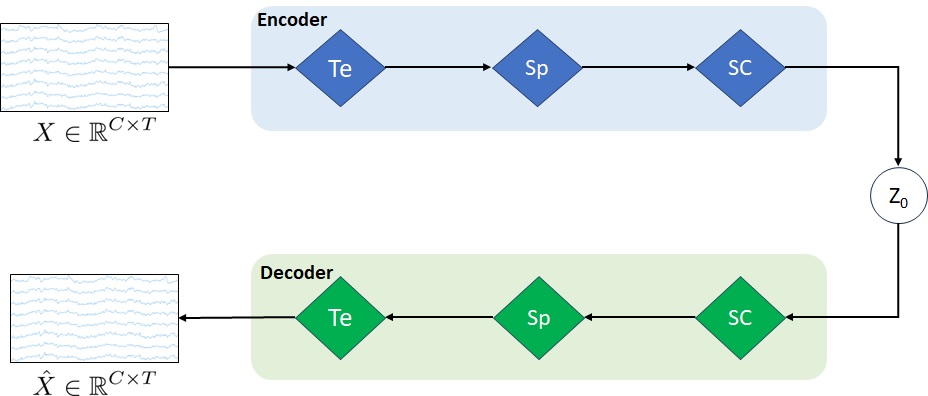} 
    \caption{Schematic architecture of our model called \textit{vEEGNet-ver3}. The encoder block is formed by three blue diamonds representing three different processing layers: i.e., Te stands for \emph{temporal convolution}, Sp stands for \emph{spatial convolution}, and SC stands for \emph{separable convolution}. The decoder block includes three green diamonds representing the same operations, in the reverse order. $z_0$ represents the latent space.}     \label{fig:veegnet3}
\end{figure}

Fig.~\ref{fig:veegnet3} represents the schematic architecture of this simple \gls{vae} model. As any conventional \gls{vae}, it consists of an encoder, a latent space, and a decoder.

However, inspired by the work of~\cite{EEGNet_paper}, we designed the encoder as the popular EEGNet architecture, i.e., with the three processing blocks: in the first block, a horizontal convolution (that imitates the conventional temporal filtering) is followed by a batch normalization. In the second block, a vertical convolution, acting as a spatial filter, is applied. This operation is then followed by an activation and an average pooling step. The third, and last, block performs a separable convolution with a horizontal kernel, followed by an activation and an average pooling step. We always used, as in~\cite{EEGNet_paper}, the \gls{elu} activation function. At the output of the third block, the obtained $D \times C \times T$ tensor is further transformed by means of a sampling layer which applies a convolution with a $1 \times 1$ kernel, thus doubling its depth size, resulting in a $2D \times C \times T$ tensor. 
Finally, the latter is projected onto the latent space (i.e., of dimension $N = D \cdot C \cdot T$).
In line with other previous works~[\cite{Zancanaro_2023_vEEGNet, VAE_ORIGINAL_PAPER_kingma}], the first $N$ elements of the depth map were intended as the marginal means ($\mathbf{\mu}$) and the second $N$ elements as the marginal log-variances ($\mathbf{\sigma}$) of the Gaussian distribution represented in the latent space.
Then, to reconstruct the EEG data, the latent space $\mathbf{z_0}$ is sampled using the reparametrization trick, as follows:
$$
    \mathbf{z_0} = \mathbf{\mu} + \mathbf{\sigma} \cdot \mathcal{N}(\mathbf{0}, \mathbf{1}),
$$
where $ \mathcal{N}(\mathbf{0}, \mathbf{1})$ is standard multivariate Gaussian noise (with dimension $N = D \cdot C \cdot T$).

To note, this architecture is very similar to other previous architectures proposed by the authors in~\cite{Zancanaro_2023_vEEGNet} and in~\cite{Zancanaro_CCIS}. However, it introduces a few significant novelties that leads this new model to perform much better than the older ones. 
%
% DTW
% REF: https://medium.com/walmartglobaltech/time-series-similarity-using-dynamic-time-warping-explained-9d09119e48ec#:~:text=DTW%20compares%20amplitude%20of%20first,similar%20shape%20and%20different%20phase
The most relevant novelty is that the reconstruction error $\mathcal{L}_{R}$ of the \gls{vae} loss function expressed by eq.~\ref{eq:vae} was here quantified by the \gls{dtw} similarity score~[\cite{Sakoe_1978_DynamicPA_DTW_Paper}], i.e., replacing the more standard \gls{mse}. \gls{dtw} leads to a more suitable measure of the similarity between two time-series~[\cite{DTWcorrelation2012}], thus allowing the model better learn to reconstruct EEG data. In fact, \gls{dtw} is known to be more robust to non-linear transformations of time-series~[\cite{DTWonEEG1985}], thus capturing the similarity between two time-series even in presence of time shrinkage or dilatation, i.e., warpings. This cannot be achieved by \gls{mse}, which is highly sensitive to noise, i.e., the error computed by \gls{mse} rapidly increases when small modifications are applied to time-series.

In brief, given two time-series $a(i)$ and $b(j)$, where $i,j = 1,2, ..., T$ (i.e., for simplicity, we consider two series with the same length), \gls{dtw} is a time-series alignment algorithm that extensively searches for the best match between them, by following a five-step procedure:
\begin{enumerate}
    \item The cost matrix $\mathcal{W}$ is initialized, with each row $i$ associated with the corresponding amplitude value of the first time-series $a(T-i+1)$, while each column $j$ associated with the corresponding amplitude value of the second time-series $b(j)$.
    \item Starting from position $\mathcal{W}(0,0)$, the value of each matrix element is computed as
          $     \mathcal{W}(i,j) = |a(i) - b(j)| + \min[\mathcal{W}(i-1,j-1), \mathcal{W}(i,j-1), \mathcal{W}(i-1,j) ]      $, if $i,j>0$, otherwise $\mathcal{W}(i,j) = |a(i) - b(j)|.$
    \item The optimal warping path is identified as the minimum cost path in $\mathcal{W}$, starting from the element $\mathcal{W}(1,T)$, i.e., the upper right corner, ending to the element $\mathcal{W}(T,1)$.
    \item the array $d$ is formed by taking the values of $\mathcal{W}$ included in the optimal warping path. Note that $d$ might have a different (i.e., typically longer) length compared to the two original time-series, as a single element of one series could be associated with multiple elements of the other.
    \item Finally, the \emph{normalized} \gls{dtw} score is computed as $$ score = \frac{ \sum_{k=1}^K d(k) }{K}  \label{eq:dtw}$$ where $K$ is the length of the array $d$. To note, normalization was not applied during the models' training (to keep this contribution in the range of the other loss function contributions). Whereas, during the performance evaluation, we used the normalized score. Nevertheless, this difference did not induce criticisms, as all segments share the same length.
    % NOI NON NORMALIZZIAMO DURANTE IL TRAINING
\end{enumerate}

Finally, the projection onto the latent space was also modified w.r.t. our previous implementations: earlier, the tensor obtained by the convolutional layers was flattened into a vector and projected through a \gls{ffnn} into a $2N$-size vector, with $N$ the dimension of the latent space. Then, the first $N$ elements were interpreted as mean values and the second $N$ elements as the log-variance values, respectively, of the distribution encoded in the latent space. Now, we apply a $1 \times 1$ convolution to the output of the encoder, thus obtaining a depth map whose first half is taken as the mean and the second half as the log-variance of the distribution of the latent space.

\subsection{hvEEGNet} \label{subsec:hveegnet}
\begin{figure}[htbp!]
    \centering
    \includegraphics[width = 0.7\textwidth]{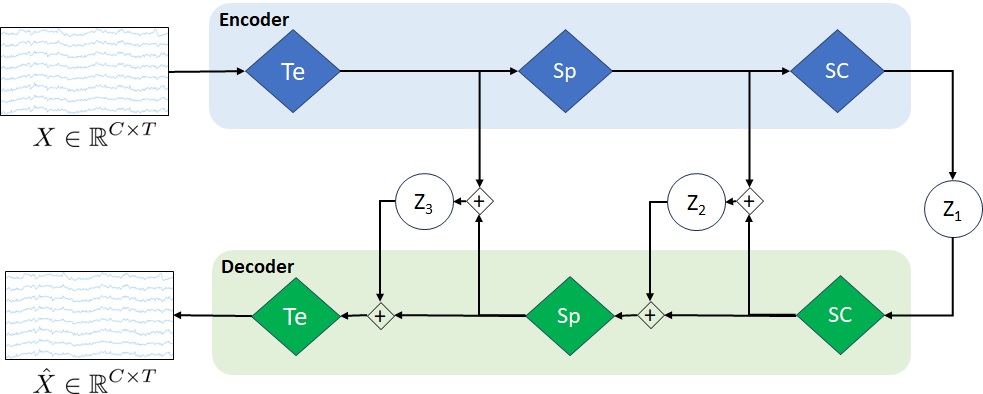}    %{Figure/hvEEGNet.jpg}
    \caption{Schematic architecture of our model called \textit{hvEEGNet}. The encoder block is formed by the same three processing layers (i.e., blue diamonds) as in \gls{vdtw} (Fig.\ref{fig:veegnet3}). The decoder block includes three green diamonds representing the same operations, in the reverse order. $z_1$, $z_2$, and $z_3$ represent the latent spaces obtained at the three different processing levels.}
    \label{fig:hvEEGNet}
\end{figure}

As we observed sub-optimal reconstruction results with \gls{vdtw} and in line with other literature on computer vision~[\cite{vahdat_2021_nvae}], we developed a new architecture, called \gls{hvdtw}, to overcome the remaining issues of \gls{vdtw}. 
The most relevant change in \gls{hvdtw} w.r.t. \gls{vdtw} is its hierarchical architecture with three different latent spaces, namely $z_1$, $z_2$, and $z_3$, with $z_1$ being the deepest one.
Each of them is located at the output of each main block of the encoder, i.e., after the temporal convolution (Te) block ($z_3$), after the spatial convolution (Sp) block ($z_2$), and after the separable convolution (SC) block ($z_1$).
The input to the decoder's Sp block is now given by the linear combination (i.e., the sum) of the SC block's output and the sampled data from $z_2$. Similarly, the input to the decoder's Te block is obtained by the sum of the Sp block's output and the sampled data from $z_3$.
Incidentally, but significantly, it is worth noting that we kept here using the \gls{dtw} algorithm to compute the reconstruction loss $\mathcal{L_R}$ (with reference to eq.~\ref{eq:dtw}).
The main structure of the model is depicted in Fig.~\ref{fig:hvEEGNet}.

%% Tabella parametri architetture proposte
Table~\ref{tab:model_parameters} reports the details of both \gls{vdtw} and \gls{hvdtw} architectures.
\begin{table}[htbp!]
\centering
\caption{Parameters' values of the \gls{vdtw}'s and \gls{hvdtw}'s encoder, respectively. In. dep. = Input depth. Out. dep. = Output depth.}
\label{tab:model_parameters}
\resizebox{\textwidth}{!}{%
\begin{tabular}{|cc|cccc|cccc|}
\hline
\multicolumn{2}{|c|}{}                                                                                                                                                                        & \multicolumn{4}{c|}{\cellcolor[HTML]{C0C0C0}\textbf{vEEGNet-ver3}}                & \multicolumn{4}{c|}{\cellcolor[HTML]{C0C0C0}\textbf{hvEEGNet}}                    \\
\multicolumn{2}{|c|}{\multirow{-2}{*}{\textbf{Parameters name}}}                                                                                                                              & \textbf{Kernel} & \textbf{In. dep.} & \textbf{Out. dep.} & \textbf{Notes}         & \textbf{Kernel} & \textbf{In. dep.} & \textbf{Out. dep.} & \textbf{Notes}         \\ \hline
\multicolumn{1}{|c|}{}                                                                                                            & Convolution 2d                                            & (1, 128)        & 1                 & 8                  & Depth-wise convolution & (1, 128)        & 1                 & 8                  & Depth-wise convolution \\
\multicolumn{1}{|c|}{\multirow{-2}{*}{\textbf{\begin{tabular}[c]{@{}c@{}}First   block \\ (temporal filter)\end{tabular}}}}       & Batch Norm 2d                                             & -               & -                 & -                  & Default parameters     & -               & -                 & -                  & Default parameters     \\ \hline
\multicolumn{1}{|c|}{}                                                                                                            & Convolution 2d                                            & (1, 22)         & 8                 & 16                 & Depth-wise convolution & (1, 22)         & 8                 & 16                 & Depth-wise convolution \\
\multicolumn{1}{|c|}{}                                                                                                            & Batch Norm 2d                                             & -               & -                 & -                  & Default Parameters     & -               & -                 & -                  & Default Parameters     \\
\multicolumn{1}{|c|}{}                                                                                                            & Activation                                                & -               & -                 & -                  & ELU                    & -               & -                 & -                  & ELU                    \\
\multicolumn{1}{|c|}{}                                                                                                            & \begin{tabular}[c]{@{}c@{}}Average\\ pooling\end{tabular} & (1, 4)          & -                 & -                  & -                      & -               & -                 & -                  & No pooling used        \\
\multicolumn{1}{|c|}{\multirow{-5}{*}{\textbf{\begin{tabular}[c]{@{}c@{}}Second   block\\ (spatial filter)\end{tabular}}}}        & Dropout                                                   & -               & -                 & -                  & p = 0.5                & -               & -                 & -                  & p = 0.5                \\ \hline
\multicolumn{1}{|c|}{}                                                                                                            & Convolution 2d                                            & (1, 32)         & 16                & 16                 & Depth-wise convolution & (1, 32)         & 16                & 16                 & Depth-wise convolution \\
\multicolumn{1}{|c|}{}                                                                                                            & Convolution 2d                                            & (1, 1)          & 16                & 16                 & Pointwise convolution  & (1, 1)          & 16                & 16                 & Pointwise convolution  \\
\multicolumn{1}{|c|}{}                                                                                                            & Batch Norm 2d                                             &                 &                   &                    & Default parameters     &                 &                   &                    & Default parameters     \\
\multicolumn{1}{|c|}{}                                                                                                            & Activation                                                & -               & -                 & -                  & ELU                    & -               & -                 & -                  & ELU                    \\
\multicolumn{1}{|c|}{}                                                                                                            & \begin{tabular}[c]{@{}c@{}}Average\\ pooling\end{tabular} & (1, 8)          & -                 & -                  & -                      & (1, 10)         & -                 & -                  & -                      \\
\multicolumn{1}{|c|}{\multirow{-6}{*}{\textbf{\begin{tabular}[c]{@{}c@{}}Third   Block \\ (Separable Convolutoin)\end{tabular}}}} & Dropout                                                   & -               & -                 & -                  & p = 0.5                & -               & -                 & -                  & p = 0.5                \\ \hline
\multicolumn{1}{|c|}{\textbf{Sample layer}}                                                                                       & Convolution 2d                                            & (1,1)           & 16                & 32                 & Pointwise convolution  & (1,1)           & 16                & 32                 & Pointwise convolution  \\ \hline
\end{tabular}%
}
\end{table}

\begin{table}[]
\caption{Parameters' values of the \gls{vdtw}'s and \gls{hvdtw}'s decoder, respectively. In. dep. = Input depth. Out. dep. = Output depth.}
\label{tab:model_parameters_DECODER}
\resizebox{\textwidth}{!}{%
\begin{tabular}{|cc|cccc|cccc|}
\hline
\multicolumn{2}{|c|}{}                                                                                                                                                                                 & \multicolumn{4}{c|}{\cellcolor[HTML]{C0C0C0}\textbf{vEEGNet-ver3}}                & \multicolumn{4}{c|}{\cellcolor[HTML]{C0C0C0}\textbf{hvEEGNet}}                    \\ \cline{3-10} 
\multicolumn{2}{|c|}{\multirow{-2}{*}{\textbf{Parameters name}}}                                                                                                                                       & \textbf{Kernel} & \textbf{In. dep.} & \textbf{Out. dep.} & \textbf{Notes}         & \textbf{Kernel} & \textbf{In. dep.} & \textbf{Out. Dep.} & \textbf{Notes}         \\ \hline
\multicolumn{1}{|c|}{}                                                                                                            & Dropout                                                            & -               & -                 & -                  & p = 0.5                & -               & -                 & -                  & p = 0.5                \\
\multicolumn{1}{|c|}{}                                                                                                            & Upsample                                                           & (1, 8)          & -                 & -                  & -                      & (1, 10)         & -                 & -                  & -                      \\
\multicolumn{1}{|c|}{}                                                                                                            & Activation                                                         & -               & -                 & -                  & ELU                    & -               & -                 & -                  & ELU                    \\
\multicolumn{1}{|c|}{}                                                                                                            & Batch Norm 2d                                                      &                 &                   &                    & Default parameters     &                 &                   &                    & Default parameters     \\
\multicolumn{1}{|c|}{}                                                                                                            & \begin{tabular}[c]{@{}c@{}}Transpose\\ Convolution 2d\end{tabular} & (1, 1)          & 16                & 16                 & Pointwise convolution  & (1, 1)          & 16                & 16                 & Pointwise convolution  \\
\multicolumn{1}{|c|}{\multirow{-6}{*}{\textbf{\begin{tabular}[c]{@{}c@{}}Third   Block \\ (Separable Convolutoin)\end{tabular}}}} & \begin{tabular}[c]{@{}c@{}}Transpose\\ Convolution 2d\end{tabular} & (1, 32)         & 16                & 16                 & Depth-wise convolution & (1, 32)         & 16                & 16                 & Depth-wise convolution \\ \hline
\multicolumn{1}{|c|}{}                                                                                                            & Dropout                                                            & -               & -                 & -                  & p = 0.5                & -               & -                 & -                  & p = 0.5                \\
\multicolumn{1}{|c|}{}                                                                                                            & Upsample                                                           & (1, 4)          & -                 & -                  & -                      & -               & -                 & -                  & No pooling used        \\
\multicolumn{1}{|c|}{}                                                                                                            & Activation                                                         & -               & -                 & -                  & ELU                    & -               & -                 & -                  & ELU                    \\
\multicolumn{1}{|c|}{}                                                                                                            & Batch Norm 2d                                                      & -               & -                 & -                  & Default parameters     & -               & -                 & -                  & Default parameters     \\
\multicolumn{1}{|c|}{\multirow{-5}{*}{\textbf{\begin{tabular}[c]{@{}c@{}}Second   block\\ (spatial filter)\end{tabular}}}}        & \begin{tabular}[c]{@{}c@{}}Transpose\\ Convolution 2d\end{tabular} & (1, 22)         & 8                 & 16                 & Depth-wise convolution & (1, 22)         & 8                 & 16                 & Depth-wise convolution \\ \hline
\multicolumn{1}{|c|}{}                                                                                                            & Batch Norm 2d                                                      & -               & -                 & -                  & Default parameters     & -               & -                 & -                  & Default parameters     \\
\multicolumn{1}{|c|}{\multirow{-2}{*}{\textbf{\begin{tabular}[c]{@{}c@{}}First   block\\ (temporal filter)\end{tabular}}}}        & \begin{tabular}[c]{@{}c@{}}Transpose\\ Convolution 2d\end{tabular} & (1, 128)        & 1                 & 8                  & Depth-wise convolution & (1, 128)        & 1                 & 8                  & Depth-wise convolution \\ \hline
\end{tabular}%
}
\end{table}

\subsection{Outlier identification} \label{subsec:outlier}

As our architectures implement completely self-supervised models, we have the opportunity to use them as anomaly detectors. In line with the vast majority of related work (as introduced in Section~\ref{sec:state_of_the_art}), in the present study, we define as an \emph{outlier} any sample (i.e., \gls{eeg} segment) that is very poorly reconstructed. This, in turn, is verified by large values of the \gls{dtw} similarity score between the reconstructed \gls{eeg} sample and the original one.
To identify such samples, we decided to use the \gls{knn} algorithm~[\cite{kNN1967}]. \gls{knn} is an unsupervised \gls{ml} algorithm that computes the distance between every sample and its $k$-th nearest neighbour (with $k$ properly chosen). All samples in the dataset are sorted w.r.t. increasing values of such distance. Those points whose distance (from their $k$-th nearest neighbour) exceeds a user-defined \emph{threshold} are labelled as outliers.

In this study, we used our \gls{hvdtw} model, only, to identify the outliers. Before applying \gls{knn}, we performed two pre-processing steps: 
we computed the \gls{dtw} similarity score for the \gls{eeg} segment (i.e., channel- and repetition-wise). To note, by definition (see eq.~\ref{eq:dtw}), the score is normalized by the number of time points in the series (even though all time-series have fixed length in this work). For each training run, we built the following matrix  $\mathbf{E}$:
\begin {equation}
\mathbf{E}(t) =   \left[
                    \begin{matrix}
                         e_{11}(t)  & e_{12}(t) & \cdots & e_{1C}(t) \\
                         e_{21}(t)  & \ddots    & \cdots & \vdots    \\
                         \vdots     & \vdots    & \ddots & \vdots    \\
                         e_{R1}(t)  & \cdots    & \cdots & e_{RC}(t) \\
                    \end{matrix}
                \right]
                \label{eq_correlation}
\end{equation}
with $t = 1,2, ..., T$, given $T$ the number of training runs, $r = 1,2, ..., R$, with $R$ the number of segments (i.e., task repetitions), and $c = 1,2, ..., C$, with $C$ the number of \gls{eeg} channels. Then, $e_{rc}(t)$ represents the normalized \gls{dtw} value obtained from the reconstruction of the $r$-th segment at the $c$-th channel after training the \gls{hvdtw} model in the $t$-th training run.
Finally, the matrices $\mathbf{E}(t)$, with $t = 1,2, ..., T$ are averaged to obtain $\overline{\mathbf{E}}$, and then \gls{knn} is applied.
Also note that \gls{knn} took every sample of the dataset as defined by a $C$-dimensional \gls{eeg} segment (i.e., one row in matrix $\overline{\mathbf{E}}$). This allowed us to identify two types of outliers: (1) repetitions where all (or, the majority of the) channels were affected by some mild to severe problem, or (2) repetitions where only one (or, a few) channel was highly anomalous. Both are very common situations that might occur during neuroscience experiments~[\cite{teplan2002fundamentals}].

\subsection*{Implementation} \label{subsec:implementation}

We employed PyTorch to implement all pre-processing steps and to design and train our models.

To implement the newly proposed loss function, i.e., including the \gls{dtw} computation, we exploited the soft-\gls{dtw} loss function CUDA time-efficient implementation\footnote{Available at: https://github.com/Maghoumi/pytorch-softdtw-cuda}~[\cite{maghoumi2020dissertation_softDTWCUDA_1, maghoumi2021deepnag_softDTWCUDA_2}]. In fact, the original \gls{dtw} algorithm is quite time-consuming and employs a minimum function that is non-differentiable.
Then, in~\cite{Cuturi_2017_SoftDTWAD}, a modification of the original algorithm was proposed to specifically be used in \gls{dl} models, i.e., to be differentiable, thus suitable as a loss function. Then, CUDA was employed to make it time-efficient, too.
Also, it is worth noting that \gls{dtw} works with 1D time-series. However, our models aimed to reconstruct multi-channel \gls{eeg} time-series. Then, during training, we computed the channel-wise \gls{dtw} similarity score between the original and the reconstructed \gls{eeg} segment. Then, in the loss function, we added the contribution coming from the sum of all channel-wise \gls{dtw} scores.

The models were trained using the free cloud service offered by Google Colab, based on Nvidia Tesla T4 GPU.
The hyperparameters were set as follows: batch size to $30$, learning rate to $0.01$, the number of epochs to $80$, an exponential learning rate scheduler with $\gamma$ set to $0.999$.
$20$ training runs for each subject, were performed, in order to better evaluate the stability of the models training and the error trend along the epochs.
The total number of parameters of the \gls{vdtw} is $4992$ and the state dictionary (i.e., including all parameter weights) is $\SI{40}{\kilo \byte}$-weight.
The total number of parameters of the \gls{hvdtw} model is 8224, with $5456$ of them to define the encoder, and the remaining $2768$ for the decoder. Note that the higher number of parameters in the decoder is due to the sampling layers that operate on the three different latent spaces. The state dictionary of the parameter weights is about $\SI{56}{\kilo \byte}$.

Finally, for the \gls{knn} algorithm for outliers detection (see Section~\ref{subsec:outlier}), we employed the well-known \emph{knee method} in the implementation given by the kneed python package~[\cite{kneed_package}] to find the \emph{threshold distance} to actually mark some samples as outliers.

To foster an \emph{open science} approach to scientific research, we made our code available on GitHub~\footnote{Available at: https://github.com/jesus-333/Variational-Autoencoder-for-EEG-analysis}.

\subsection{Performance evaluation} \label{subsec:evaluation}

In this work, we evaluated our models in a within-subject scenario~[\cite{zancanaro_CIBCB_article}], only. Cross-subject evaluations, even though possible, are left for future developments as they deserve an entire new campaign of experiments and analyses.

The evaluation was carried on based on two different approaches: first, visual inspection of the reconstructed data in both the time and frequency domains (with the most convenient frequency range selected figure by figure); second, the quantification of the average reconstruction quality using the normalized \gls{dtw} similarity score, as defined in eq.~\ref{eq:dtw}.

For visual inspection, we compared in a single plot the time domain representations of the original \gls{eeg} segment and its corresponding reconstructed one. Also, we computed the Welch's spectrogram~[\cite{welch1967}] (in the implementation provided by the Python Scipy package\footnote{Available at: \url{https://docs.scipy.org/doc/scipy/reference/generated/scipy.signal.welch.html}}) with the following parameters: Hann's window of $500$ time points with $250$ time points overlap between consecutive segments.

Then, to train and test our models (both \gls{vdtw} and \gls{hvdtw}), we inherited the same split proposed by~\cite{dataset_BCI_competition}: for every subject, $50\%$ data were used for the training and the remaining $50\%$ (i.e., a later experimental session) for the test. Furthermore, we applied cross-validation using $90\%$ of the training data for the actual models' training and $10\%$ for the validation.
With the aim of investigating the training behaviour of our models w.r.t. the particular input dataset, we repeated $20$ training runs for each subject (i.e., each run started from a different random seed, thus ensuring a different training/validation split in the overall training set). This allowed us to provide a more robust evaluation of the training curve along the training epochs.
We reported the models' performance in terms of descriptive statistics (mean and standard deviation across multiple training runs) of the reconstruction error along the training epochs (i.e., in other words, the training time).
To note, in some rare cases where the loss function's gradient could not be minimized, we excluded those training runs from our final evaluation and training visualization.

Finally, the reconstruction ability of our models, after proper training (i.e., $80$ epochs), was evaluated on the test set, too, by means of the same normalized \gls{dtw} similarity score.

\section{Results and Discussion} \label{sec:results}

%% --- DATASET --- 
\subsection{Dataset} \label{subsec:dataset}
\begin{figure}
    \centering
    \includegraphics[width = \linewidth]{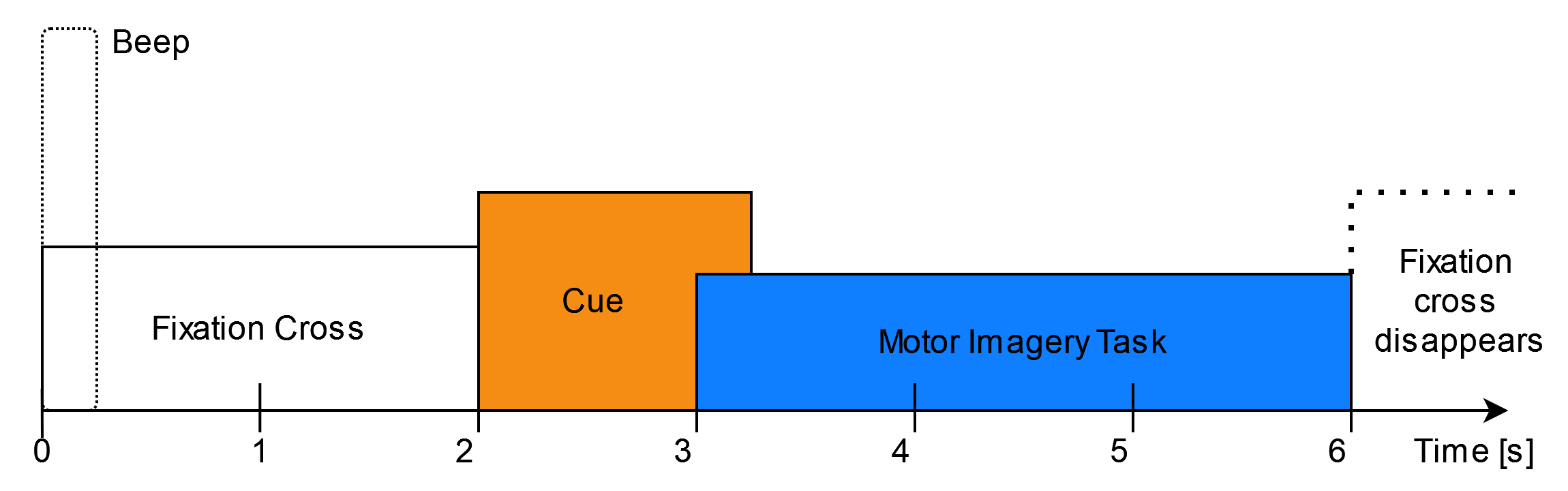}
    \caption{The timeline of the experimental paradigm (modified from \cite{dataset_BCI_competition}).}
    \label{fig:dataset_paradigm}
\end{figure}

To validate our model we used the Dataset 2a of BCI Competition IV~[\cite{dataset_BCI_competition}]. The dataset was downloaded using the MOABB tool~[\cite{moabb_2018}] and it is composed by the 22-channel \gls{eeg} recordings of $9$ subjects while they repeatedly performed four different \gls{mi} tasks: imagining the movement of the right hand, left hand, feet or tongue. Each \emph{repetition} consists of about \SI{2}{\second} fixation cross task, where a white cross appeared on a black screen and the subject needed to fix it and relax (as much as possible). Then, a \SI{1.25}{\second} cue allowed the subject to start imagining the required movement. The cue was displayed as an arrow pointing either left, right, up, or down, to indicate the corresponding task to perform, i.e., either left hand, right hand, tongue, or feet \gls{mi}. \gls{mi} was mantained until the fixation cross disappeared from the screen (for \SI{3}{\second}). A random inter-trial interval of a few seconds was applied (to avoid subjects habituation and expectation). Then, several repetitions of each type of \gls{mi} were required to be performed. The order to repeat the different \gls{mi} tasks was randomized to avoid habituation. The timeline of the experimental paradigm is depicted in Fig.~\ref{fig:dataset_paradigm}.
A total number of $576$ trials (or, repetitions) was collected from each individual subject and made available as a public dataset, namely the \emph{Dataset 2a of BCI Competition IV}.
The \gls{eeg} data were recorded with a sampling frequency of \SI{250}{\hertz} and the authors filtered the data with a $0.5-100\SI{}{\hertz}$ band-pass filter and a notch filter at \SI{50}{\hertz} (accordingly to the experimental records associated with the public dataset). We kept these settings as they were, to be in line with the literature~[\cite{EEGNet_paper}] and to be consistent with our previous studies~[\cite{zancanaro_CIBCB_article, Zancanaro_2023_vEEGNet, Zancanaro_CCIS}].

As explained in Section~\ref{subsec:evaluation}, we adopted the pre-defined 50/50 training/test split on the dataset and thus, for each subject, we obtained $260$ \gls{eeg} segments for the training set, $28$ for the validation set, and $288$ for the test set.
To note, the dataset was perfectly balanced in terms of stratification of the different subjects in all splits.

\begin{figure}[htbp!]
    \centering
    \begin{subfigure}[b]{\textwidth}
        \includegraphics[width=0.49\textwidth]{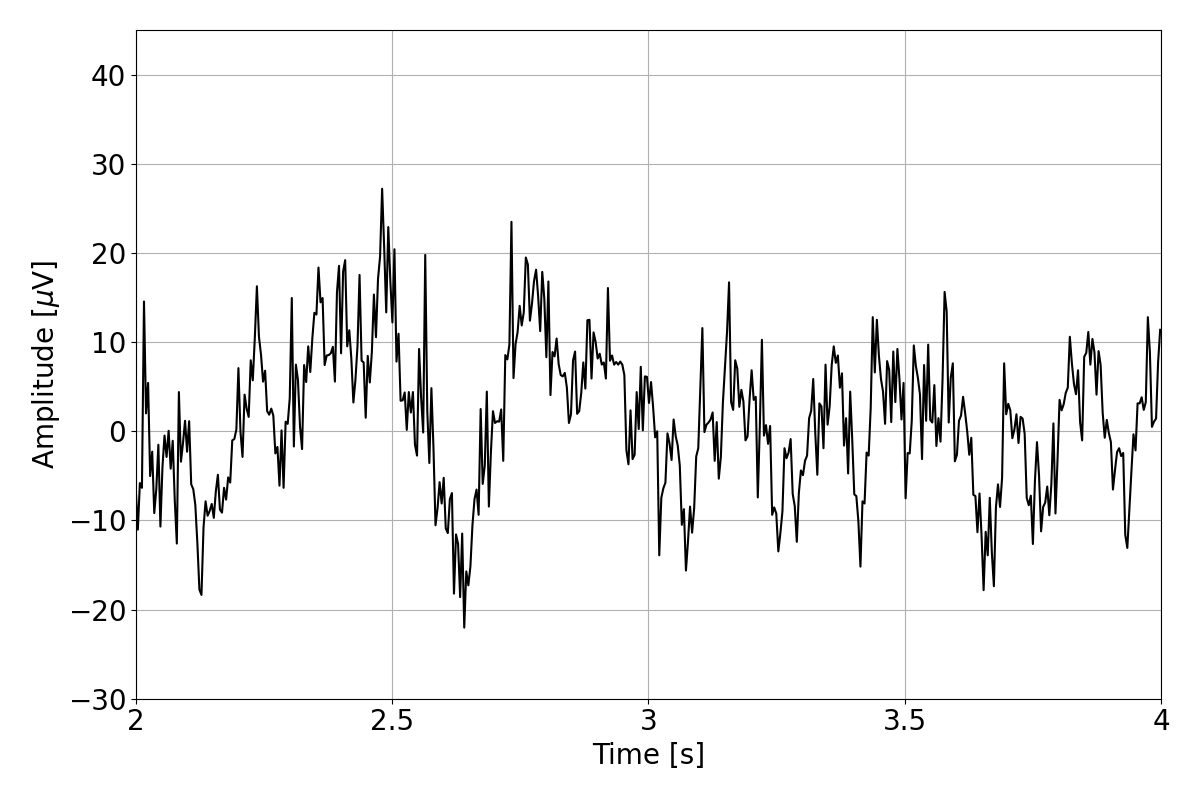}
        \hfill
        \includegraphics[width=0.49\textwidth]{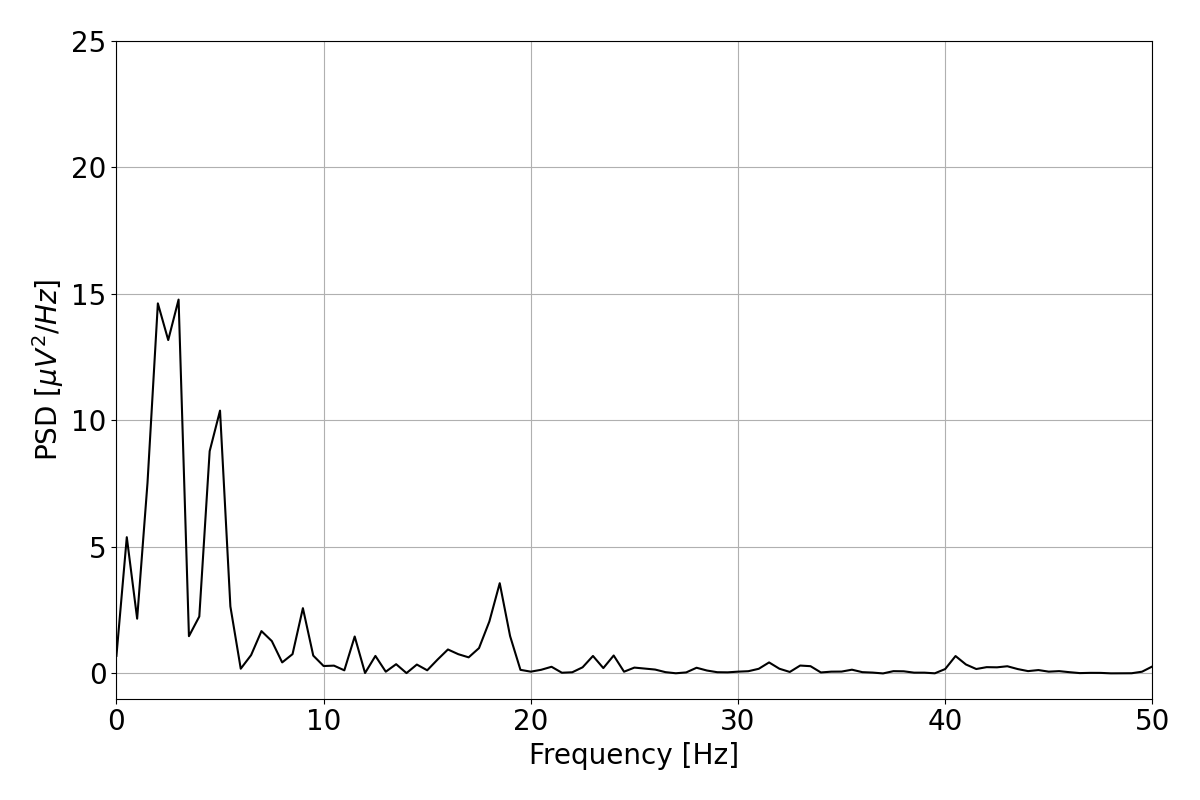}
        \caption{Segment no.$1$, ch. C4, left hand movement.}
        \label{fig:example_trials_time_1}
    \end{subfigure}

    \begin{subfigure}[b]{\textwidth}
        \includegraphics[width=0.49\textwidth]{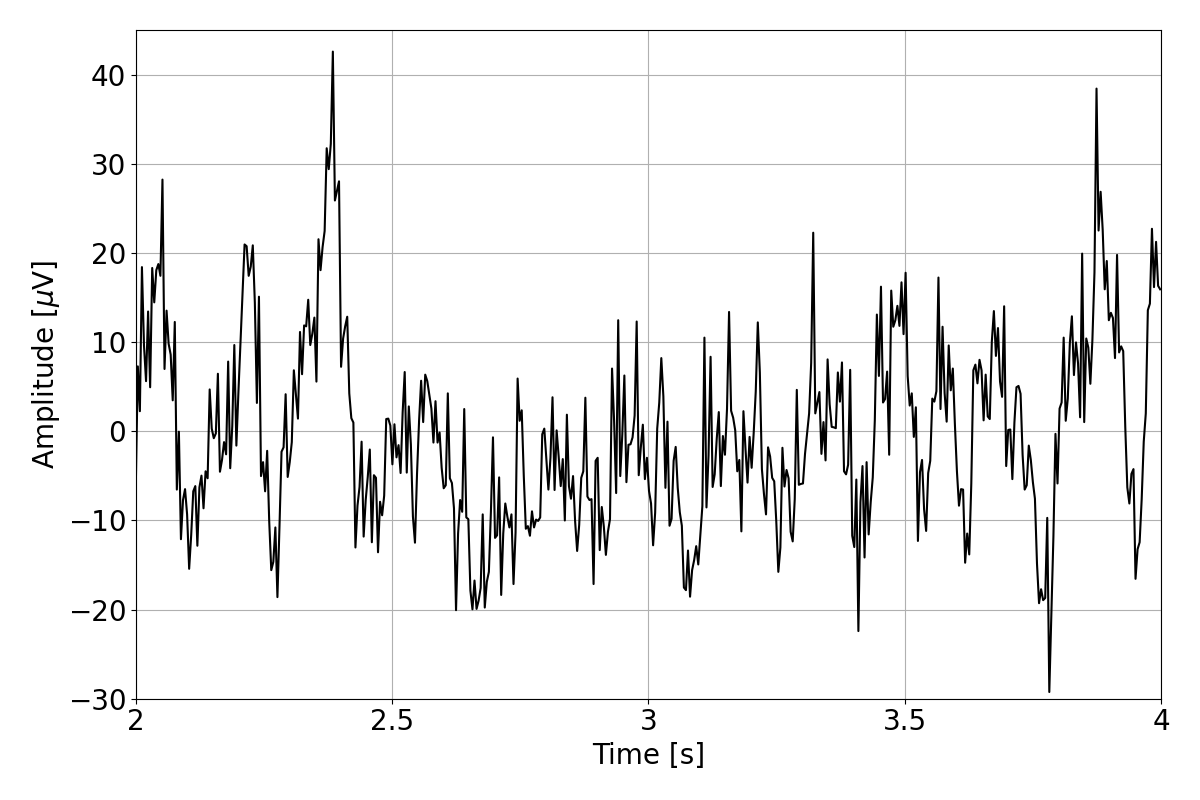}
        \hfill
        \includegraphics[width=0.49\textwidth]{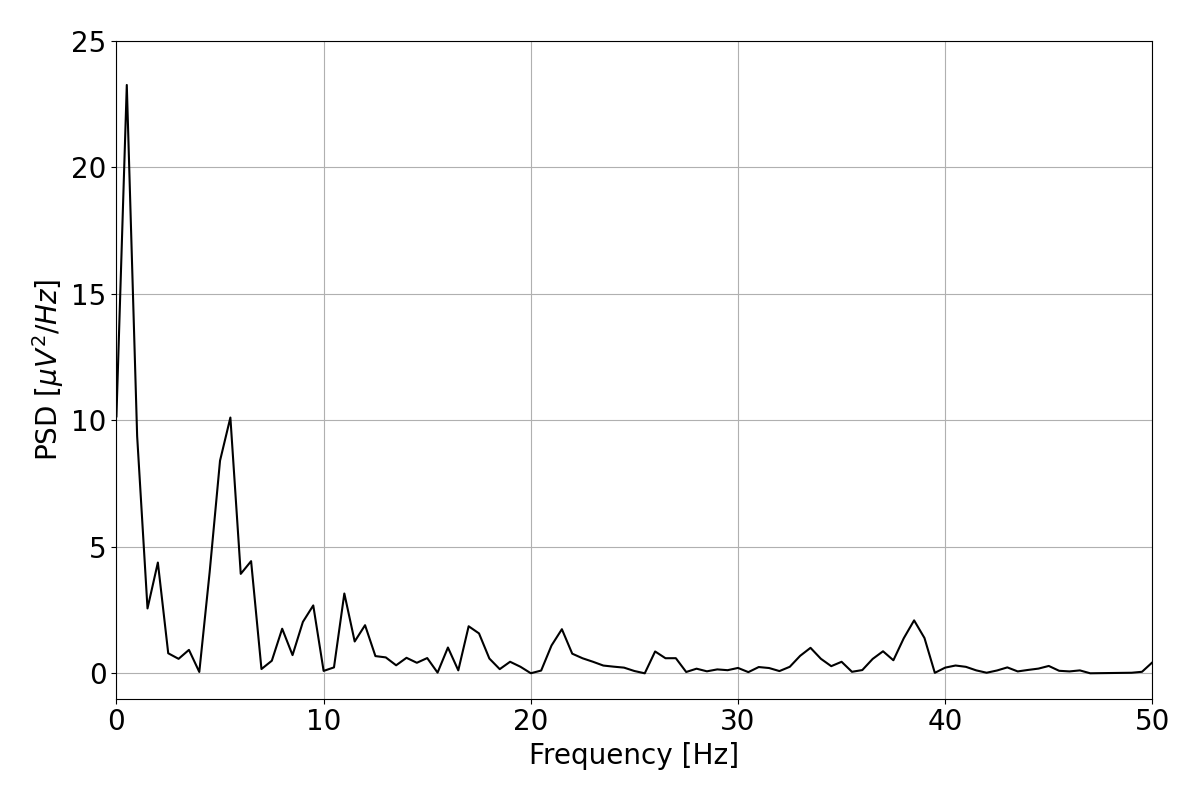}
        \caption{Segment no.$10$, ch. Fz, tongue movement.}
        \label{fig:example_trials_time_2}
    \end{subfigure}

    \caption{Two representative \gls{eeg} segments from the Dataset 2a from S3 (time range limited from \SI{2}{\second} to \SI{4}{\second}, frequency range limited to \SI{50}{\hertz}). Left panels: time domain representation. Right panels: frequency domain representation.}
    \label{fig:example_trial}
\end{figure}

We performed segmentation and, for each \gls{mi} repetition, we extracted a $\SI{4}{\second}$ (22-channel) \gls{eeg} \emph{segment}. The piece of \gls{eeg} was selected in the most \emph{active} \gls{mi} part of the repetition, i.e., from $2$ to $\SI{6}{\second}$, in order to isolate the most apparent brain behaviour related to the \gls{mi} process. Fig.~\ref{fig:example_trial} shows an example of two raw \gls{eeg} signals, represented both in the time domain and in the frequency domain (with the frequency range limited to $50$~Hz for visualization purposes). To note, to improve visualization in the time domain, the signals are shown in the limited time range from $2$ to $\SI{4}{\second}$. However, in the frequency domain, the entire \SI{4}{\second} segment was used to compute the power spectrum (via Welch method, as described in Section~\ref{subsec:evaluation}).
Then, a total of $1000$ time points are included in each \gls{eeg} segment.

\subsection{Reconstruction performance}
\label{subsec:vEEGNet_vs_hvEEGNet}

In this section we show the performance of \gls{vdtw} and \gls{hvdtw}, and we discuss to what extent the new loss function (with the \gls{dtw} contribution) and the hierarchical architecture influenced the reconstruction performance.

\begin{figure}[htbp!]
    \begin{subfigure}[b]{0.99\textwidth}
        \includegraphics[width=\textwidth]{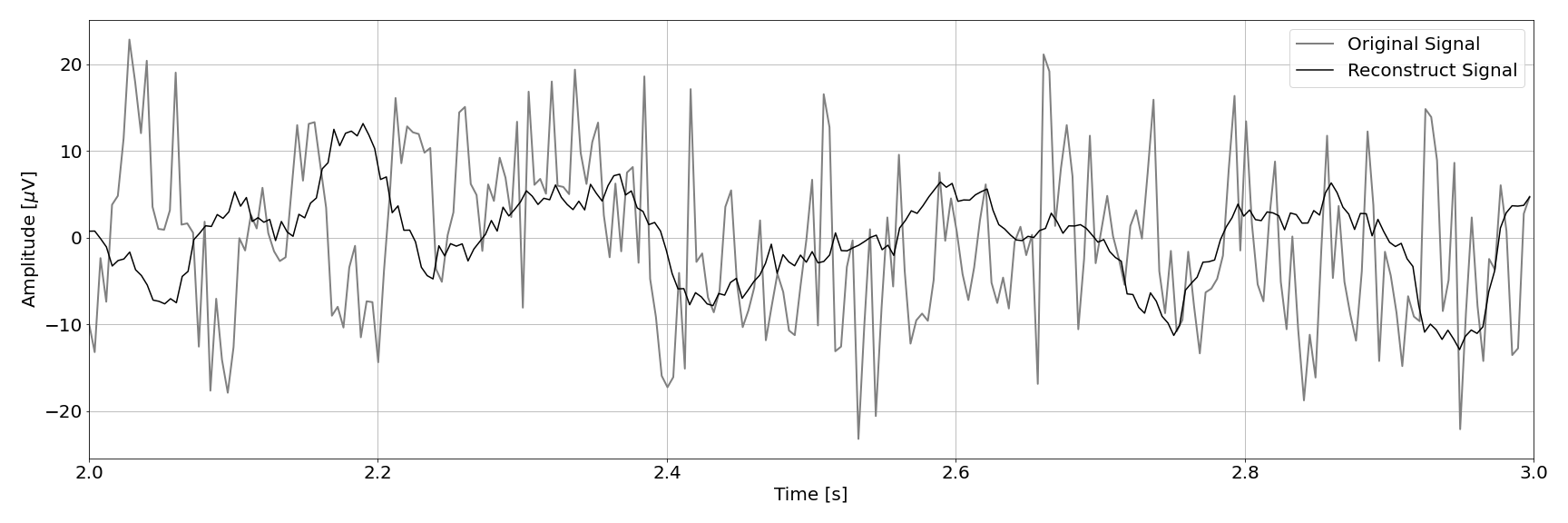}
        \caption{Reconstruction via \gls{vdtw}.}
        %\label{fig:vEEGNet_recon_time}
    \end{subfigure}

    \begin{subfigure}[b]{0.99\textwidth}
        \includegraphics[width=\textwidth]{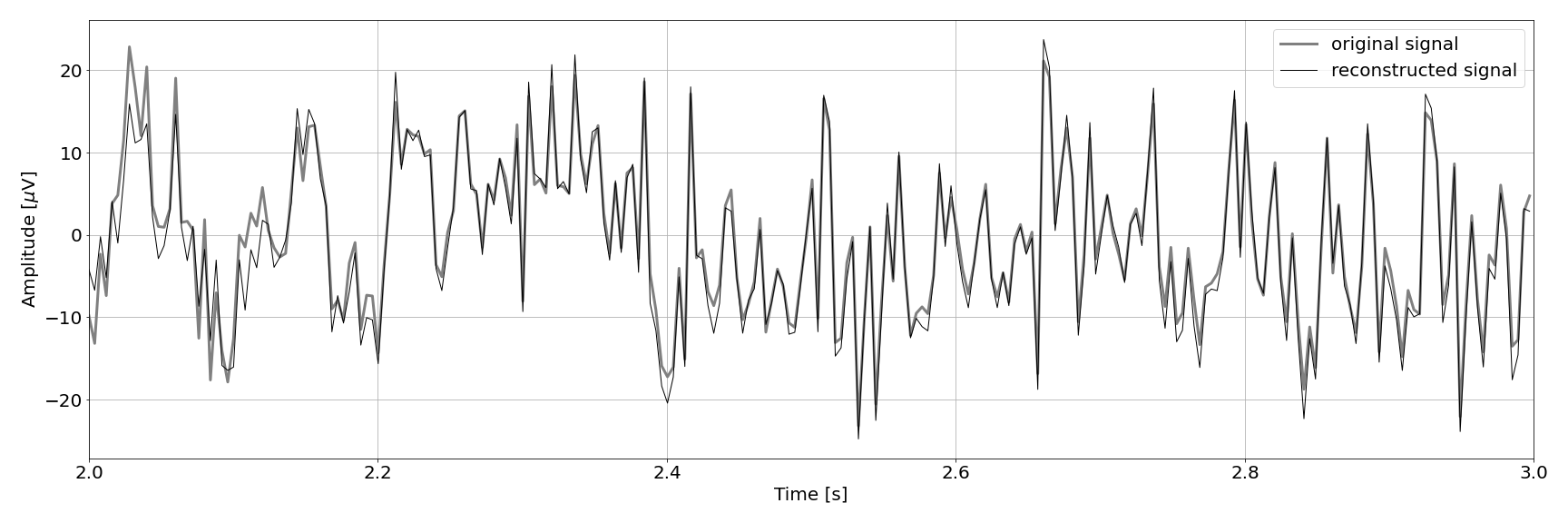}
        \caption{Reconstruction via \gls{hvdtw}.}
        %\label{fig:hvEEGNet_recon_time}
    \end{subfigure}

    \begin{subfigure}[b]{0.49\textwidth}
        \includegraphics[width=\textwidth]{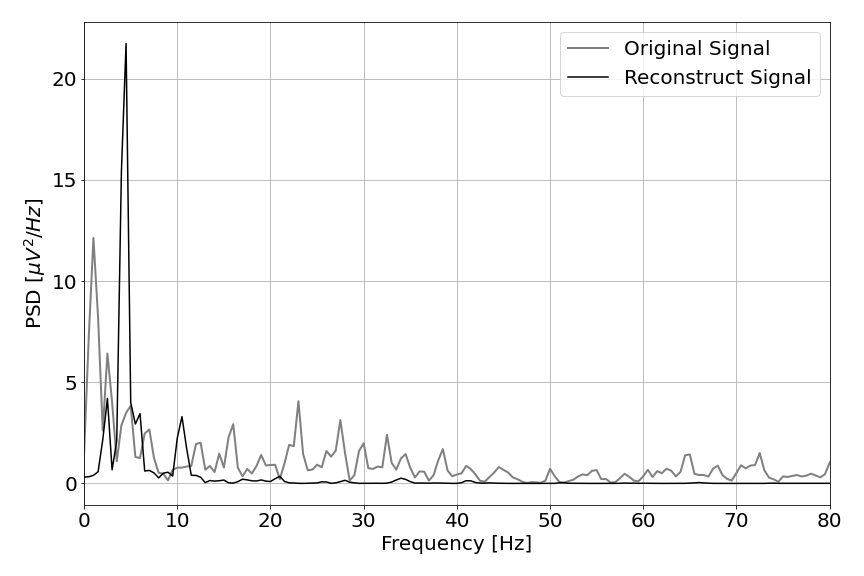}
        \caption{Frequency domain \gls{vdtw}.}
        %\label{fig:vEEGNet_recon_freq}
    \end{subfigure}
    \begin{subfigure}[b]{0.49\textwidth}
        \includegraphics[width=\textwidth]{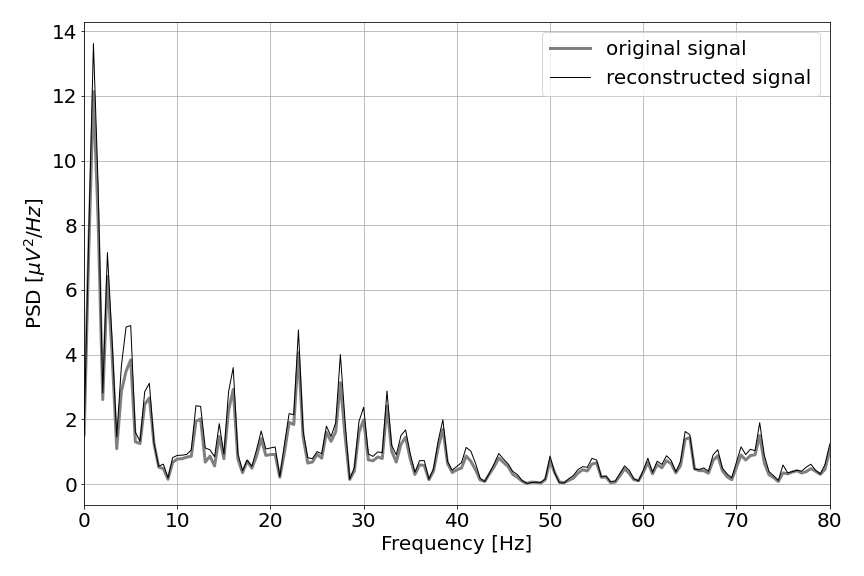}
        \caption{Frequency domain \gls{hvdtw}.}
        %\label{fig:hvEEGNet_recon_freq}
    \end{subfigure}

    \caption{Comparison of the reconstruction performance in the time and frequency domain (in the test phase) in a representative subject (S3), task repetition (no.$1$, corresponding to \gls{lh} \gls{mi}), channel (C3). Time range was limited from \SI{2}{\second} to \SI{3}{\second}, frequency range extended to \SI{80}{\hertz}.} 
    \label{fig:vEEGNet_vs_hvEEGNet}

\end{figure}

First, we visually inspect the output from our two models. Fig.~\ref{fig:vEEGNet_vs_hvEEGNet} shows a representative example of one \gls{eeg} segment as reconstructed by \gls{vdtw} and \gls{hvdtw}, respectively, in both the time and the frequency domain (with the frequency range extended to $80$~Hz for visualization purposes). 
%
% dtw contribution
As it can be observed, \gls{hvdtw} is much better in reconstructing the \gls{eeg} segment, and this can be very clearly appreciated in both domains. However, it is worth mentioning that \gls{vdtw} brought a large improvement w.r.t. its previous versions (i.e., vEEGNet-1~[\cite{Zancanaro_2023_vEEGNet}] and vEEGNet-2~[\cite{Zancanaro_CCIS}]) as well as to other recently proposed architectures in the literature~[\cite{Bethge_2022_EEG2VEC}].
In fact, from our previous work~[\cite{Zancanaro_2023_vEEGNet}], we noticed that the model trained with a reconstruction error based on \gls{mse} was capable of reconstructing slow components, only, while now with \gls{vdtw} the reconstructed signal has a much broader spectrum, with higher frequency components.
Therefore, we can conclude that our choice to train the models using a loss function where the reconstruction error is quantified by the \gls{dtw} made a significant improvement.
%
%
%
%
% architecture contribution
Nevertheless, we can also infer that the hierarchical architecture has a relevant influence in the ability of the model to reconstruct the signal with high-fidelity, as one might expect from the literature on \gls{vae}s as applied to reduce blurry effects in the reconstructed images~[\cite{vahdat_2021_nvae}].

To more systematically compare the results from the two architectures, we filled Table~\ref{tab:recon_error_table} with all subject-wise performance of both models, after training (i.e., at the $80$-th epoch) and in the test phase. 
Here, the mean values represent the average across channels and repetitions of the \emph{normalized} \gls{dtw} similarity score between every original \gls{eeg} segment and its corresponding reconstructed one. Whereas, the standard deviation values were computed as the standard deviation of all mean values obtained by averaging across repetitions, only. 
The \emph{grand}-average and the \emph{grand}-standard deviation (i.e., the last two rows of the Table~\ref{tab:recon_error_table}) are the mean and the standard deviation, respectively, taken across (the mean values of) all $9$ subjects.
As we can observe, \gls{hvdtw} largely outperforms \gls{vdtw} in all subjects, both during training and during test.
It is worth noting that the data coming from two individuals, i.e., S2 and S5, resulted as particularly difficult to be reconstructed for both architectures. Later, we will deepen the investigation of these two cases providing a reasonable explanation for this problem.

\begin{table}[]
\centering
\centering
\caption{Average ($\pm$ standard deviation) reconstruction error for \gls{vdtw} and \gls{hvdtw}, as expressed in terms of \emph{normalized} \gls{dtw} similarity score.} \label{tab:recon_error_table}
\resizebox{0.85\textwidth}{!}{%
\begin{tabular}{|c|cc|cc|}
\hline
 & \multicolumn{2}{c|}{\cellcolor[HTML]{C0C0C0}\textbf{vEEGNet-ver3}} & \multicolumn{2}{c|}{\cellcolor[HTML]{C0C0C0}\textbf{hvEEGNet}} \\
\multirow{-2}{*}{\textbf{Subject id.}} & \textbf{Train} & \textbf{Test} & \textbf{Train} & \textbf{Test} \\ \hline
\textbf{1} & 18.41±6.26 & 22.99±22.52 & 1.16±0.36 & 2.3±1.84 \\
\textbf{2} & 18.06±8.88 & 128.05±162.16 & 1.7±0.81 & 60.81±65.84 \\
\textbf{3} & 48.34±14.19 & 41.35±34.16 & 1.87±0.62 & 4.96±5.81 \\
\textbf{4} & 18.1±21.66 & 18.01±12.51 & 3.59±7.44 & 1.51±1.13 \\
\textbf{5} & 17.38±15.16 & 49.09±9.92 & 1.01±0.45 & 15.67±3.95 \\
\textbf{6} & 32.92±13.28 & 29.01±12.49 & 1.76±0.72 & 1.87±0.61 \\
\textbf{7} & 13.49±3.67 & 12.37±2.85 & 1.02±0.28 & 0.9±0.33 \\
\textbf{8} & 42.13±21.19 & 48.5±12.36 & 4.07±1.61 & 5.46±1.65 \\
\textbf{9} & 36.45±21.33 & 33.87±8.12 & 2.01±1.23 & 1.91±0.57 \\ \hline
\textbf{AVG} & 27.25 & 42.58 & 2.02 & 10.6 \\ \hline
\textbf{STD} & 13.96 & 30.79 & 1.5 & 9.08 \\ \hline
\end{tabular}%
}
\end{table}

% \AZ{Dobbiamo specificare come calcoliamo la std? (Perchè avevamo ideato 3 possibili modi di farlo e alla fine abbiamo scelto di fare la media lungo i canali e poi la std di queste medie)}

\begin{figure} [htbp!]
    \centering
    \begin{subfigure}[b]{0.99\textwidth}
        \includegraphics[width=\textwidth]{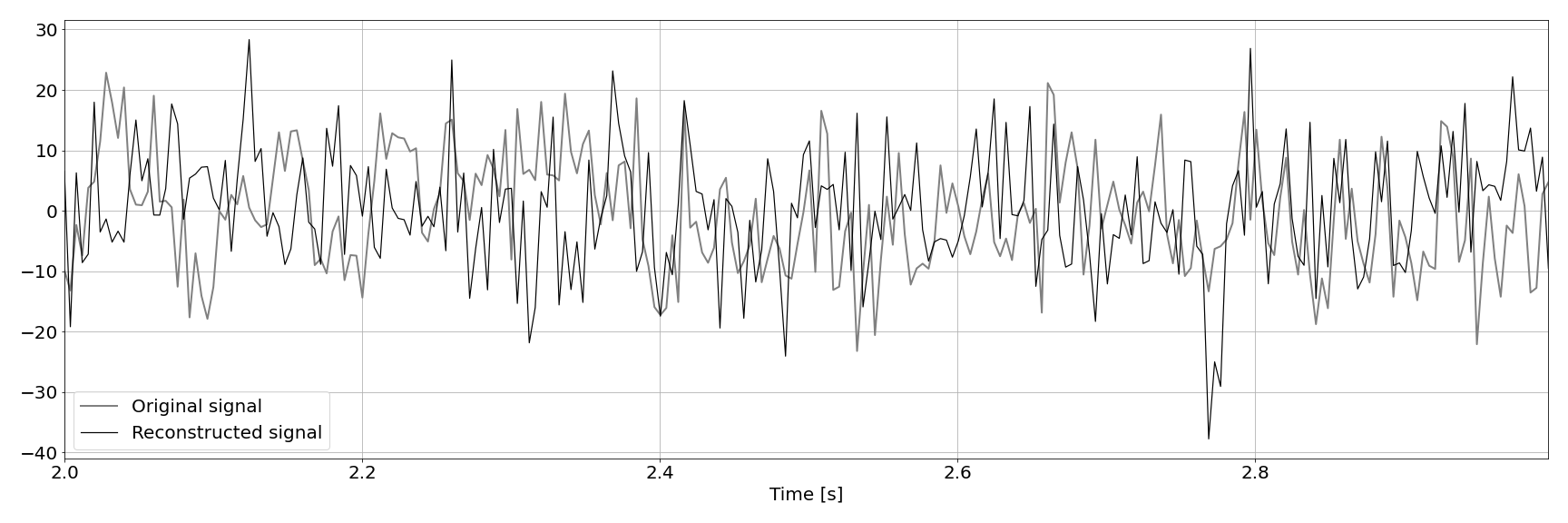}
        \caption{Time-domain reconstruction at the output of $z_1$.}
    \end{subfigure}
    
    \begin{subfigure}[b]{0.99\textwidth}
        \includegraphics[width=\textwidth]{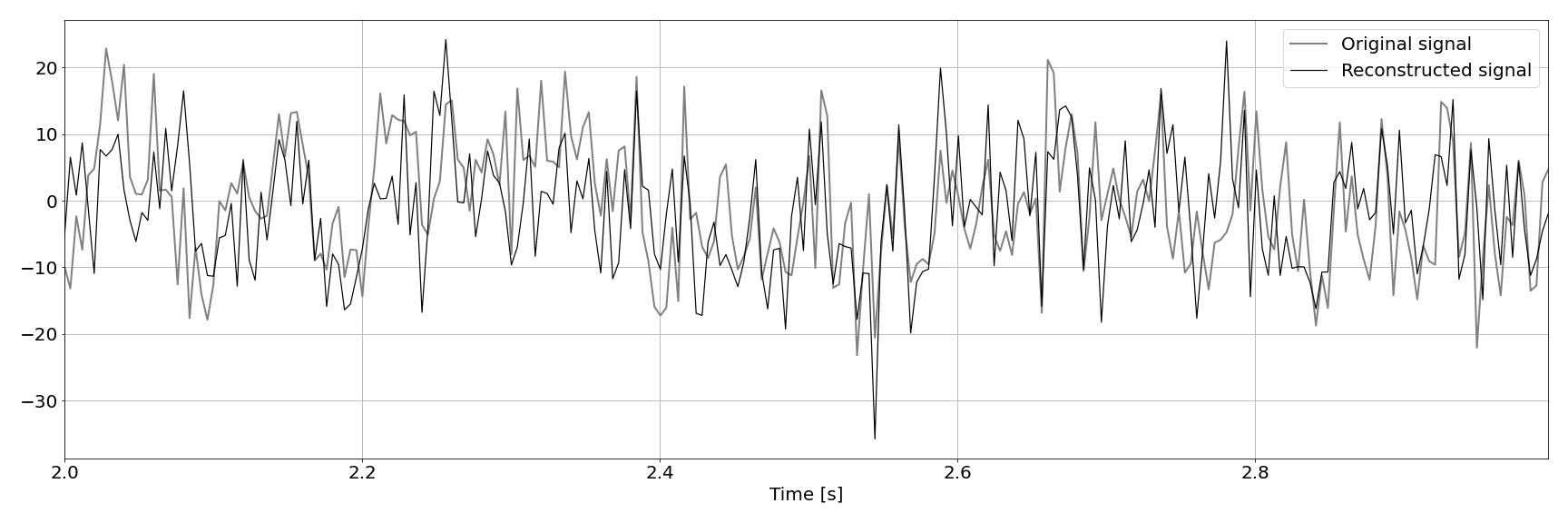}
        \caption{Time-domain reconstruction at the output of $z_2$ (including information from $z_1$).}
    \end{subfigure}
    
    \begin{subfigure}[b]{0.99\textwidth}
        \includegraphics[width=\textwidth]{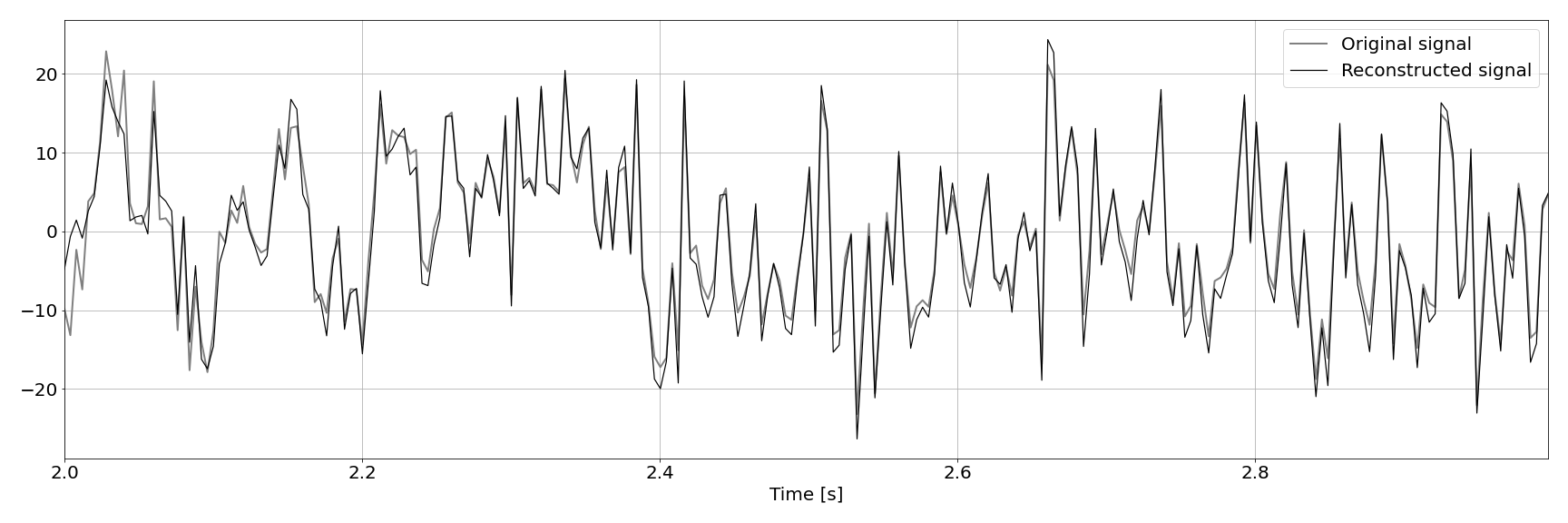}
        \caption{Time-domain reconstruction at the output of $z_3$ (including information from $z_2$ and $z_1$).}
    \end{subfigure}

    \begin{subfigure}[b]{0.32\textwidth}
        \includegraphics[width=\textwidth]{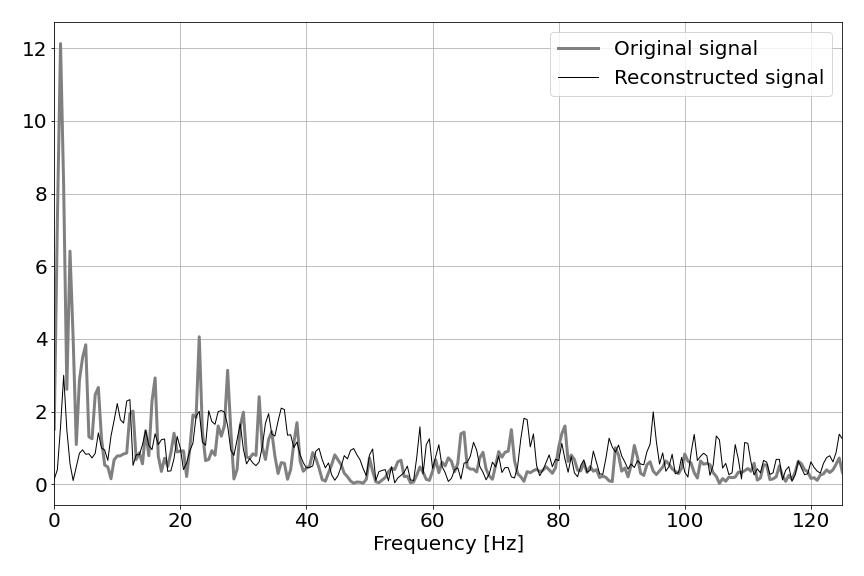}
        \caption{Output from $z_1$.}
    \end{subfigure}  
    \begin{subfigure}[b]{0.32\textwidth}
        \includegraphics[width=\textwidth]{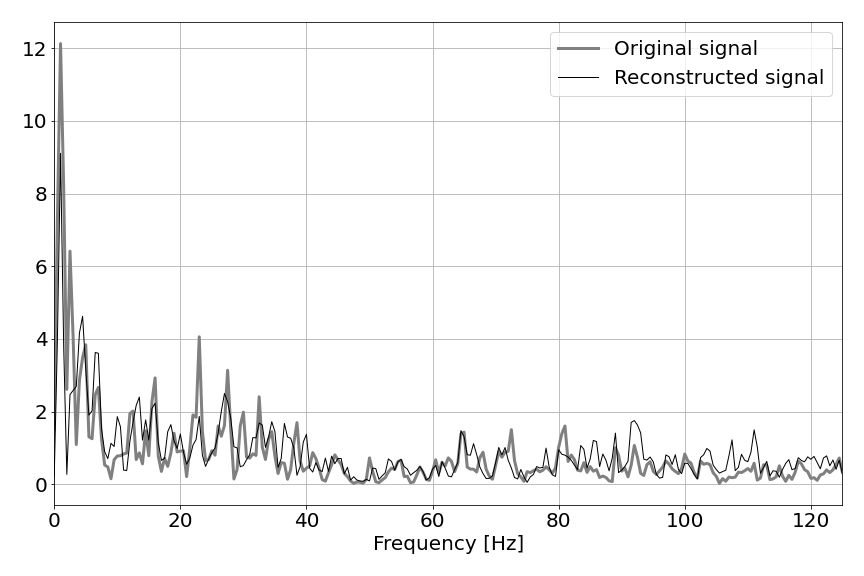}
        \caption{Output from $z_2$.}        
    \end{subfigure}
    \begin{subfigure}[b]{0.32\textwidth}
        \includegraphics[width=\textwidth]{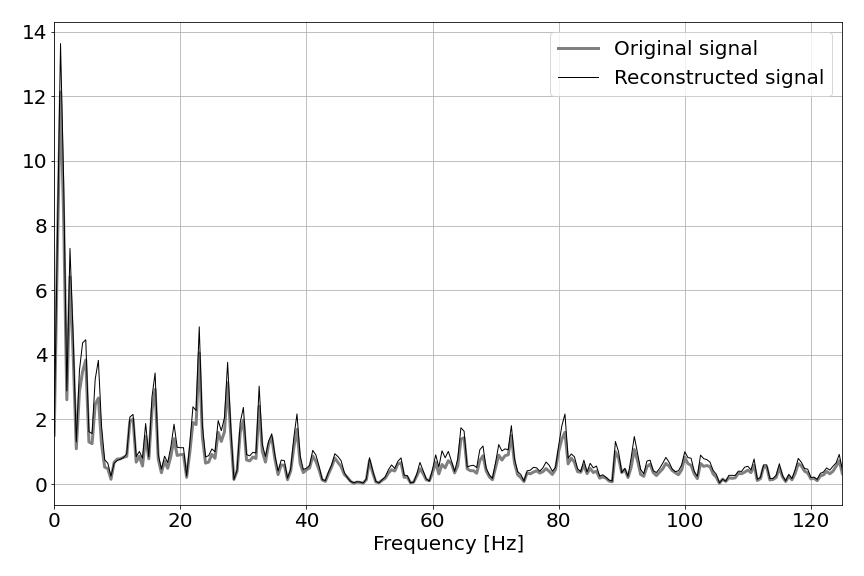}
        \caption{Output from $z_3$.}        
    \end{subfigure}
    
    \caption{Reconstruction as obtained at different points of the hierarchy both in time and frequency domains for one representative subject (S3), task repetition (no.1, corresponding to left hand \gls{mi}), and channel (C3), i.e., the same as in Fig.~\ref{fig:vEEGNet_vs_hvEEGNet}. The first three rows represent time domain reconstructions, the last row reports power spectra in the three different points of the hierarchy.}
    \label{fig:hvEEGNet_comparison_latent_space}
\end{figure}

% contributions from each latent space
Computer vision literature has already shown that the hierarchical architecture made the \gls{vae} models able to generate more detailed images, i.e., more effective in learning and generating high frequency components~[\cite{razavi_NEURIPS2019_vqVAE2, prost_2022_super_resolution_hierarchical_vae}].
Here, the use of more than one latent space seemed to have similarly allowed \gls{hvdtw} to better learn the underlying distribution of the data, and consequently greatly improved the reconstruction performance.
This is also confirmed by Fig.~\ref{fig:hvEEGNet_comparison_latent_space}, where it is possible to see how the contributions of the three different latent spaces influenced the reconstruction performance of the model.
As it can be observed, the \emph{deepest} latent space ($z_1$) can quite follow the original signal, has a similar dynamics (check also the power spectrum in Fig.~\ref{fig:hvEEGNet_comparison_latent_space}), but suffers from some time shifts and amplitude mismatches. Still, this result is better than the \gls{vdtw} output, even though sampling from $z_1$ in \gls{hvdtw} could have similarities with sampling from $z_0$ in \gls{vdtw} (e.g., much faster components can be recovered from $z_1$, but not from $z_0$).
Then, sampling from more \emph{superficial} (i.e., detailed) latent spaces produces an increasingly better reconstruction quality: when sampling from $z_2$ (including the effect from the deepest latent space $z_1$), amplitude mismatches are less frequent compared to the previous case, and the power spectrum is very similar to the original one. Finally, when sampling from $z_2$, the reconstruction is almost perfect, with minimal amplitude incongruences and time shifts.

However, we found cases where \gls{hvdtw} dramatically failed in reconstructing the original \gls{eeg} data. Also, there were cases in which the same number of training epochs was not enough for the \gls{hvdtw} model to reconstruct a particular subject. These two issues are discussed in the following, with additional investigations.

\subsection{Training behaviour vs training set: investigations on \gls{hvdtw}} \label{subsec: train_behav}

\begin{figure}[htpb!]
    \centering
    
    \begin{subfigure}[b]{0.32\textwidth}
        \includegraphics[width=\textwidth]{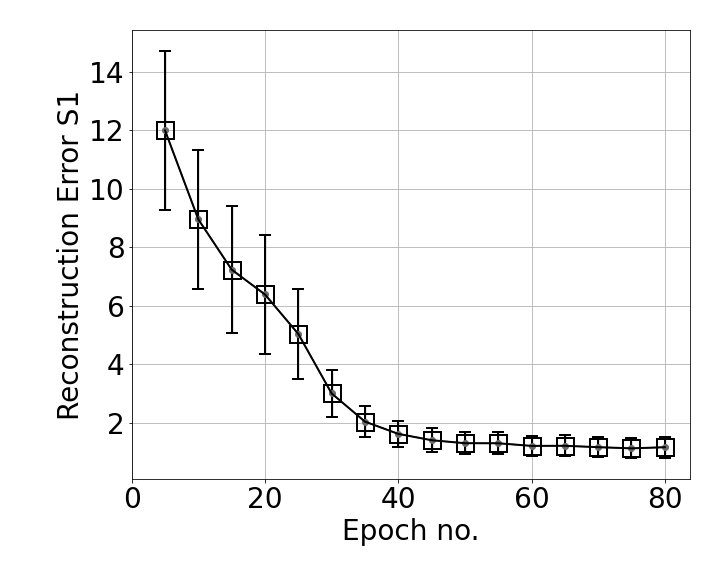}
        \label{fig:error_plus_std_across_epoch_subj_1}
    \end{subfigure}
    \begin{subfigure}[b]{0.32\textwidth}
        \includegraphics[width=\textwidth]{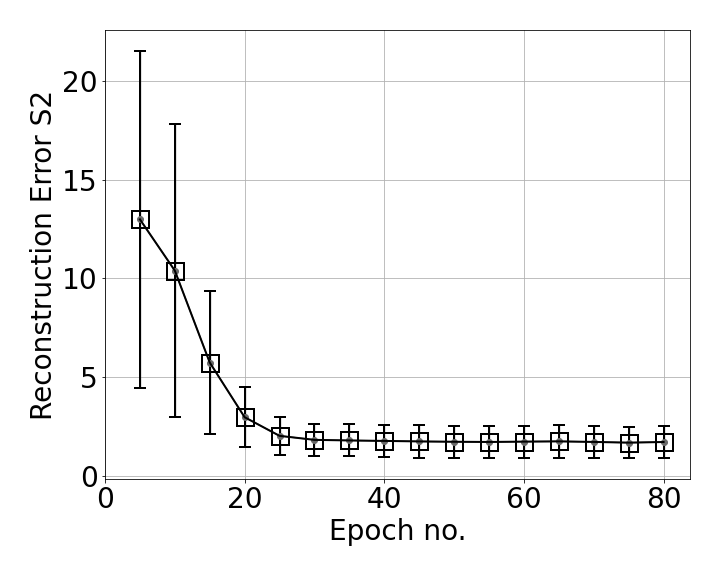}
        \label{fig:error_plus_std_across_epoch_subj_2}
    \end{subfigure}
    \begin{subfigure}[b]{0.32\textwidth}
        \includegraphics[width=\textwidth]{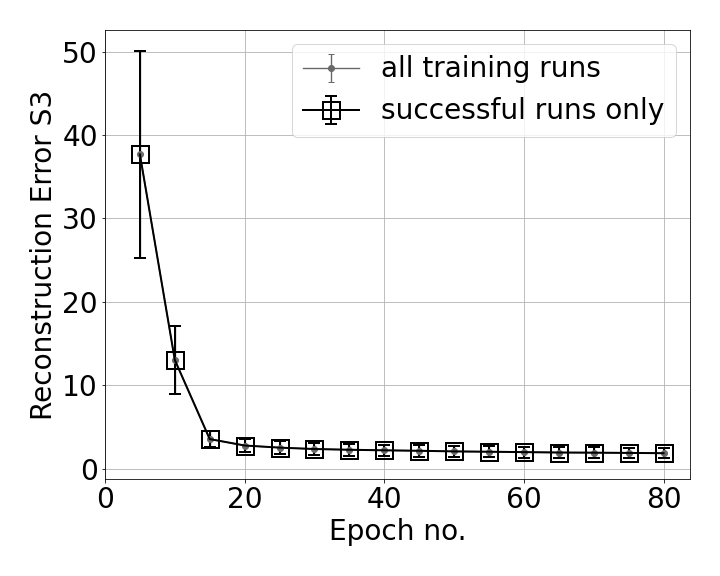}
        \label{fig:error_plus_std_across_epoch_subj_3}
    \end{subfigure}

    \begin{subfigure}[b]{0.32\textwidth}
        \includegraphics[width=\textwidth]{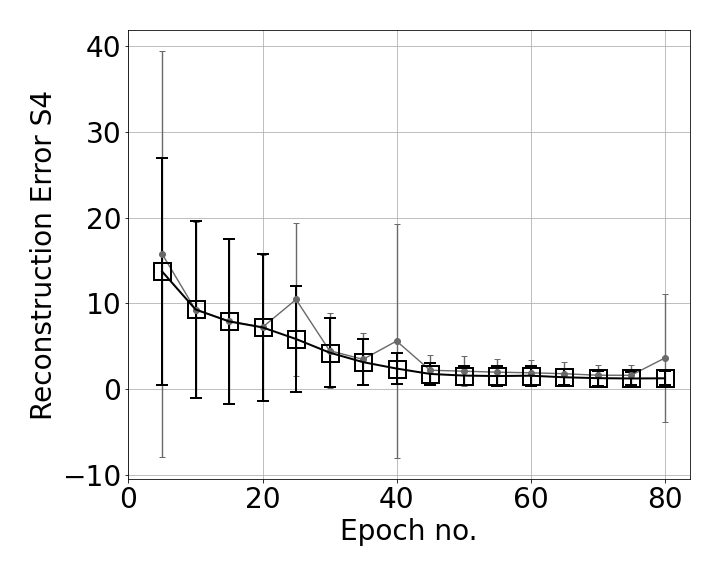}
        \label{fig:error_plus_std_across_epoch_subj_4}
    \end{subfigure}
    \begin{subfigure}[b]{0.32\textwidth}
        \includegraphics[width=\textwidth]{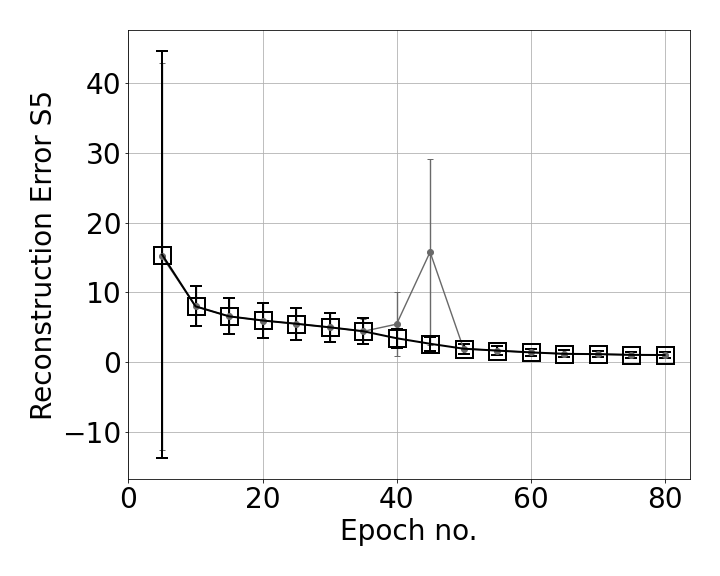}
        \label{fig:error_plus_std_across_epoch_subj_5}
    \end{subfigure}
    \begin{subfigure}[b]{0.32\textwidth}
        \includegraphics[width=\textwidth]{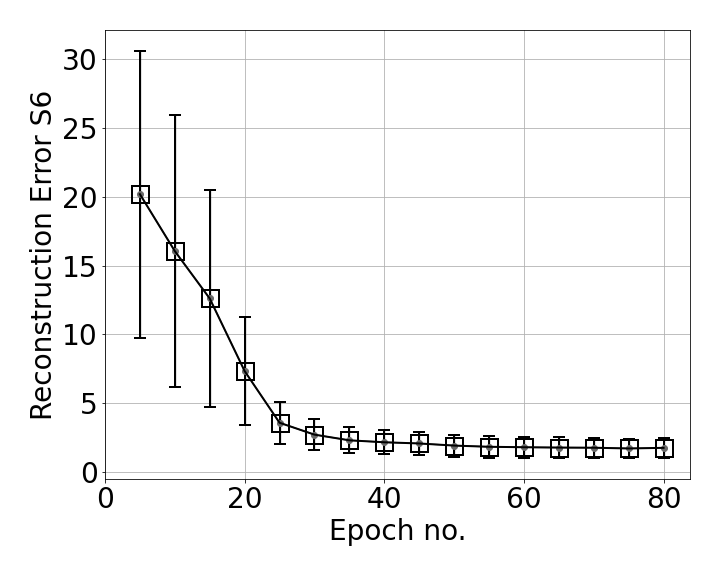}
        \label{fig:error_plus_std_across_epoch_subj_6}
    \end{subfigure}

    \begin{subfigure}[b]{0.32\textwidth}
        \includegraphics[width=\textwidth]{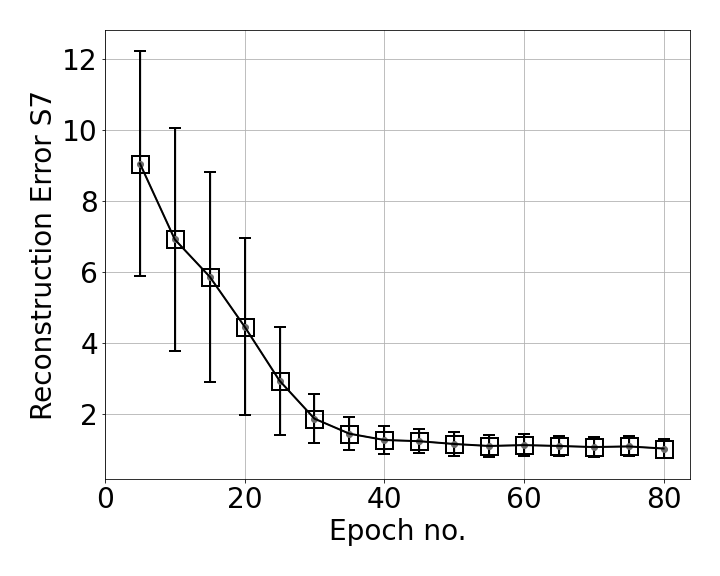}
        \label{fig:error_plus_std_across_epoch_subj_7}
    \end{subfigure}
    \begin{subfigure}[b]{0.32\textwidth}
        \includegraphics[width=\textwidth]{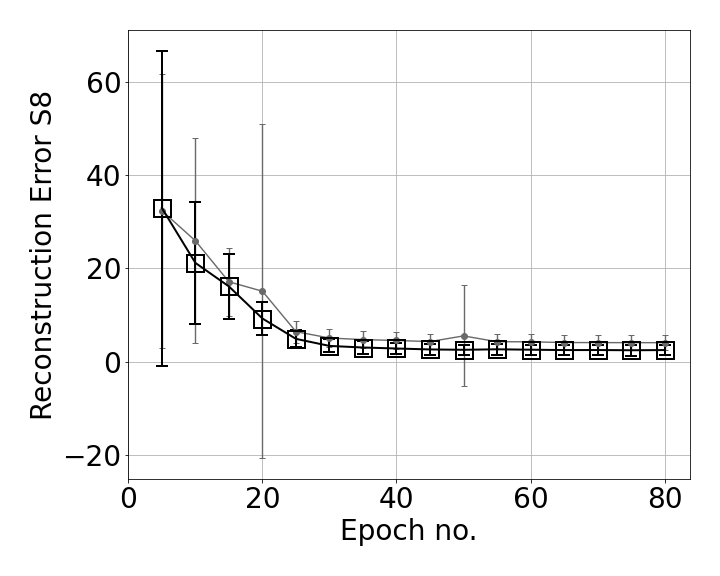}
        \label{fig:error_plus_std_across_epoch_subj_8}
    \end{subfigure}
    \begin{subfigure}[b]{0.32\textwidth}
        \includegraphics[width=\textwidth]{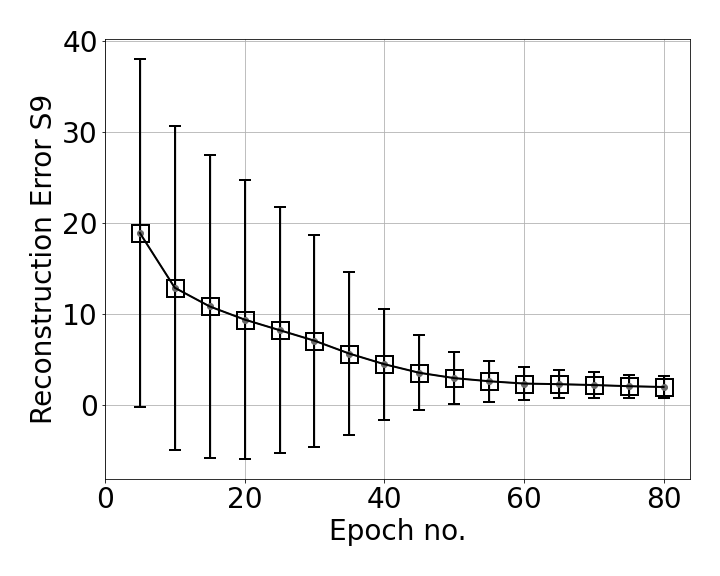}
        \label{fig:error_plus_std_across_epoch_subj_9}
    \end{subfigure}

    \caption{Reconstruction error (within-subject, across-EEG channels). Box markers show the average across multiple training runs, bars represent the standard deviation. Grey is used to show results obtained with all training runs, while black is when unsuccessful training runs (a few for three subjects out of nine, only) were removed from the analysis. Note that, for the sake of a better visualization, y-axes might have different ranges.}
    \label{fig:error_plus_std_across_epoch_subj_divided}
\end{figure}

% training course
\gls{hvdtw} should be trained until the \gls{dtw} is small enough to guarantee optimal reconstruction. We performed multiple (about $20$) training runs with $80$ epochs each, to evaluate the statistical behaviour of the model's training in different subjects.
We also computed the average normalized \gls{dtw} similarity score and its standard deviation across multiple runs and could show, for each subject, separately, the number of epochs at which that average is low enough and the standard deviation stabilizes, at the same time.
Fig.~\ref{fig:error_plus_std_across_epoch_subj_divided} displays the average (and standard deviation) \gls{dtw}-based error for an increasing number of epochs for each subject during training. 
We can observe that the \gls{dtw}-based error clearly decreases as the number of epochs increases, as expected. Then, for all subjects, $80$ epochs are enough to obtain almost perfect reconstruction. However, we also clearly noted that the time (no. epochs) needed to reach that point highly varies from subject to subject.
For example, S3 reaches an optimal model very rapidly, in about $15$ epochs: we can see that the training of an \gls{hvdtw} model starts with an average \gls{dtw} error of $38$ and a large standard deviation of $12$, then it fastly decreases in its mean and variability, reaching a stable average of $5$ and a very small standard deviation in $15$ epochs.
A completely different case is represented by S9: here, the average beginning error is smaller than the S3's one, but the standard deviation is much larger. Also, it takes much more - on average - to the model to adapt to this subject and reach a stable and optimal model (at about $60$ epochs).
%
%
% E' FIGURA COMPOSITA!
Therefore, we have just empirically proved that there is a  relationship between the training time (i.e., the number of epochs needed to reach an optimal model) and the distribution of the input training set that cannot be overlooked~[\cite{Gyori_2022_data_distribution_impacts_ml}].

\begin{figure}
    \begin{subfigure}[b]{\textwidth}
        \includegraphics[width=0.49\textwidth]{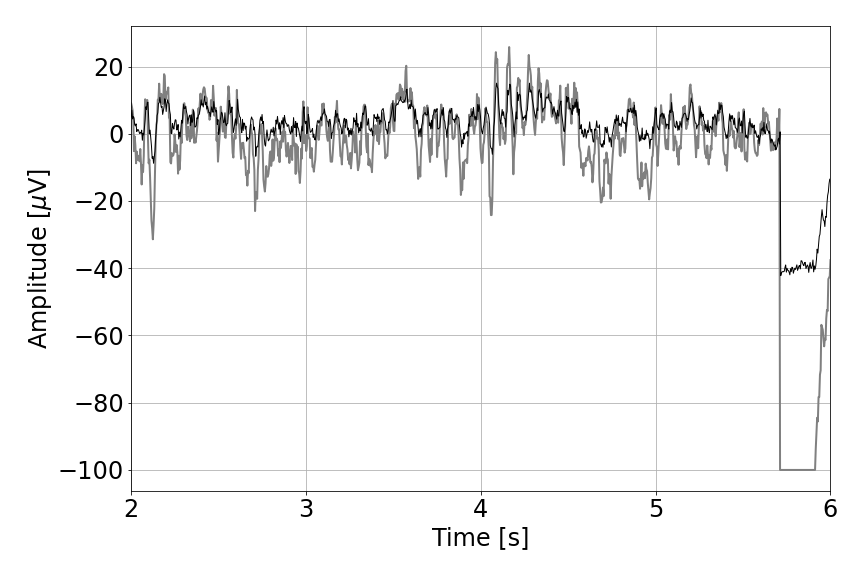}
        \hfill
        \includegraphics[width=0.49\textwidth]{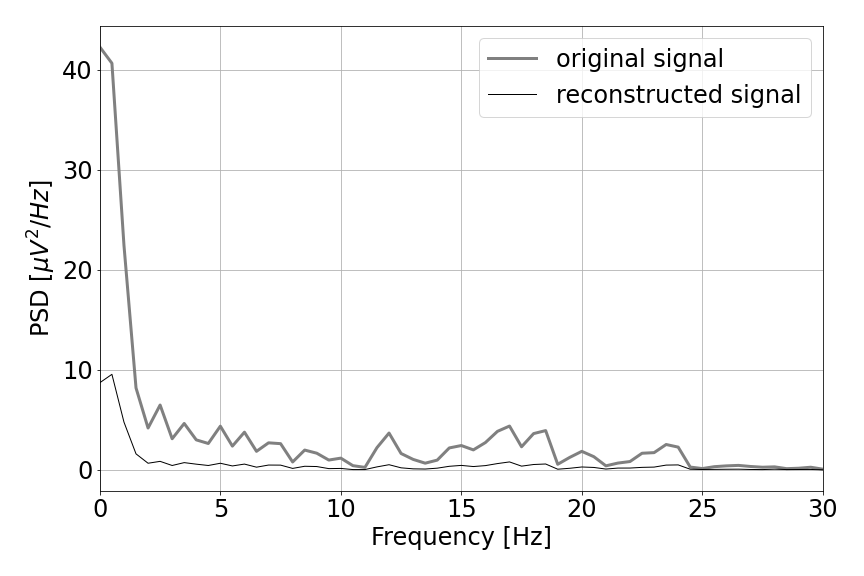}
        \caption{S4, segment no. $180$, ch. C3.}
        \label{fig:saturated_trials_subj_4_time}
    \end{subfigure}
    
    \begin{subfigure}[b]{\textwidth}
        \includegraphics[width=0.49\textwidth]{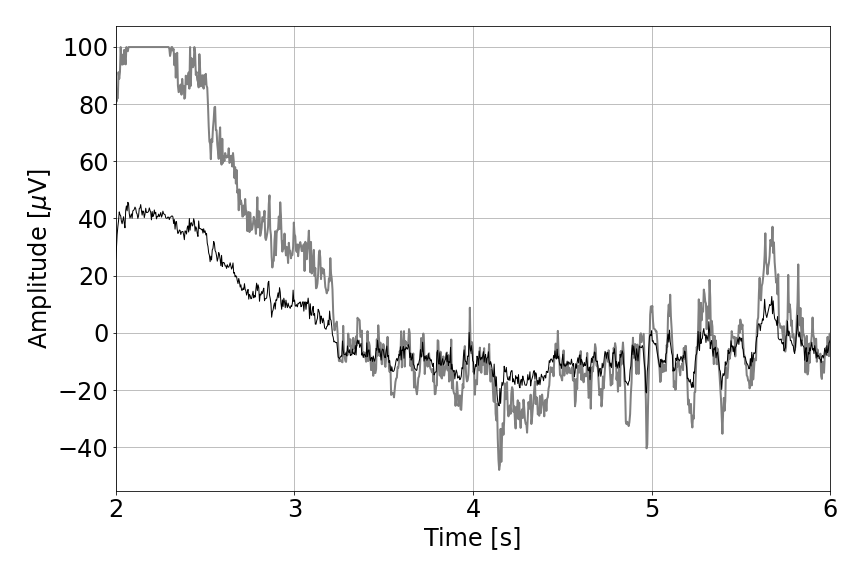}
        \hfill
        \includegraphics[width=0.49\textwidth]{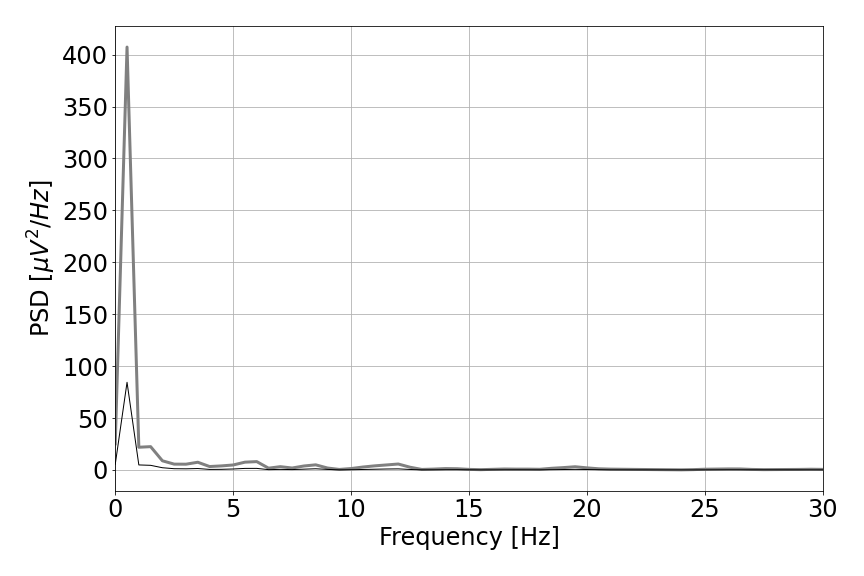}
        % \caption{Subject 5, trial 221, channel Cz. Time domain.}
        \caption{S5, segment no. $221$, ch. Cz.}
        \label{fig:saturated_trials_subj_8_time}
    \end{subfigure}

    \begin{subfigure}[b]{\textwidth}
        \includegraphics[width=0.49\textwidth]{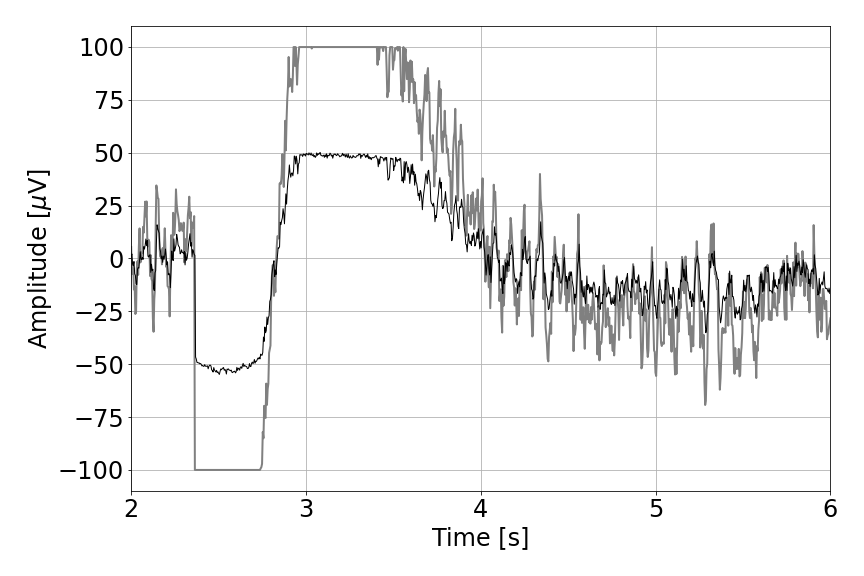}
        \hfill
        \includegraphics[width=0.49\textwidth]{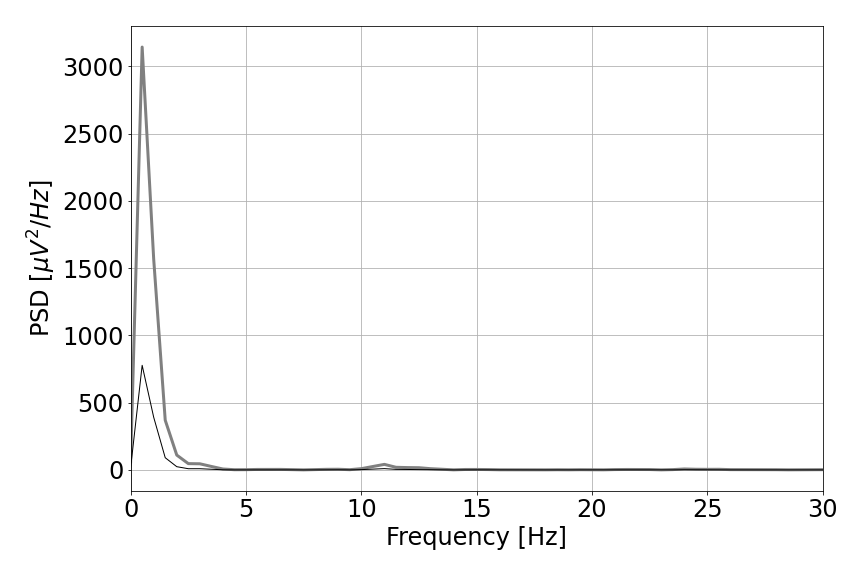}
        \caption{S8, segment no. $82$, ch. CP4.}
        \label{fig:saturated_trials_subj_8_time}
    \end{subfigure}

    \caption{Example of saturated trials in the training set leading to very high \gls{dtw} values. Left panels: time domain representation. Right panels: frequency domain representation.}
    \label{fig:saturated_trials}
\end{figure}

% saturated training runs
Another relevant case to discuss is the dramatic fail of the \gls{hvdtw} model in reconstructing some - rare - specific \gls{eeg} segments.
We found four anomalous training runs where the model failed, i.e., two for S4, one for S5, and another for S8. We further analyzed all segments in these three subjects and discovered that the model fail was due to problems of saturation that happened during the acquisition step of the \gls{eeg} data (those segments had not been removed from the public available dataset). This, in turn, led the \gls{dtw} score to assume extremely high values, i.e., the model to significantly fail the reconstruction.
Fig.~\ref{fig:saturated_trials} shows one example of \gls{eeg} segment for each subject (S4, S5, and S8) where signal saturation was identified during the \gls{hvdtw} model training.
No matter where saturation occurs, i.e., soon or later in the segment, its effect on the model training is to dramatically increase the \gls{dtw}-based error.
These events, in turn, are responsible for that sudden increase of the standard deviation, as it can be noticed at epochs $25$ and $40$ for S4, at epoch $45$ for S5, and at epoch $20$ for S8.
On the other hand, we also checked that the vast majority of the other S4, S5, and S8's segments led to \gls{dtw} score values in a range similar to the other subjects.
Thus, we decided to exclude those training runs where the \gls{hvdtw} model suffered from the disruptive effect of acquisition saturation problems, namely unsuccessful training runs. For this reason, for S4, S5, and S8, Fig.~\ref{fig:error_plus_std_across_epoch_subj_divided} shows the model training behaviour along the epochs both with and without the unsuccessful training runs (grey and black line, respectively).
Nonetheless, we cannot assess that saturation during acquisition is the only possible cause of training inaccuracy for the \gls{hvdtw} model.

However, we provided some further insights that there is a correspondence between the training behaviour and the quality of the input training set, thus highlighting the importance to preliminary evaluate the quality and the distribution of the input dataset.
More importantly, we should also remind that the vast majority of the related work uses this dataset as it is, as input to a wide variety of \gls{dl} models with no questions on the quality and distribution of the input dataset~[\cite{Schirrmeister_EEG_CNN,EEGNet_paper,Sakhavi_AC_KS_TAB_4_5,DFNN_AC_TAB_7,MI_EEGNet_AC_TAB_10}]. All of them have shown a large variability in the classification results (i.e., classification of the different \gls{mi} tasks), but there is no study - as far as the authors know - reporting a systematic investigation of the relationship between the models training and the characteristics of the input data.

In the following, we investigate the performance of the \gls{hvdtw} model, when \emph{sufficiently trained} (i.e., for a number of epochs that varies from subject to subject), and we explore its ability to identify other anomalies as well as reconstructing clean EEG segments.

\subsection{hvEEGNet as anomaly detector}

% OUTLIER IDENTIFICATION WITH AN EXTENSIVELY TRAINED MODEL
Once the \gls{hvdtw} model is properly trained, we can look into its ability to identify outliers. To be conservative, for all subjects, we searched for outliers in the test set with the \gls{hvdtw} model trained for $80$ epochs.
As described in Section~\ref{subsec:outlier}, we employed the \gls{knn} algorithm on the matrix $\mathbf{\overline{E}}$ given by all values obtained by averaging the normalized \gls{dtw} scores across the training runs for each pair task repetition-channel (subject-wise).
Here, $\mathbf{\overline{E}}$ is a $288 \times 22$ matrix. Then, we applied \gls{knn} over $\mathbf{\overline{E}}$ to find out any possible outliers. We implemented the algorithm using the scikit-learn Python package~[\cite{scikit-learn}], with default settings and the number of nearest neighbours ($k$) equal to $15$. We empirically found that $15$ was a good trade-off between the stability of the results and the expected proximity among all samples in the dataset. % \hl{DA CHIARIRE}
Also, note that each sample of this matrix is characterized by $22$ dimensions, and the \gls{knn} algorithm worked in such high-dimensional space to find proximity among points as well as outliers.
%
% FIGURA: esempi di outlier=artefatti su S2, S4, S9 in test set
\begin{figure}

% Time domain
    \begin{subfigure}[b]{\textwidth}
        \includegraphics[width=0.49\textwidth]{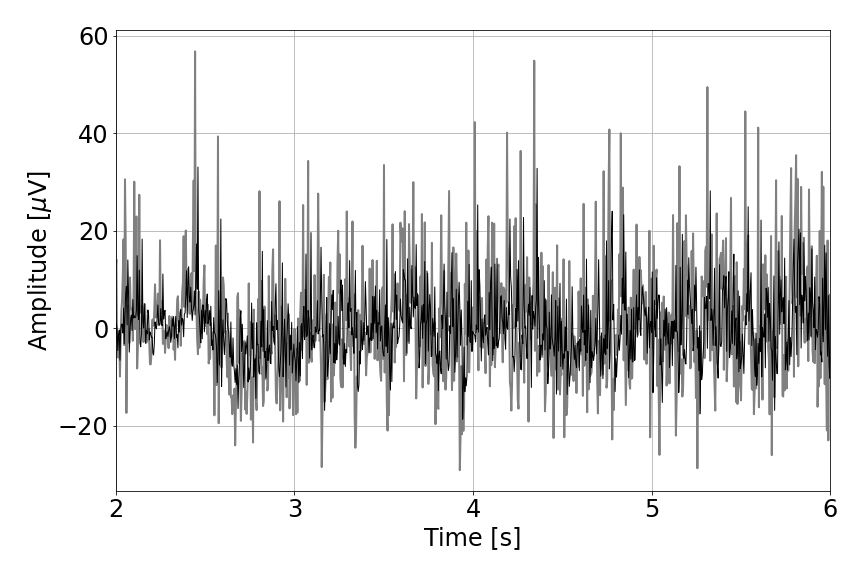}
        \hfill
        \includegraphics[width=0.49\textwidth]{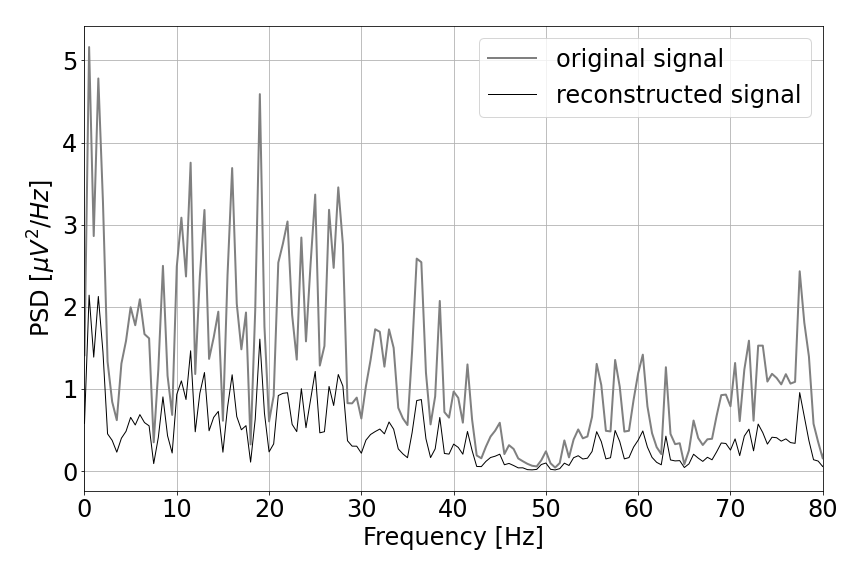}
        \caption{S2, segment no.$1$, ch. FC3.}
        % \label{fig:outliers_trials_subj_2_time}
    \end{subfigure}
    
    \begin{subfigure}[b]{\textwidth}
        \includegraphics[width=0.49\textwidth]{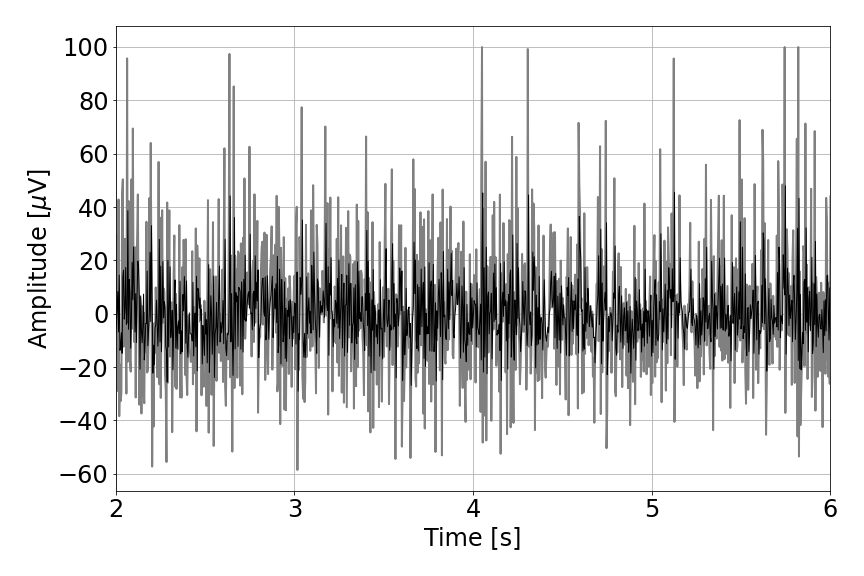}
        \hfill
        \includegraphics[width=0.49\textwidth]{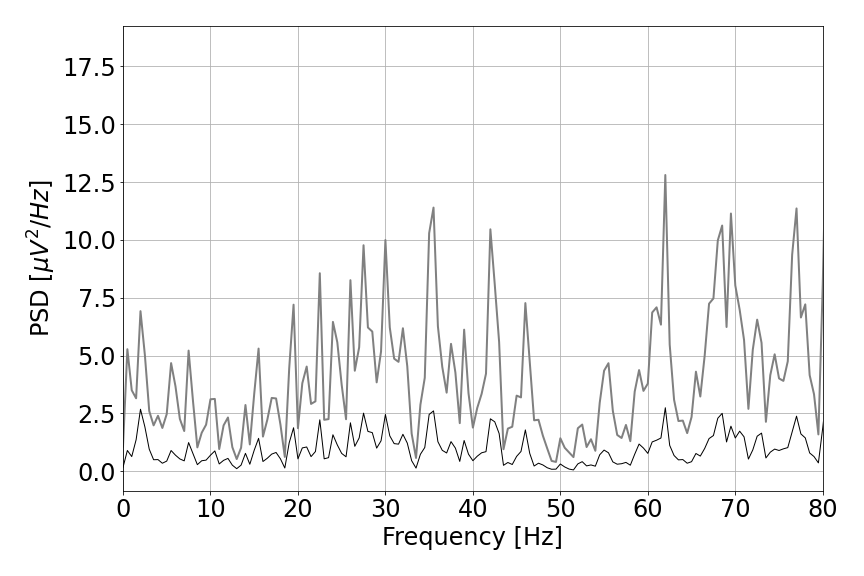}
        \caption{S4, segment no.$146$, ch. C5.}
        \label{fig:outliers_trials_subj_4_time}
    \end{subfigure}
    
    \begin{subfigure}[b]{\textwidth}
        \includegraphics[width=0.49\textwidth]{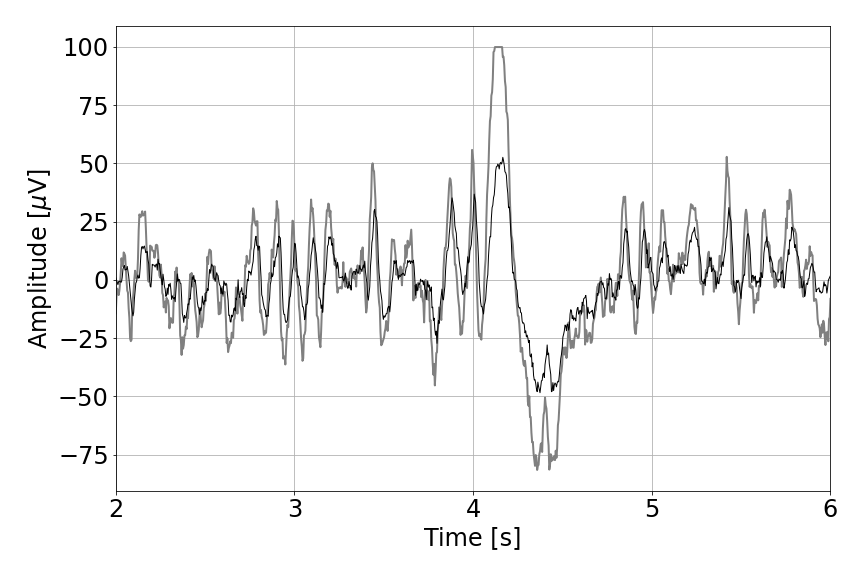}
        \includegraphics[width=0.49\textwidth]{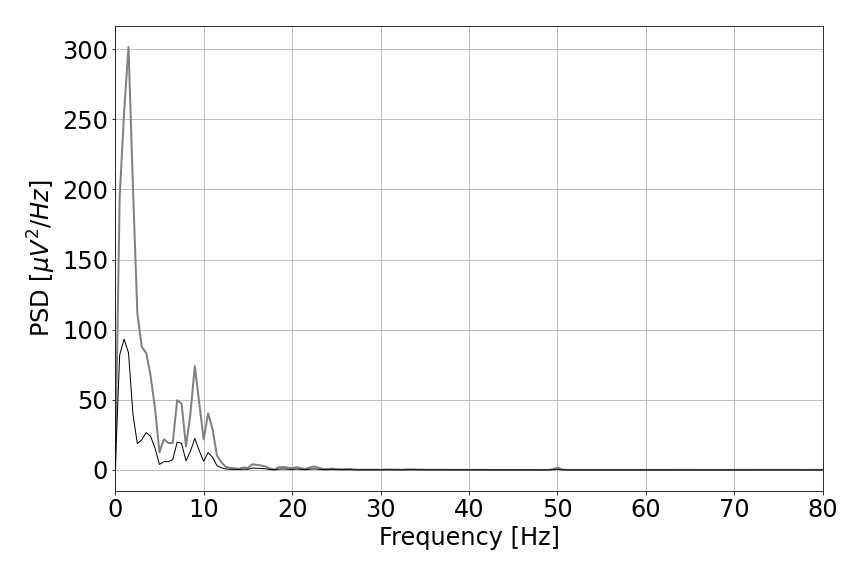}
        \caption{S9, segment no.$251$, ch. C1.}
        %\label{fig:outliers_trials_subj_9_time}
    \end{subfigure}
    
    \caption{Three representative examples of \gls{eeg} segments (belonging to the test set) marked as outlier by the \gls{hvdtw} model extensively trained. Left panels: time domain representation. Right panels: frequency domain representation.}
    % (a)-(c) Time domain. (d)-(f) Frequency domain.
    
    \label{fig:outliers_trials}
    
\end{figure}
Fig.~\ref{fig:outliers_trials} shows three representative examples of \gls{eeg} segments that were marked as outliers by our extensively trained model. By visually inspecting them (in both time and frequency domain) and based on previous expertise~[\cite{cisotto2015displacement}] as well as well-established literature~[\cite{durka2003simple, gao2010automatic}], we can easily confirm that those segments have a frequency characterization similar to a muscular artefact or eye blink activity.

\begin{figure}[htbp!]
    \centering
    
    \begin{subfigure}[b]{0.32\textwidth}
        \includegraphics[width=\textwidth]{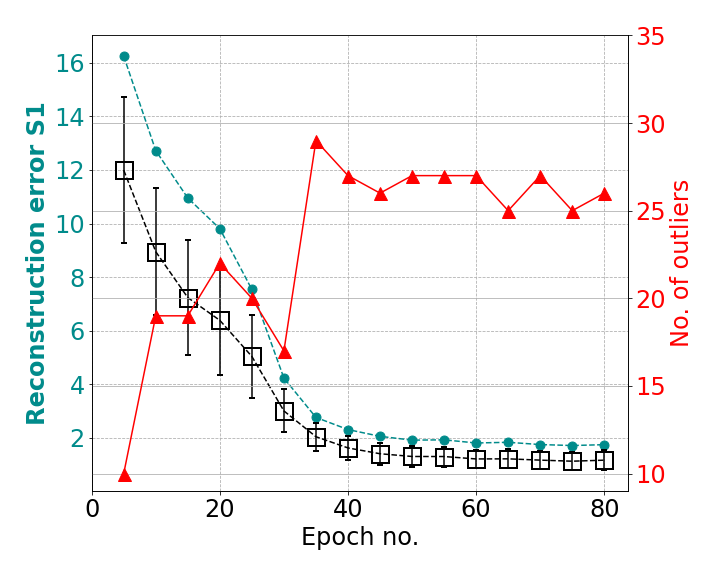}
        %\caption{Subject 1.}
        %\label{fig:outliers_neighborhood_15_subj_1}
    \end{subfigure}
    \begin{subfigure}[b]{0.32\textwidth}
        \includegraphics[width=\textwidth]{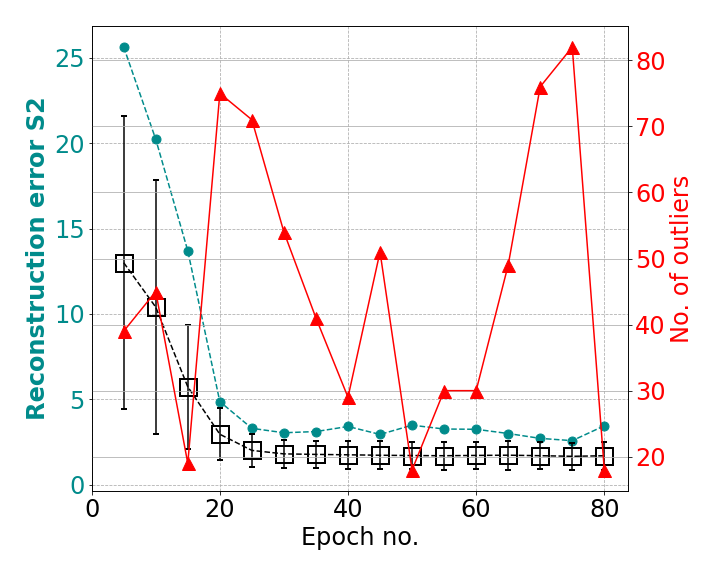}
        %\caption{Subject 2.}
        %\label{fig:outliers_neighborhood_15_subj_2}
    \end{subfigure}
    \begin{subfigure}[b]{0.32\textwidth}
        \includegraphics[width=\textwidth]{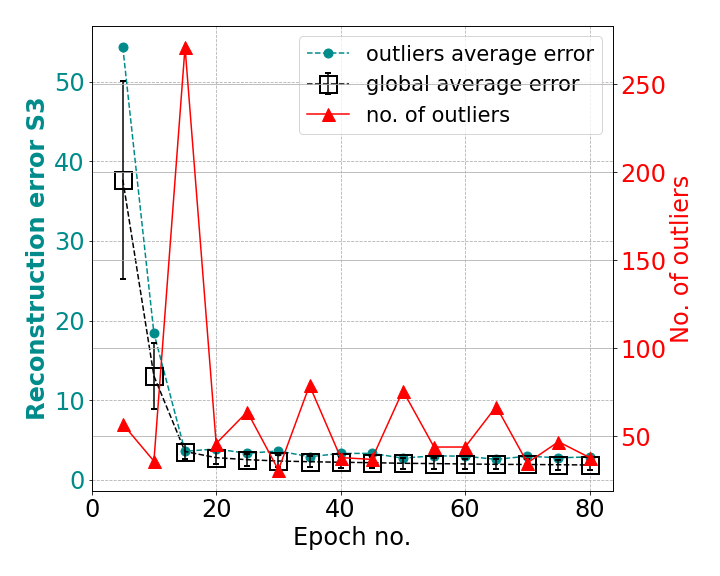}
        %\caption{Subject 3.}
        %\label{fig:outliers_neighborhood_15_subj_3}
    \end{subfigure}

    \begin{subfigure}[b]{0.32\textwidth}
        \includegraphics[width=\textwidth]{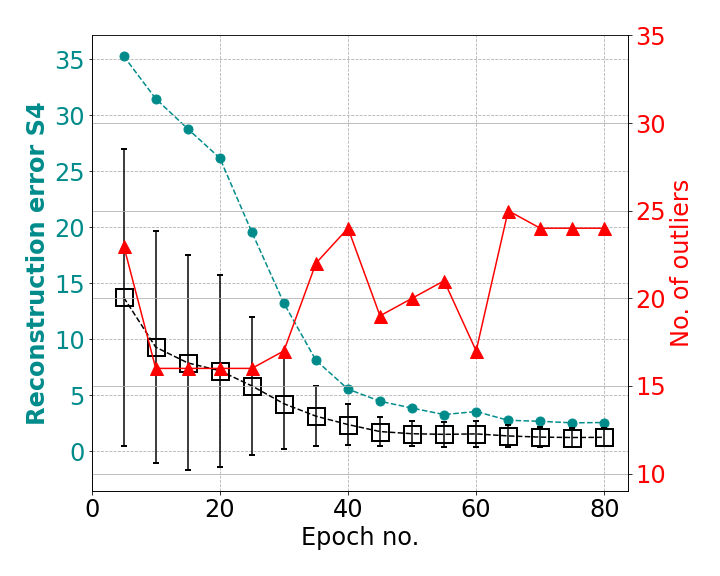}
        %\caption{Subject 4.}
        %\label{fig:outliers_neighborhood_15_subj_4}
    \end{subfigure}
    \begin{subfigure}[b]{0.32\textwidth}
        \includegraphics[width=\textwidth]{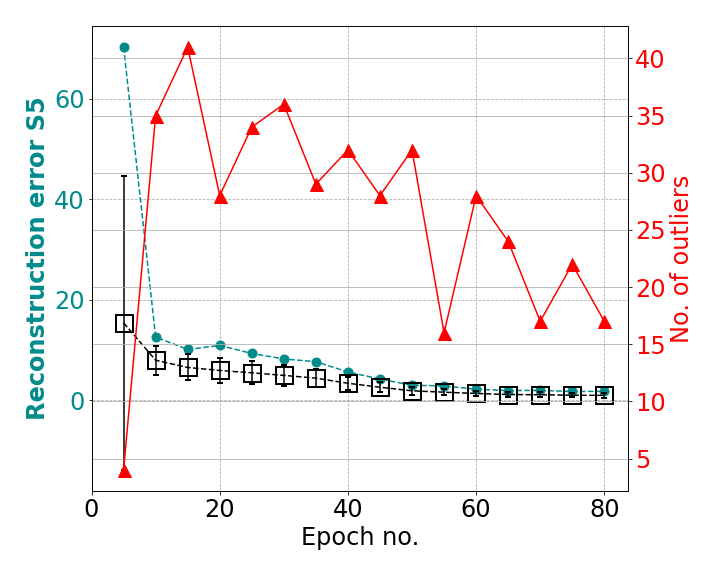}
        %\caption{Subject 5.}
        %\label{fig:outliers_neighborhood_15_subj_5}
    \end{subfigure}
    \begin{subfigure}[b]{0.32\textwidth}
        \includegraphics[width=\textwidth]{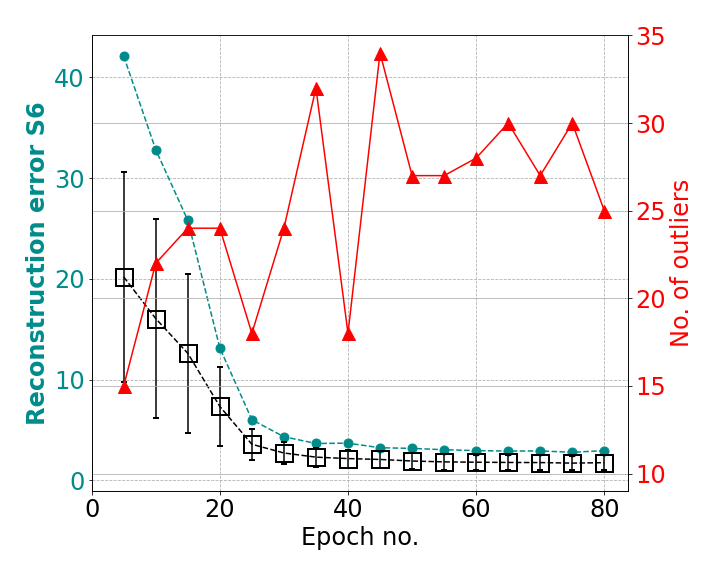}
        %\caption{Subject 6.}
        %\label{fig:outliers_neighborhood_15_subj_6}
    \end{subfigure}

    \begin{subfigure}[b]{0.32\textwidth}
        \includegraphics[width=\textwidth]{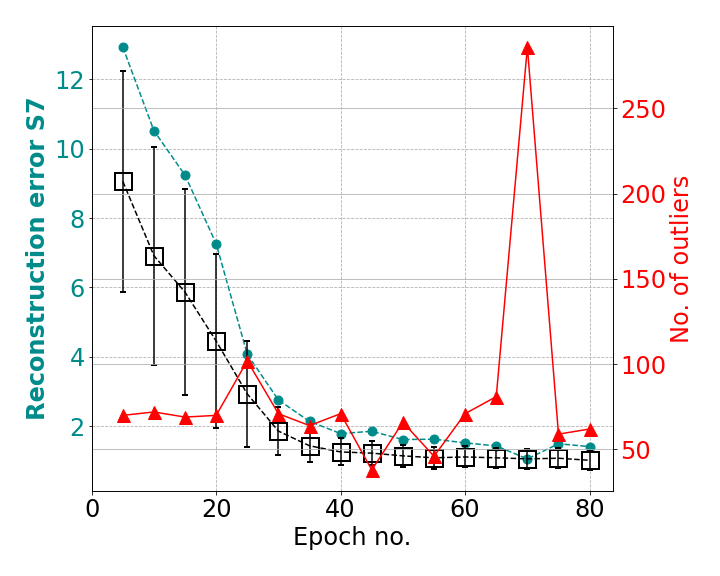}
        %\caption{Subject 7.}
        %\label{fig:outliers_neighborhood_15_subj_7}
    \end{subfigure}
    \begin{subfigure}[b]{0.32\textwidth}
        \includegraphics[width=\textwidth]{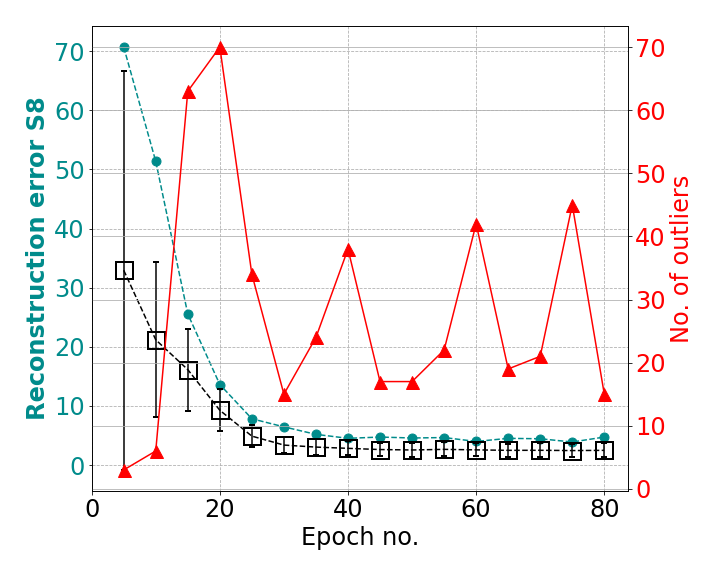}
        %\caption{Subject 8.}
        %\label{fig:outliers_neighborhood_15_subj_8}
    \end{subfigure}
    \begin{subfigure}[b]{0.32\textwidth}
        \includegraphics[width=\textwidth]{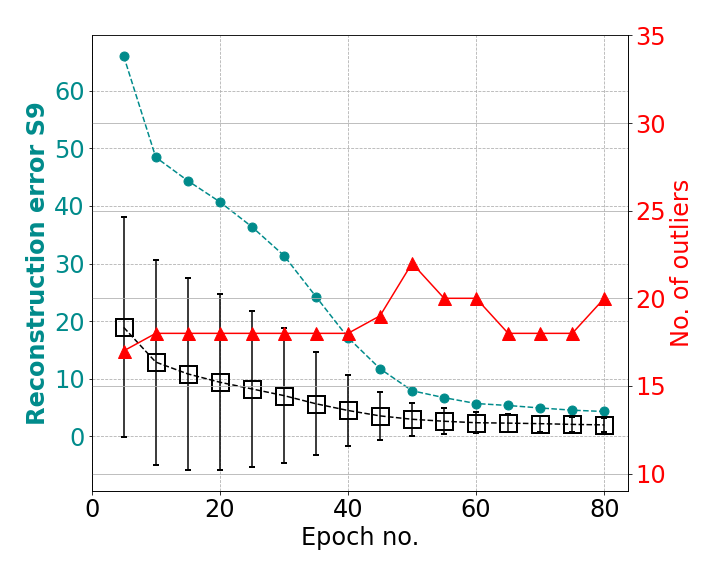}
        %\caption{Subject 9.}
        %\label{fig:outliers_neighborhood_15_subj_9}
    \end{subfigure}

    \caption{Relationship between training effectiveness and outliers identification ability w.r.t. individual subjects. Each panel reports the number of outliers (red line), the average error exclusively due to the
outliers (dark green line), and the global average error of the whole training set (black line) for each subject (specified on the left y-axis), at each $5$-epoch step during training.}
    \label{fig:outliers_neighborhood_15}
\end{figure}

% PASSAGGIO
However, Fig.~\ref{fig:error_plus_std_across_epoch_subj_divided} has shown that the \gls{hvdtw} model can reach very low average errors (with very small standard deviations) in a number of epochs typically lower than $80$. Moreover, this time highly depends on the specific subject to analyze.
%
%
% OUTLIER IDENTIFICATION WITH A SUFFICIENTLY TRAINED MODEL
To systematically investigate the relationship between the \emph{training effectiveness} and the outliers identification ability w.r.t. individual subjects, we plotted Fig.~\ref{fig:outliers_neighborhood_15}, where the global average error of the whole training set, the number of detected outliers, and the average error exclusively due to the outliers are reported for every subject, separately, at each $5$-epoch step during training.

In all panels, i.e., for all subjects, from Fig.~\ref{fig:outliers_neighborhood_15} we can easily distinguish two phases in the training behaviour: in the first part of the training, as expected, the average reconstruction error is progressively reduced (black line). This typically corresponds to a few outliers (red line) significantly contributing to the global average error (dark green line).
In the second phase, i.e., after the model has reached a stable performance (in terms of average reconstruction error), the number of outliers starts to vary with the average global error and the outliers average error remaining quite small. This can be intuitively explained by the fact that the \gls{eeg} segments are generally well reconstructed and small variations on the error are enough to make the corresponding \gls{eeg} segment to be considered an outlier by the \gls{knn} algorithm.

Therefore, we decided to deepen the investigation on the \emph{transition point} to check if a subject-independent characterization of the training behaviour can be obtained, and to verify the opportunity to stop the training at that point.
First, we empirically defined the \emph{transition point} as the number of epochs where the global average error showed the maximal slope (an elbow point), with low standard deviation, and the number of identified outliers was about to suddenly increase. For example, for S1 the transition point was identified at $30$ epochs, while for S9 at $40$ epochs. To note, the earliest transition point was found in S3 and S5 at $20$ epochs, while the latest transition point was found in S4 at $45$ epochs.
Second, we re-evaluated the performance of our \gls{hvdtw} model on the test set with the training stopped at the \emph{transition point}. Table~\ref{tab:recon_error_table_TRANSPOINT} reports the average (and standard deviation) reconstruction error at the subject-specific transition point.
We can observe that the reconstruction performance are similar to the performance obtained for an extensively trained model (see Table~\ref{tab:recon_error_table} for the comparison).
Thus, we can conclude that our \gls{hvdtw} model could reach very high-fidelity reconstruction in a short time, lower than 30 minutes (approx. time needed to train the model for 50 epochs, as reported in Section~\ref{subsec:computational_complexity}).
%
% \begin{table}[htbp!]
% \centering
% \caption{Average ($\pm$ standard deviation) reconstruction error for \gls{hvdtw} in the test set, with the model training stopped at the subject-dependent transition point.} \label{tab:recon_error_table_TRANSPOINT}
% \begin{tabular}{c|c|cc|}
% \cline{2-4}
%                         & Transition Point & \multicolumn{2}{c|}{hvEEGNet} \\
% \cline{2-4} 
%                         & no. epochs     & Train   & Test       \\ \hline
% \multicolumn{1}{|c|}{1} &  30              &         3.0±0.81  &   4.68±3.55\\
% \multicolumn{1}{|c|}{2} &  45              &         1.73±0.82 &   60.01±65.89\\
% \multicolumn{1}{|c|}{3} &  20              &         2.79±0.79 &   3.58±1.53\\
% \multicolumn{1}{|c|}{4} &  45              &         2.18±1.76 &   2.19±1.71\\
% \multicolumn{1}{|c|}{5} &  20              &         5.97±2.5  &   36.29±6.44\\
% \multicolumn{1}{|c|}{6} &  25              &         2.01±0.95 &   3.37±1.02\\
% \multicolumn{1}{|c|}{7} &  30              &         1.86±0.68 &   1.42±0.43\\
% \multicolumn{1}{|c|}{8} &  30              &         5.06±1.92 &   6.49±1.82\\
% \multicolumn{1}{|c|}{9} &  40              &         4.51±6.1 &    3.58±0.8\\ \hline
% \end{tabular}
% \end{table}

\begin{table}[htbp!]
\centering
\caption{Average ($\pm$ standard deviation) reconstruction error for \gls{hvdtw} in the test set, with the model training stopped at the subject-dependent transition point.} 
\label{tab:recon_error_table_TRANSPOINT}
\resizebox{0.85\textwidth}{!}{%
\begin{tabular}{|c|c|cc|}
\hline
                                       &                                                                                                        & \multicolumn{2}{c|}{\textbf{Reconstruction error}} \\
\multirow{-2}{*}{\textbf{Subject id.}} & \multirow{-2}{*}{\textbf{\begin{tabular}[c]{@{}c@{}}Transition point \\ {[}epoch no.{]}\end{tabular}}} & \textbf{Train}                 & \textbf{Test}                 \\ \hline
\textbf{1}                             & 30                                                                                                     & 3.0±0.81                       & 4.68±3.55                     \\
\textbf{2}                             & 45                                                                                                     & 1.73±0.82                      & 60.01±65.89                   \\
\textbf{3}                             & 20                                                                                                     & 2.79±0.79                      & 3.58±1.53                     \\
\textbf{4}                             & 45                                                                                                     & 2.18±1.76                      & 2.19±1.71                     \\
\textbf{5}                             & 20                                                                                                     & 5.97±2.5                       & 36.29±6.44                    \\
\textbf{6}                             & 25                                                                                                     & 2.01±0.95                      & 3.37±1.02                     \\
\textbf{7}                             & 30                                                                                                     & 1.86±0.68                      & 1.42±0.43                     \\
\textbf{8}                             & 30                                                                                                     & 5.06±1.92                      & 6.49±1.82                     \\
\textbf{9}                             & 40                                                                                                     & 4.51±6.1                       & 3.58±0.8                      \\ \hline
\end{tabular}%
}
\end{table}

\begin{figure}[htbp!]
    \centering
    
    \begin{subfigure}[b]{0.49\textwidth}
        \includegraphics[width=\textwidth]{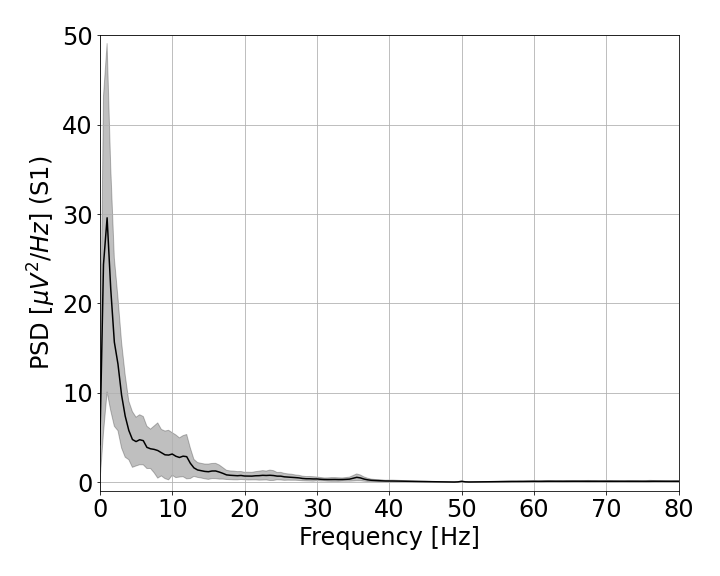}
        %\caption{Subject 1.}
        %\label{fig:outliers_neighborhood_15_subj_1}
    \end{subfigure}
    \begin{subfigure}[b]{0.49\textwidth}
        \includegraphics[width=\textwidth]{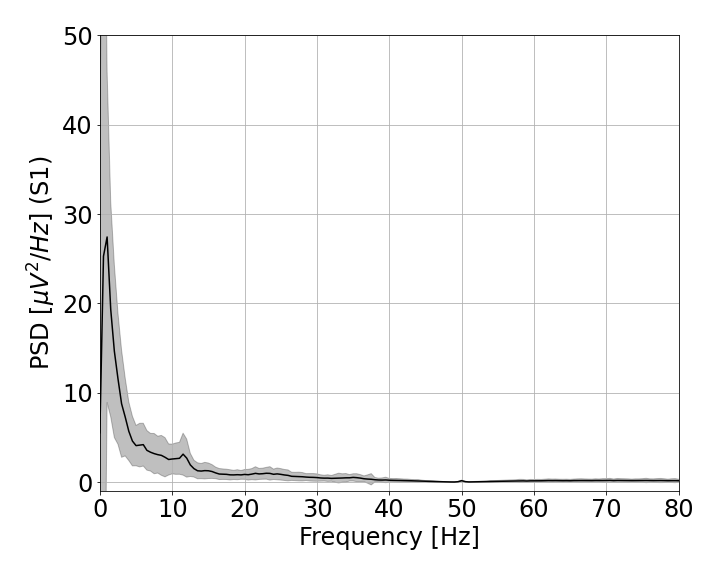}
        %\caption{Subject 2.}
        %\label{fig:outliers_neighborhood_15_subj_2}
    \end{subfigure}

    \begin{subfigure}[b]{0.49\textwidth}
        \includegraphics[width=\textwidth]{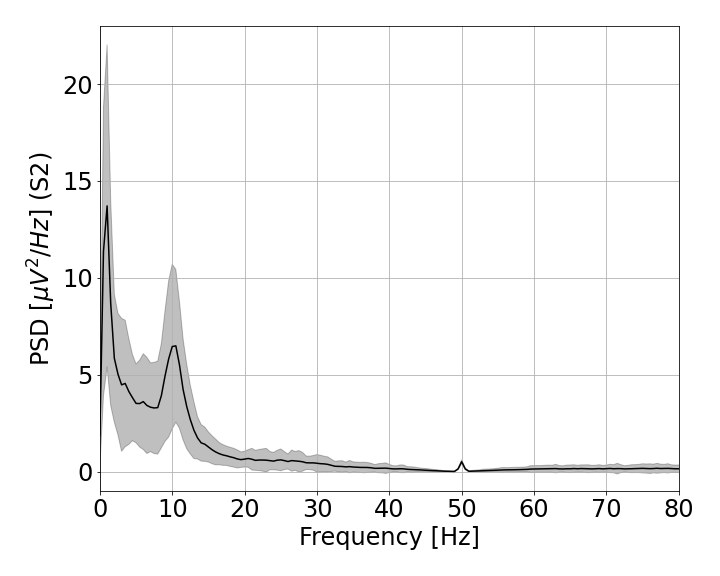}
        %\caption{Subject 1.}
        %\label{fig:outliers_neighborhood_15_subj_1}
    \end{subfigure}
    \begin{subfigure}[b]{0.49\textwidth}
        \includegraphics[width=\textwidth]{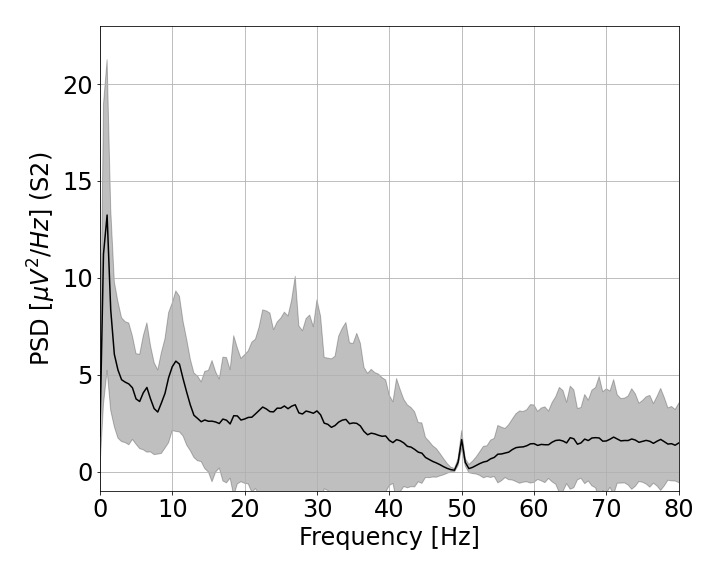}
        %\caption{Subject 2.}
        %\label{fig:outliers_neighborhood_15_subj_2}
    \end{subfigure}

    \begin{subfigure}[b]{0.49\textwidth}
        \includegraphics[width=\textwidth]{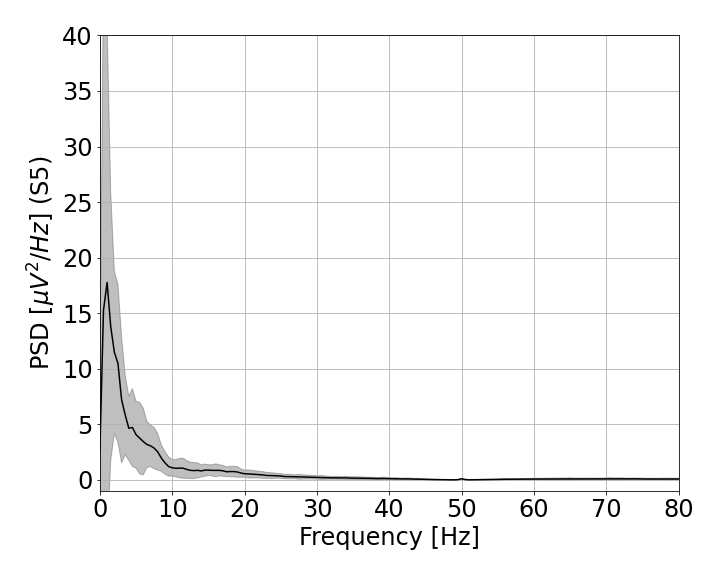}
    %    \caption{Training set.}
    %    \label{fig:average_spectra_train}
    \end{subfigure}
    \begin{subfigure}[b]{0.49\textwidth}
        \includegraphics[width=\textwidth]{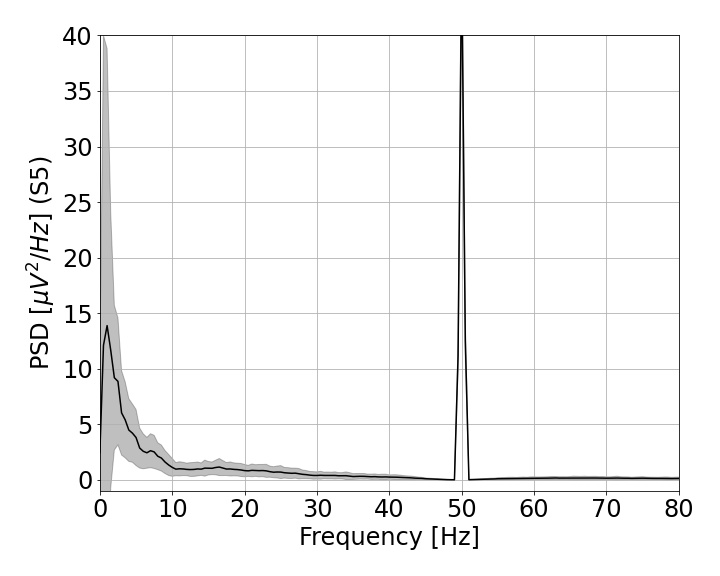}
    %      \caption{Test set.}
    %      \label{fig:average_spectra_test}
    \end{subfigure}

    \caption{Average power spectra for training set (panels on the left side) and test set (panels on the right side) for three subjects (S1, S2, and S5) at channel Cz.}
    \label{fig:psS1S2}
\end{figure}

Usually, as already discussed in Section~\ref{sec:state_of_the_art}, when using autoencoder architectures to identify outliers, the model is trained on \emph{normal} data~[\cite{ortiz_2020_dyslexia_autoencoder_anomaly}] and anomalies result from the model's largest errors~[\cite{Guansong_2021_anomaly_detection_review}]. Anyway, in more ecological acquisition scenarios~[\cite{muharemi2019machine}] and, frequently, when the human is \emph{in-the-loop}~[\cite{straetmans2022neural}], anomaly detectors can be successfully trained on a mixture of clean and noisy data, too~[\cite{al2021review}].
%
%
%
% GIUSTIFICAZIONE DEL NOSTRO MODO DI PROCEDERE
Anyway, for \gls{eeg} data, \emph{normality} cannot be easily defined and it is quite challenging to ensure a dataset to be anomaly-free. For example, this public dataset was supposed to be fully \emph{normal}, including a group of $9$ healthy subjects, acquired via a research-grade \gls{eeg} equipment, thus providing high data quality. Therefore, one might have expected to be able to build a robust anomaly detector based on this dataset.
However, we showed that other kinds of \emph{anomaly} are present and have been found by our \gls{hvdtw} model: e.g., artefactual data, that affect the training set and the test set in different rates, thus making challenging the design of a traditional anomaly detector on these data.
To support our claim and to deepen the investigation on those subjects having an out-of-normality distribution (i.e., S2 and S5, as already mentioned in Section~\ref{subsec:vEEGNet_vs_hvEEGNet}), we provide Fig.~\ref{fig:psS1S2}. It shows the average power spectra for the training set and the test set, separately, for three subjects at channel Cz. By inspecting this figure (and all other power spectra, not reported for space compactness), we realized how S2 and S5 are the only two individuals whose test sets were significantly different from all other data of this dataset. More specifically, we found out that the test set of S2 and of S5 (but not their training sets) are highly corrupted by noise and (muscular) artefacts. In fact, it is well-known~[\cite{buzsaki2004neuronal}] that the typical power spectrum of a clean \gls{eeg} acquired from a healthy subject follows a $1/f$ shape, with other relevant components (contributing as visible peaks) at the center of band of the $\alpha$ band (approx. $10$~Hz) and of the $\beta$ band (approx. $20$~Hz, generally less visible). In its upper panels, Fig.~\ref{fig:psS1S2} shows an example of clean dataset (from S1). Whereas, the lower panels report the power spectra of S2 and S5. It was decisive to visualize these spectra to realize that S2 has a \emph{normal} (average) power spectrum in his/her training set, while a highly noisy power spectrum in his/her test set. Furthermore, it could be easily recognized that the large power contribution in other frequency ranges (e.g., higher than $50$~Hz) is possibly be due to muscular activity that was simultaneously recorded by the \gls{eeg} electrodes during the test session~[\cite{chen2019removal}]. A similar situation was found for S5: again, all data coming from the test set were clearly corrupted by the $50$~Hz power supply. We might only guess that, for some reason, the notch filter at $50$~Hz (see Section~\ref{subsec:dataset}) was not actually applied for this subject during the second recording session, i.e., the test session.

This finally explained why our \gls{hvdtw} model dramatically failed at reconstructing S2 and S5 in their test sets, while keeping very satisfactory performance in the training phase.
Furthermore, this might also motivate why the large majority of the related work classifying (i.e., with \gls{dl} models, at least) this public dataset found the worst results on S2 and S5~[\cite{Zancanaro_2023_vEEGNet}]. 

\subsection{Computational complexity}
\label{subsec:computational_complexity}

Finally, we provide some reference measurements of the time spent in training and inference by our \gls{hvdtw} model.
We measured a training time, for each subject, of approximately $5$ minutes for $10$ epochs, on hardware freely available on Google Colab. This implies that a training run of $80$ epochs for a single subject approximately takes $40$ minutes. Note that the \gls{dtw} is computationally heavier than \gls{mse}, thus increasing the training times. On the other hand, we proved that a \gls{dtw}-based loss function leads to a significantly lower number of epochs needed for the training. Future improvements of our model might include the approximation of the \gls{dtw} function itself with a neural network, as recently proposed  by~\cite{Lerogeron_2023_dtw_1,Lerogeron_2023_dtw_2}.

Table~\ref{tab:inference_time} shows the inference time, i.e. the time the model used to encode and reconstruct 1, 10, 100, or 288 (all) \gls{eeg} segments, respectively, using four different machines, namely CPU~1, GPU~1, CPU~2, and GPU~2.
The details of the four machines are reported in the table's description.
Results refer to the average (and standard deviation) time needed to perform $20$ training runs.
\begin{table}[htbp!]
    \centering
    \caption{\gls{hvdtw} inference time on four different machines. CPU~1: Intel(R) Core(TM) i7-10750H CPU 2.60GHz (DELL G15 Laptop, 2020). GPU~1: NVIDIA GeForce RTX 2070 (DELL G15 Laptop, 2020). CPU~2: Intel(R) Xeon(R) CPU 2.20GHz (Google Colab). GPU~2: NVIDIA Tesla T4 (Google Colab).}.
    \label{tab:inference_time}
    \resizebox{\textwidth}{!}{%
    \begin{tabular}{|c|c|c|c|c|}
    \hline
    \textbf{\begin{tabular}[c]{@{}c@{}}Batch size \\ {[}no. of EEG segments{]}\end{tabular}} & \textbf{\begin{tabular}[c]{@{}c@{}}CPU~1\\ {[}s{]}\end{tabular}} & \textbf{\begin{tabular}[c]{@{}c@{}}GPU~1\\ {[}ms{]}\end{tabular}} & \textbf{\begin{tabular}[c]{@{}c@{}}CPU~2\\ {[}s{]}\end{tabular}} & \textbf{\begin{tabular}[c]{@{}c@{}}GPU~2\\ {[}ms{]}\end{tabular}} \\ \hline
    \textbf{1}                                                                               & 0.13 ± 0.05                                                           & 8 ± 0.9                                                                                        & 0.05 ± 0.003                                                          & 3.24 ± 0.26                                                            \\
    \textbf{10}                                                                              & 0.83 ± 0.13                                                           & 23.3 ± 17.2                                                                                    & 0.33 ± 0.03                                                           & 10.45 ±1.58                                                            \\
    \textbf{100}                                                                             & 5.91 ± 1.45                                                           & 63.7 ± 10.3                                                                                    & 5.02 ± 0.49                                                           & 62.15 ± 4.95                                                           \\
    \textbf{288 (all)}                                                                       & 10.69 ± 1.87                                                          & 182.7 ± 30.1                                                                                   & 14.73 ± 0.44                                                          & 179.16 ± 18.22                                                         \\ \hline
    \end{tabular}%
    }
\end{table}

% \begin{table}[htbp!]
%     \centering
%     \caption{\gls{hvdtw} inference time on four different machines. CPU~1: Intel(R) Core(TM) i7-10750H CPU 2.60GHz (DELL G15 Laptop, 2020). GPU~1: NVIDIA GeForce RTX 2070 (DELL G15 Laptop, 2020). CPU~2: Intel(R) Xeon(R) CPU 2.20GHz (Google Colab). GPU~2: NVIDIA Tesla T4 (Google Colab).}.
%     \label{tab:inference_time}
%     \resizebox{\linewidth}{!}{%
%     \begin{tabular}{|c||c|c||c|c|}
%         \cline{1-5}
%         No. \gls{eeg} segments     & CPU~1       & GPU~1          & CPU~2          & GPU~2              \\ \hline
%         \multicolumn{1}{|c||}{1}    & 0.13s±0.05  & 8ms±0.9ms      & 0.05s ± 0.003s & 3.24ms ± 0.26ms    \\
%         \multicolumn{1}{|c||}{10}  & 0.83s±0.13  & 23.3ms±17.2ms  & 0.33s ± 0.03s  & 10.45ms ±1.58ms    \\
%         \multicolumn{1}{|c||}{100} & 5.91s±1.45  & 63.7ms±10.3ms  & 5.02s ± 0.49s  & 62.15ms ± 4.95ms   \\
%         \multicolumn{1}{|c||}{288 (all)} & 10.69s±1.87 & 182.7ms±30.1ms & 14.73s ± 0.44s & 179.16ms ± 18.22ms \\ \hline
%     \end{tabular}
%     }
% \end{table}

%% LIMITATIONS

The present work still suffers from a number of limitations, e.g., the investigation of the balance between the different components of the training loss and their impact of the training course and quality. Their investigation and solution have been left to further studies in favor of a few relevant take-home messages that can be robustly supported by the results available so far. As one of many possible future perspectives, \gls{hvdtw} will be tested on different \gls{eeg} datasets, including different types of anomalies, to prove its generalizability and the extent to which it can identify either artefactual or pathological \gls{eeg} data.

%% CONCLUSIONS
\section{Conclusions} \label{sec:conclusions}

In this paper, we targeted the problem of \gls{eeg} reconstruction using two \gls{vae}-based models, which inherently provide a latent representation of the data that could then be used to enhance other \gls{dl}-based models for classication or anomaly detection tasks. Specifically, we proposed \gls{vdtw} and \gls{hvdtw} where we used a new loss function based on the \gls{dtw} similarity score which allowed us to significant improve over previous models based on the standard \gls{mse} loss function. Furthermore, with \gls{hvdtw}, we provided a \gls{vae} model which encodes the multi-channel \gls{eeg} data using a \gls{cnn} (i.e., the architecture of the popular EEGNet) and a hierarchical structure where three different latent spaces capture complementary information of the data.

We tested our models on the public, and very popular, \emph{Dataset 2a - BCI Competition IV}, where a $22$-channel \gls{eeg} dataset was collected from $9$ subjects repeatedly performing \gls{mi} of the right hand, the left hand, the feet, and the tongue. Our results showed that \gls{hvdtw} brings to a very high-fidelity reconstruction, outperforming our previous solutions as well as the other state-of-the-art models.
This outcome was also consistent across all subjects and repetitions.
Furthermore, we deeply investigated the reconstruction fidelity across every individual subject of the dataset, and found that \gls{hvdtw} failed to reconstruct some \gls{eeg} data belonging to the test sessions of S2 and S5. However, we realized that this failure could not be addressed to the poorness of the model, but rather to the input data quality. Interestingly, we were able to identify specific \gls{eeg} segments and channels where the raw \gls{eeg} data were corrupted (e.g., by saturation during the acquisition phase). Therefore, \gls{hvdtw} could be effectively employed as an automatic detector for noisy (e.g., artefactual) \gls{eeg} segments, thus supporting the domain experts. % in the time-consuming data cleaning phase as well as to identify  detection phases.
Also, with this deep investigation, we have finally provided an explanation for the high variability of classification results that are regularly found in the literature, when any \gls{ml} and \gls{dl}-based models are applied to this dataset.
%
% Then, we claim that are not accompanied by a robust validation or visual inspection.

This work opens new fundamental research questions, i.e., regarding the relationship between the \gls{dl}-based models training (including its effectiveness and possible biases) and the quality of the input data, still very poorly investigated in this application domain. This approach intends to adhere to the best scientific methodological practises of new AI methods applied to the medical domain, in line with~\cite{cabitza2021need}. In the future, \gls{hvdtw} and the proposed investigation methods could be applied to other \gls{eeg} datasets and to other multi-channel time-series to help in the modeling of complex dynamic systems, possibly not limited to neuroscience.

\subsection*{CRediT authorship contribution statement}
\textbf{Giulia Cisotto:} Conceptualization, Investigation, Methodology, Supervision, Validation, Visualization, Writing - original draft, Writing - review \& editing.
\textbf{Alberto Zancanaro:} Conceptualization, Data curation, Formal Analysis, Investigation, Software, Validation, Visualization, Writing - original draft, Writing - review \& editing.
\textbf{Italo F. Zoppis:} Funding acquisition, Supervision, Conceptualization, Writing - review \& editing.
\textbf{Sara L. Manzoni:} Project administration, Writing - review \& editing.

\subsection*{Acknowledgements}
This work was partially supported by the Italian Ministry of University and Research under the grant “Dipartimenti di Eccellenza 2023-2027” of the Department of Informatics, Systems and Communication of the University of Milano-Bicocca, Italy.
G. Cisotto also acknowledges the financial support of PON "Green and Innovation" 2014-2020 action IV.6 funded by the Italian Ministry of University and Research to the University of Milan-Bicocca (Milan, Italy).
A. Zancanaro acknowledges the financial support of PON "Green and Innovation" 2014-2020 action IV.5 funded by the Italian Ministry of University and Research to the MUR at the University of Padova (Padova, Italy).

% ---- Bibliography ----
%
%
%% For ESWA journal you need to use APA style
% \bibliographystyle{model5-names}\biboptions{authoryear}
\bibliography{biblio}

\begin{thebibliography}{75}
\expandafter\ifx\csname natexlab\endcsname\relax\def\natexlab#1{#1}\fi
\providecommand{\url}[1]{\texttt{#1}}
\providecommand{\href}[2]{#2}
\providecommand{\path}[1]{#1}
\providecommand{\DOIprefix}{doi:}
\providecommand{\ArXivprefix}{arXiv:}
\providecommand{\URLprefix}{URL: }
\providecommand{\Pubmedprefix}{pmid:}
\providecommand{\doi}[1]{\href{http://dx.doi.org/#1}{\path{#1}}}
\providecommand{\Pubmed}[1]{\href{pmid:#1}{\path{#1}}}
\providecommand{\bibinfo}[2]{#2}
\ifx\xfnm\relax \def\xfnm[#1]{\unskip,\space#1}\fi
%Type = Article
\bibitem[{Al-amri et~al.(2021)Al-amri, Murugesan, Man, Abdulateef, Al-Sharafi \& Alkahtani}]{al2021review}
\bibinfo{author}{Al-amri, R.}, \bibinfo{author}{Murugesan, R.~K.}, \bibinfo{author}{Man, M.}, \bibinfo{author}{Abdulateef, A.~F.}, \bibinfo{author}{Al-Sharafi, M.~A.}, \& \bibinfo{author}{Alkahtani, A.~A.} (\bibinfo{year}{2021}).
\newblock \bibinfo{title}{A review of machine learning and deep learning techniques for anomaly detection in iot data}.
\newblock {\it \bibinfo{journal}{Applied Sciences}\/},  {\it \bibinfo{volume}{11}\/}, \bibinfo{pages}{5320}.
%Type = Inproceedings
\bibitem[{Al-Marridi et~al.(2018)Al-Marridi, Mohamed \& Erbad}]{Marridi2018_AE_reconstruction_and_compression_d2a}
\bibinfo{author}{Al-Marridi, A.~Z.}, \bibinfo{author}{Mohamed, A.}, \& \bibinfo{author}{Erbad, A.} (\bibinfo{year}{2018}).
\newblock \bibinfo{title}{Convolutional autoencoder approach for eeg compression and reconstruction in m-health systems}.
\newblock In {\it \bibinfo{booktitle}{2018 14th International Wireless Communications \& Mobile Computing Conference (IWCMC)}\/} (pp. \bibinfo{pages}{370--375}).
\newblock \DOIprefix\doi{10.1109/IWCMC.2018.8450511}.
%Type = Article
\bibitem[{Anders \& Arnrich(2022)}]{anders2022wearable}
\bibinfo{author}{Anders, C.}, \& \bibinfo{author}{Arnrich, B.} (\bibinfo{year}{2022}).
\newblock \bibinfo{title}{Wearable electroencephalography and multi-modal mental state classification: A systematic literature review}.
\newblock {\it \bibinfo{journal}{Computers in Biology and Medicine}\/},  (p. \bibinfo{pages}{106088}).
%Type = Article
\bibitem[{Andrzejak et~al.(2002)Andrzejak, Lehnertz, Mormann, Rieke, David \& Elger}]{yang_2018_DSAE_EEG_dataset}
\bibinfo{author}{Andrzejak, R.}, \bibinfo{author}{Lehnertz, K.}, \bibinfo{author}{Mormann, F.}, \bibinfo{author}{Rieke, C.}, \bibinfo{author}{David, P.}, \& \bibinfo{author}{Elger, C.} (\bibinfo{year}{2002}).
\newblock \bibinfo{title}{Indications of nonlinear deterministic and finite-dimensional structures in time series of brain electrical activity: Dependence on recording region and brain state}.
\newblock {\it \bibinfo{journal}{Physical review. E, Statistical, nonlinear, and soft matter physics}\/},  {\it \bibinfo{volume}{64}\/}, \bibinfo{pages}{061907}. \DOIprefix\doi{10.1103/PhysRevE.64.061907}.
%Type = Inproceedings
\bibitem[{Ang et~al.(2008)Ang, Chin, Zhang \& Guan}]{ang2008filter}
\bibinfo{author}{Ang, K.~K.}, \bibinfo{author}{Chin, Z.~Y.}, \bibinfo{author}{Zhang, H.}, \& \bibinfo{author}{Guan, C.} (\bibinfo{year}{2008}).
\newblock \bibinfo{title}{Filter bank common spatial pattern (fbcsp) in brain-computer interface}.
\newblock In {\it \bibinfo{booktitle}{2008 IEEE international joint conference on neural networks (IEEE world congress on computational intelligence)}\/} (pp. \bibinfo{pages}{2390--2397}).
\newblock \bibinfo{organization}{IEEE}.
%Type = Article
\bibitem[{Bank{\'o} \& Abonyi(2012)}]{DTWcorrelation2012}
\bibinfo{author}{Bank{\'o}, Z.}, \& \bibinfo{author}{Abonyi, J.} (\bibinfo{year}{2012}).
\newblock \bibinfo{title}{Correlation based dynamic time warping of multivariate time series}.
\newblock {\it \bibinfo{journal}{Expert Systems with Applications}\/},  {\it \bibinfo{volume}{39}\/}, \bibinfo{pages}{12814--12823}.
%Type = Article
\bibitem[{Beraldo et~al.(2022)Beraldo, Tonin, Millán \& Menegatti}]{Beraldo2022}
\bibinfo{author}{Beraldo, G.}, \bibinfo{author}{Tonin, L.}, \bibinfo{author}{Millán, J. d.~R.}, \& \bibinfo{author}{Menegatti, E.} (\bibinfo{year}{2022}).
\newblock \bibinfo{title}{Shared intelligence for robot teleoperation via bmi}.
\newblock {\it \bibinfo{journal}{IEEE Transactions on Human-Machine Systems}\/},  {\it \bibinfo{volume}{52}\/}, \bibinfo{pages}{400--409}. \DOIprefix\doi{10.1109/THMS.2021.3137035}.
%Type = Article
\bibitem[{Berger(1929)}]{berger1929eeg}
\bibinfo{author}{Berger, H.} (\bibinfo{year}{1929}).
\newblock \bibinfo{title}{On the eeg in humans}.
\newblock {\it \bibinfo{journal}{Arch. Psychiatr. Nervenkr}\/},  {\it \bibinfo{volume}{87}\/}, \bibinfo{pages}{527--570}.
%Type = Inproceedings
\bibitem[{Bethge et~al.(2022)Bethge, Hallgarten, Grosse-Puppendahl, Kari, Chuang, Özdenizci \& Schmidt}]{Bethge_2022_EEG2VEC}
\bibinfo{author}{Bethge, D.}, \bibinfo{author}{Hallgarten, P.}, \bibinfo{author}{Grosse-Puppendahl, T.}, \bibinfo{author}{Kari, M.}, \bibinfo{author}{Chuang, L.~L.}, \bibinfo{author}{Özdenizci, O.}, \& \bibinfo{author}{Schmidt, A.} (\bibinfo{year}{2022}).
\newblock \bibinfo{title}{Eeg2vec: Learning affective eeg representations via variational autoencoders}.
\newblock In {\it \bibinfo{booktitle}{2022 IEEE International Conference on Systems, Man, and Cybernetics (SMC)}\/} (pp. \bibinfo{pages}{3150--3157}).
\newblock \DOIprefix\doi{10.1109/SMC53654.2022.9945517}.
%Type = Article
\bibitem[{Blankertz et~al.(2007)Blankertz, Dornhege, Krauledat, Müller \& Curio}]{dataset_BCI_competition}
\bibinfo{author}{Blankertz, B.}, \bibinfo{author}{Dornhege, G.}, \bibinfo{author}{Krauledat, M.}, \bibinfo{author}{Müller, K.-R.}, \& \bibinfo{author}{Curio, G.} (\bibinfo{year}{2007}).
\newblock \bibinfo{title}{The non-invasive {B}erlin {B}rain-{C}omputer {I}nterface: Fast acquisition of effective performance in untrained subjects}.
\newblock {\it \bibinfo{journal}{NeuroImage}\/},  {\it \bibinfo{volume}{37}\/}, \bibinfo{pages}{539--50}. \DOIprefix\doi{10.1016/j.neuroimage.2007.01.051}.
%Type = Article
\bibitem[{Blei et~al.(2017)Blei, Kucukelbir \& McAuliffe}]{blei2017variational}
\bibinfo{author}{Blei, D.~M.}, \bibinfo{author}{Kucukelbir, A.}, \& \bibinfo{author}{McAuliffe, J.~D.} (\bibinfo{year}{2017}).
\newblock \bibinfo{title}{Variational inference: A review for statisticians}.
\newblock {\it \bibinfo{journal}{Journal of the American statistical Association}\/},  {\it \bibinfo{volume}{112}\/}, \bibinfo{pages}{859--877}.
%Type = Article
\bibitem[{Bressan et~al.(2021)Bressan, Cisotto, M{\"u}ller-Putz \& Wriessnegger}]{Bressan2021}
\bibinfo{author}{Bressan, G.}, \bibinfo{author}{Cisotto, G.}, \bibinfo{author}{M{\"u}ller-Putz, G.~R.}, \& \bibinfo{author}{Wriessnegger, S.~C.} (\bibinfo{year}{2021}).
\newblock \bibinfo{title}{Deep learning-based classification of fine hand movements from low frequency {eeg}}.
\newblock {\it \bibinfo{journal}{Future Internet}\/},  {\it \bibinfo{volume}{13}\/}, \bibinfo{pages}{103}.
%Type = Article
\bibitem[{Buzsaki \& Draguhn(2004)}]{buzsaki2004neuronal}
\bibinfo{author}{Buzsaki, G.}, \& \bibinfo{author}{Draguhn, A.} (\bibinfo{year}{2004}).
\newblock \bibinfo{title}{Neuronal oscillations in cortical networks}.
\newblock {\it \bibinfo{journal}{science}\/},  {\it \bibinfo{volume}{304}\/}, \bibinfo{pages}{1926--1929}.
%Type = Misc
\bibitem[{Cabitza \& Campagner(2021)}]{cabitza2021need}
\bibinfo{author}{Cabitza, F.}, \& \bibinfo{author}{Campagner, A.} (\bibinfo{year}{2021}).
\newblock \bibinfo{title}{The need to separate the wheat from the chaff in medical informatics: Introducing a comprehensive checklist for the (self)-assessment of medical ai studies}.
%Type = Article
\bibitem[{Chen et~al.(2019)Chen, Xu, Liu, Lee, Chen, Zhang, McKeown \& Wang}]{chen2019removal}
\bibinfo{author}{Chen, X.}, \bibinfo{author}{Xu, X.}, \bibinfo{author}{Liu, A.}, \bibinfo{author}{Lee, S.}, \bibinfo{author}{Chen, X.}, \bibinfo{author}{Zhang, X.}, \bibinfo{author}{McKeown, M.~J.}, \& \bibinfo{author}{Wang, Z.~J.} (\bibinfo{year}{2019}).
\newblock \bibinfo{title}{Removal of muscle artifacts from the eeg: A review and recommendations}.
\newblock {\it \bibinfo{journal}{IEEE Sensors Journal}\/},  {\it \bibinfo{volume}{19}\/}, \bibinfo{pages}{5353--5368}.
%Type = Inproceedings
\bibitem[{Cisotto et~al.(2015)Cisotto, Silvano et~al.}]{cisotto2015displacement}
\bibinfo{author}{Cisotto, G.}, \bibinfo{author}{Silvano, P.} et~al. (\bibinfo{year}{2015}).
\newblock \bibinfo{title}{Real-time detection of eeg electrode displacement for brain-computer interface applications}.
\newblock In {\it \bibinfo{booktitle}{Proceedings of 5th International Conference on Wireless Communications, Vehicular Technology, Information Theory and Aerospace \& Electronic Systems (Wireless VITAE)}\/} (pp. \bibinfo{pages}{1--5}).
%Type = Article
\bibitem[{Cover \& Hart(1967)}]{kNN1967}
\bibinfo{author}{Cover, T.}, \& \bibinfo{author}{Hart, P.} (\bibinfo{year}{1967}).
\newblock \bibinfo{title}{Nearest neighbor pattern classification}.
\newblock {\it \bibinfo{journal}{IEEE transactions on information theory}\/},  {\it \bibinfo{volume}{13}\/}, \bibinfo{pages}{21--27}.
%Type = Inproceedings
\bibitem[{Cuturi \& Blondel(2017)}]{Cuturi_2017_SoftDTWAD}
\bibinfo{author}{Cuturi, M.}, \& \bibinfo{author}{Blondel, M.} (\bibinfo{year}{2017}).
\newblock \bibinfo{title}{Soft-dtw: a differentiable loss function for time-series}.
\newblock In {\it \bibinfo{booktitle}{International Conference on Machine Learning}\/}.
\newblock \URLprefix \url{https://api.semanticscholar.org/CorpusID:9566599}.
%Type = Article
\bibitem[{Dasan \& Gnanaraj(2022)}]{Dasan_2022_ECG_EMG_EEG_compression_and_reconstruction}
\bibinfo{author}{Dasan, E.}, \& \bibinfo{author}{Gnanaraj, R.} (\bibinfo{year}{2022}).
\newblock \bibinfo{title}{Joint ecg--emg--eeg signal compression and reconstruction with incremental multimodal autoencoder approach}.
\newblock {\it \bibinfo{journal}{Circuits, Systems, and Signal Processing}\/},  {\it \bibinfo{volume}{41}\/}, \bibinfo{pages}{6152--6181}. \URLprefix \url{https://doi.org/10.1007/s00034-022-02071-x}. \DOIprefix\doi{10.1007/s00034-022-02071-x}.
%Type = Article
\bibitem[{{De Vos} et~al.(2017){De Vos}, Vanvooren, Vanderauwera, Ghesquière \& Wouters}]{vos_2017_dataset_used_for_ortiz_2020}
\bibinfo{author}{{De Vos}, A.}, \bibinfo{author}{Vanvooren, S.}, \bibinfo{author}{Vanderauwera, J.}, \bibinfo{author}{Ghesquière, P.}, \& \bibinfo{author}{Wouters, J.} (\bibinfo{year}{2017}).
\newblock \bibinfo{title}{A longitudinal study investigating neural processing of speech envelope modulation rates in children with (a family risk for) dyslexia}.
\newblock {\it \bibinfo{journal}{Cortex}\/},  {\it \bibinfo{volume}{93}\/}, \bibinfo{pages}{206--219}. \URLprefix \url{https://www.sciencedirect.com/science/article/pii/S0010945217301624}. \DOIprefix\doi{https://doi.org/10.1016/j.cortex.2017.05.007}.
%Type = Article
\bibitem[{Dong \& Japkowicz(2018)}]{dong_2018_threaded_autoencoder_streaming}
\bibinfo{author}{Dong, Y.}, \& \bibinfo{author}{Japkowicz, N.} (\bibinfo{year}{2018}).
\newblock \bibinfo{title}{Threaded ensembles of autoencoders for stream learning}.
\newblock {\it \bibinfo{journal}{Computational Intelligence}\/},  {\it \bibinfo{volume}{34}\/}, \bibinfo{pages}{261--281}.
%Type = Article
\bibitem[{Durka et~al.(2003)Durka, Klekowicz, Blinowska, Szelenberger \& Niemcewicz}]{durka2003simple}
\bibinfo{author}{Durka, P.~J.}, \bibinfo{author}{Klekowicz, H.}, \bibinfo{author}{Blinowska, K.~J.}, \bibinfo{author}{Szelenberger, W.}, \& \bibinfo{author}{Niemcewicz, S.} (\bibinfo{year}{2003}).
\newblock \bibinfo{title}{A simple system for detection of eeg artifacts in polysomnographic recordings}.
\newblock {\it \bibinfo{journal}{IEEE transactions on biomedical engineering}\/},  {\it \bibinfo{volume}{50}\/}, \bibinfo{pages}{526--528}.
%Type = Article
\bibitem[{Emami et~al.(2019{\natexlab{a}})Emami, Kunii, Matsuo, Shinozaki, Kawai \& Takahashi}]{emami_2019_autencoder_epileptic_seizure}
\bibinfo{author}{Emami, A.}, \bibinfo{author}{Kunii, N.}, \bibinfo{author}{Matsuo, T.}, \bibinfo{author}{Shinozaki, T.}, \bibinfo{author}{Kawai, K.}, \& \bibinfo{author}{Takahashi, H.} (\bibinfo{year}{2019}{\natexlab{a}}).
\newblock \bibinfo{title}{Autoencoding of long-term scalp electroencephalogram to detect epileptic seizure for diagnosis support system}.
\newblock {\it \bibinfo{journal}{Computers in Biology and Medicine}\/},  {\it \bibinfo{volume}{110}\/}, \bibinfo{pages}{227--233}. \URLprefix \url{https://www.sciencedirect.com/science/article/pii/S0010482519301933}. \DOIprefix\doi{https://doi.org/10.1016/j.compbiomed.2019.05.025}.
%Type = Article
\bibitem[{Emami et~al.(2019{\natexlab{b}})Emami, Kunii, Matsuo, Shinozaki, Kawai \& Takahashi}]{emamai_2019_dataset}
\bibinfo{author}{Emami, A.}, \bibinfo{author}{Kunii, N.}, \bibinfo{author}{Matsuo, T.}, \bibinfo{author}{Shinozaki, T.}, \bibinfo{author}{Kawai, K.}, \& \bibinfo{author}{Takahashi, H.} (\bibinfo{year}{2019}{\natexlab{b}}).
\newblock \bibinfo{title}{Seizure detection by convolutional neural network-based analysis of scalp electroencephalography plot images}.
\newblock {\it \bibinfo{journal}{NeuroImage: Clinical}\/},  {\it \bibinfo{volume}{22}\/}, \bibinfo{pages}{101684}. \URLprefix \url{https://www.sciencedirect.com/science/article/pii/S2213158219300348}. \DOIprefix\doi{https://doi.org/10.1016/j.nicl.2019.101684}.
%Type = Article
\bibitem[{Gabardi et~al.(2023)Gabardi, Saibene, Gasparini, Rizzo \& Stella}]{gabardi2023multi}
\bibinfo{author}{Gabardi, M.}, \bibinfo{author}{Saibene, A.}, \bibinfo{author}{Gasparini, F.}, \bibinfo{author}{Rizzo, D.}, \& \bibinfo{author}{Stella, F.~A.} (\bibinfo{year}{2023}).
\newblock \bibinfo{title}{A multi-artifact eeg denoising by frequency-based deep learning}.
\newblock {\it \bibinfo{journal}{arXiv preprint arXiv:2310.17335}\/}, .
%Type = Article
\bibitem[{Gao et~al.(2010)Gao, Yang, Sun \& Yu}]{gao2010automatic}
\bibinfo{author}{Gao, J.}, \bibinfo{author}{Yang, Y.}, \bibinfo{author}{Sun, J.}, \& \bibinfo{author}{Yu, G.} (\bibinfo{year}{2010}).
\newblock \bibinfo{title}{Automatic removal of various artifacts from eeg signals using combined methods}.
\newblock {\it \bibinfo{journal}{Journal of Clinical Neurophysiology}\/},  {\it \bibinfo{volume}{27}\/}, \bibinfo{pages}{312--320}.
%Type = Article
\bibitem[{Gyori et~al.(2022)Gyori, Palombo, Clark, Zhang \& Alexander}]{Gyori_2022_data_distribution_impacts_ml}
\bibinfo{author}{Gyori, N.~G.}, \bibinfo{author}{Palombo, M.}, \bibinfo{author}{Clark, C.~A.}, \bibinfo{author}{Zhang, H.}, \& \bibinfo{author}{Alexander, D.~C.} (\bibinfo{year}{2022}).
\newblock \bibinfo{title}{Training data distribution significantly impacts the estimation of tissue microstructure with machine learning}.
\newblock {\it \bibinfo{journal}{Magnetic Resonance in Medicine}\/},  {\it \bibinfo{volume}{87}\/}, \bibinfo{pages}{932--947}. \URLprefix \url{https://onlinelibrary.wiley.com/doi/abs/10.1002/mrm.29014}. \DOIprefix\doi{https://doi.org/10.1002/mrm.29014}. \href{http://arxiv.org/abs/https://onlinelibrary.wiley.com/doi/pdf/10.1002/mrm.29014}{\tt arXiv:https://onlinelibrary.wiley.com/doi/pdf/10.1002/mrm.29014}.
%Type = Article
\bibitem[{Hosseini et~al.(2020)Hosseini, Hosseini \& Ahi}]{hosseini2020review}
\bibinfo{author}{Hosseini, M.-P.}, \bibinfo{author}{Hosseini, A.}, \& \bibinfo{author}{Ahi, K.} (\bibinfo{year}{2020}).
\newblock \bibinfo{title}{A review on machine learning for eeg signal processing in bioengineering}.
\newblock {\it \bibinfo{journal}{IEEE reviews in biomedical engineering}\/},  {\it \bibinfo{volume}{14}\/}, \bibinfo{pages}{204--218}.
%Type = Article
\bibitem[{Huang \& Jansen(1985)}]{DTWonEEG1985}
\bibinfo{author}{Huang, H.-C.}, \& \bibinfo{author}{Jansen, B.} (\bibinfo{year}{1985}).
\newblock \bibinfo{title}{Eeg waveform analysis by means of dynamic time-warping}.
\newblock {\it \bibinfo{journal}{International journal of bio-medical computing}\/},  {\it \bibinfo{volume}{17}\/}, \bibinfo{pages}{135--144}.
%Type = Article
\bibitem[{Jayaram \& Barachant(2018)}]{moabb_2018}
\bibinfo{author}{Jayaram, V.}, \& \bibinfo{author}{Barachant, A.} (\bibinfo{year}{2018}).
\newblock \bibinfo{title}{Moabb: trustworthy algorithm benchmarking for bcis}.
\newblock {\it \bibinfo{journal}{Journal of neural engineering}\/},  {\it \bibinfo{volume}{15}\/}, \bibinfo{pages}{066011}.
%Type = Article
\bibitem[{Kaiser et~al.(2012)Kaiser, Daly, Pichiorri, Mattia, M{\"u}ller-Putz \& Neuper}]{kaiser2012relationship}
\bibinfo{author}{Kaiser, V.}, \bibinfo{author}{Daly, I.}, \bibinfo{author}{Pichiorri, F.}, \bibinfo{author}{Mattia, D.}, \bibinfo{author}{M{\"u}ller-Putz, G.~R.}, \& \bibinfo{author}{Neuper, C.} (\bibinfo{year}{2012}).
\newblock \bibinfo{title}{Relationship between electrical brain responses to motor imagery and motor impairment in stroke}.
\newblock {\it \bibinfo{journal}{Stroke}\/},  {\it \bibinfo{volume}{43}\/}, \bibinfo{pages}{2735--2740}.
%Type = Article
\bibitem[{Khan et~al.(2023)Khan, Khan, Altaf \& Abbasi}]{khan_2023_autoenceder_seizures}
\bibinfo{author}{Khan, G.~H.}, \bibinfo{author}{Khan, N.~A.}, \bibinfo{author}{Altaf, M. A.~B.}, \& \bibinfo{author}{Abbasi, Q.} (\bibinfo{year}{2023}).
\newblock \bibinfo{title}{A shallow autoencoder framework for epileptic seizure detection in eeg signals}.
\newblock {\it \bibinfo{journal}{Sensors}\/},  {\it \bibinfo{volume}{23}\/}. \URLprefix \url{https://www.mdpi.com/1424-8220/23/8/4112}. \DOIprefix\doi{10.3390/s23084112}.
%Type = Article
\bibitem[{Kingma \& Welling(2013)}]{VAE_ORIGINAL_PAPER_kingma}
\bibinfo{author}{Kingma, D.~P.}, \& \bibinfo{author}{Welling, M.} (\bibinfo{year}{2013}).
\newblock \bibinfo{title}{Auto-encoding variational bayes}.
\newblock {\it \bibinfo{journal}{arXiv preprint arXiv:1312.6114}\/}, .
%Type = Article
\bibitem[{Kingma \& Welling(2019)}]{kingma2019introduction}
\bibinfo{author}{Kingma, D.~P.}, \& \bibinfo{author}{Welling, M.} (\bibinfo{year}{2019}).
\newblock \bibinfo{title}{An introduction to variational autoencoders}.
\newblock {\it \bibinfo{journal}{arXiv preprint arXiv:1906.02691}\/}, .
%Type = Article
\bibitem[{Kodama et~al.(2023)Kodama, Iwama, Morishige \& Ushiba}]{kodama2023thirty}
\bibinfo{author}{Kodama, M.}, \bibinfo{author}{Iwama, S.}, \bibinfo{author}{Morishige, M.}, \& \bibinfo{author}{Ushiba, J.} (\bibinfo{year}{2023}).
\newblock \bibinfo{title}{Thirty-minute motor imagery exercise aided by eeg sensorimotor rhythm neurofeedback enhances morphing of sensorimotor cortices: a double-blind sham-controlled study}.
\newblock {\it \bibinfo{journal}{Cerebral Cortex}\/},  {\it \bibinfo{volume}{33}\/}, \bibinfo{pages}{6573--6584}.
%Type = Article
\bibitem[{Koelstra et~al.(2012)Koelstra, Muhl, Soleymani, Lee, Yazdani, Ebrahimi, Pun, Nijholt \& Patras}]{koelstra_2012_DEAP_dataset}
\bibinfo{author}{Koelstra, S.}, \bibinfo{author}{Muhl, C.}, \bibinfo{author}{Soleymani, M.}, \bibinfo{author}{Lee, J.-S.}, \bibinfo{author}{Yazdani, A.}, \bibinfo{author}{Ebrahimi, T.}, \bibinfo{author}{Pun, T.}, \bibinfo{author}{Nijholt, A.}, \& \bibinfo{author}{Patras, I.} (\bibinfo{year}{2012}).
\newblock \bibinfo{title}{Deap: A database for emotion analysis ;using physiological signals}.
\newblock {\it \bibinfo{journal}{IEEE Transactions on Affective Computing}\/},  {\it \bibinfo{volume}{3}\/}, \bibinfo{pages}{18--31}. \DOIprefix\doi{10.1109/T-AFFC.2011.15}.
%Type = Article
\bibitem[{Lascano et~al.(2017)Lascano, Lalive, Hardmeier, Fuhr \& Seeck}]{lascano2017clinical}
\bibinfo{author}{Lascano, A.~M.}, \bibinfo{author}{Lalive, P.~H.}, \bibinfo{author}{Hardmeier, M.}, \bibinfo{author}{Fuhr, P.}, \& \bibinfo{author}{Seeck, M.} (\bibinfo{year}{2017}).
\newblock \bibinfo{title}{Clinical evoked potentials in neurology: a review of techniques and indications}.
\newblock {\it \bibinfo{journal}{Journal of Neurology, Neurosurgery \& Psychiatry}\/},  {\it \bibinfo{volume}{88}\/}, \bibinfo{pages}{688--696}.
%Type = Article
\bibitem[{Lawhern et~al.(2016)Lawhern, Solon, Waytowich, Gordon, Hung \& Lance}]{EEGNet_paper}
\bibinfo{author}{Lawhern, V.}, \bibinfo{author}{Solon, A.}, \bibinfo{author}{Waytowich, N.}, \bibinfo{author}{Gordon, S.}, \bibinfo{author}{Hung, C.}, \& \bibinfo{author}{Lance, B.} (\bibinfo{year}{2016}).
\newblock \bibinfo{title}{{EEGN}et: A compact convolutional network for {EEG}-based {B}rain-{C}omputer {I}nterfaces}.
\newblock {\it \bibinfo{journal}{Journal of Neural Engineering}\/},  {\it \bibinfo{volume}{15}\/}. \DOIprefix\doi{10.1088/1741-2552/aace8c}.
%Type = Article
\bibitem[{Lerogeron et~al.(2023{\natexlab{a}})Lerogeron, Picot-Clémente, Heutte \& Rakotomamonjy}]{Lerogeron_2023_dtw_1}
\bibinfo{author}{Lerogeron, H.}, \bibinfo{author}{Picot-Clémente, R.}, \bibinfo{author}{Heutte, L.}, \& \bibinfo{author}{Rakotomamonjy, A.} (\bibinfo{year}{2023}{\natexlab{a}}).
\newblock \bibinfo{title}{Learning an autoencoder to compress eeg signals via a neural network based approximation of dtw}.
\newblock {\it \bibinfo{journal}{Procedia Computer Science}\/},  {\it \bibinfo{volume}{222}\/}, \bibinfo{pages}{448--457}. \URLprefix \url{https://www.sciencedirect.com/science/article/pii/S1877050923009481}. \DOIprefix\doi{https://doi.org/10.1016/j.procs.2023.08.183}.
\newblock \bibinfo{note}{International Neural Network Society Workshop on Deep Learning Innovations and Applications (INNS DLIA 2023)}.
%Type = Article
\bibitem[{Lerogeron et~al.(2023{\natexlab{b}})Lerogeron, Picot-Clémente, Rakotomamonjy \& Heutte}]{Lerogeron_2023_dtw_2}
\bibinfo{author}{Lerogeron, H.}, \bibinfo{author}{Picot-Clémente, R.}, \bibinfo{author}{Rakotomamonjy, A.}, \& \bibinfo{author}{Heutte, L.} (\bibinfo{year}{2023}{\natexlab{b}}).
\newblock \bibinfo{title}{Approximating dynamic time warping with a convolutional neural network on eeg data}.
\newblock {\it \bibinfo{journal}{Pattern Recognition Letters}\/},  {\it \bibinfo{volume}{171}\/}, \bibinfo{pages}{162--169}. \URLprefix \url{https://www.sciencedirect.com/science/article/pii/S0167865523001460}. \DOIprefix\doi{https://doi.org/10.1016/j.patrec.2023.05.012}.
%Type = Article
\bibitem[{Li et~al.(2019)Li, Wang, Xu \& Fang}]{DFNN_AC_TAB_7}
\bibinfo{author}{Li, D.}, \bibinfo{author}{Wang, J.}, \bibinfo{author}{Xu, J.}, \& \bibinfo{author}{Fang, X.} (\bibinfo{year}{2019}).
\newblock \bibinfo{title}{Densely feature fusion based on convolutional neural networks for motor imagery {EEG} classification}.
\newblock {\it \bibinfo{journal}{IEEE Access}\/},  {\it \bibinfo{volume}{7}\/}, \bibinfo{pages}{132720--132730}. \DOIprefix\doi{10.1109/ACCESS.2019.2941867}.
%Type = Misc
\bibitem[{Li et~al.(2022)Li, Zhao, Zhang, Sun, Chen, Wen, Ma \& Pei}]{li_2022_dataset_kpi}
\bibinfo{author}{Li, Z.}, \bibinfo{author}{Zhao, N.}, \bibinfo{author}{Zhang, S.}, \bibinfo{author}{Sun, Y.}, \bibinfo{author}{Chen, P.}, \bibinfo{author}{Wen, X.}, \bibinfo{author}{Ma, M.}, \& \bibinfo{author}{Pei, D.} (\bibinfo{year}{2022}).
\newblock \bibinfo{title}{Constructing large-scale real-world benchmark datasets for aiops}.
\newblock \href{http://arxiv.org/abs/2208.03938}{\tt arXiv:2208.03938}.
%Type = Article
\bibitem[{Liu et~al.(2020)Liu, Wu, Luo, Qiu, Yang, Li \& Bi}]{liu_2020_eeg_emotion_SAE_CNN}
\bibinfo{author}{Liu, J.}, \bibinfo{author}{Wu, G.}, \bibinfo{author}{Luo, Y.}, \bibinfo{author}{Qiu, S.}, \bibinfo{author}{Yang, S.}, \bibinfo{author}{Li, W.}, \& \bibinfo{author}{Bi, Y.} (\bibinfo{year}{2020}).
\newblock \bibinfo{title}{Eeg-based emotion classification using a deep neural network and sparse autoencoder}.
\newblock {\it \bibinfo{journal}{Frontiers in Systems Neuroscience}\/},  {\it \bibinfo{volume}{14}\/}. \URLprefix \url{https://www.frontiersin.org/articles/10.3389/fnsys.2020.00043}. \DOIprefix\doi{10.3389/fnsys.2020.00043}.
%Type = Article
\bibitem[{Lotte et~al.(2018)Lotte, Bougrain, Cichocki, Clerc, Congedo, Rakotomamonjy \& Yger}]{congedo2018review}
\bibinfo{author}{Lotte, F.}, \bibinfo{author}{Bougrain, L.}, \bibinfo{author}{Cichocki, A.}, \bibinfo{author}{Clerc, M.}, \bibinfo{author}{Congedo, M.}, \bibinfo{author}{Rakotomamonjy, A.}, \& \bibinfo{author}{Yger, F.} (\bibinfo{year}{2018}).
\newblock \bibinfo{title}{A review of classification algorithms for eeg-based brain--computer interfaces: a 10 year update}.
\newblock {\it \bibinfo{journal}{Journal of neural engineering}\/},  {\it \bibinfo{volume}{15}\/}, \bibinfo{pages}{031005}.
%Type = Phdthesis
\bibitem[{Maghoumi(2020)}]{maghoumi2020dissertation_softDTWCUDA_1}
\bibinfo{author}{Maghoumi, M.} (\bibinfo{year}{2020}).
\newblock {\it \bibinfo{title}{{Deep Recurrent Networks for Gesture Recognition and Synthesis}}\/}.
\newblock Ph.D. thesis University of Central Florida Orlando, Florida.
%Type = Inproceedings
\bibitem[{Maghoumi et~al.(2021)Maghoumi, Taranta \& LaViola}]{maghoumi2021deepnag_softDTWCUDA_2}
\bibinfo{author}{Maghoumi, M.}, \bibinfo{author}{Taranta, E.~M.}, \& \bibinfo{author}{LaViola, J.} (\bibinfo{year}{2021}).
\newblock \bibinfo{title}{Deepnag: Deep non-adversarial gesture generation}.
\newblock In {\it \bibinfo{booktitle}{26th International Conference on Intelligent User Interfaces}\/} (pp. \bibinfo{pages}{213--223}).
%Type = Article
\bibitem[{Muharemi et~al.(2019)Muharemi, Logof{\u{a}}tu \& Leon}]{muharemi2019machine}
\bibinfo{author}{Muharemi, F.}, \bibinfo{author}{Logof{\u{a}}tu, D.}, \& \bibinfo{author}{Leon, F.} (\bibinfo{year}{2019}).
\newblock \bibinfo{title}{Machine learning approaches for anomaly detection of water quality on a real-world data set}.
\newblock {\it \bibinfo{journal}{Journal of Information and Telecommunication}\/},  {\it \bibinfo{volume}{3}\/}, \bibinfo{pages}{294--307}.
%Type = Article
\bibitem[{Munari et~al.(2023)Munari, Cola \& Badia}]{munari2023local}
\bibinfo{author}{Munari, A.}, \bibinfo{author}{Cola, T.}, \& \bibinfo{author}{Badia, L.} (\bibinfo{year}{2023}).
\newblock \bibinfo{title}{Local or edge/cloud processing for data freshness}, .
%Type = Article
\bibitem[{Ofner et~al.(2019)Ofner, Schwarz, Pereira, Wyss, Wildburger \& M{\"u}ller-Putz}]{MullerPutz2019}
\bibinfo{author}{Ofner, P.}, \bibinfo{author}{Schwarz, A.}, \bibinfo{author}{Pereira, J.}, \bibinfo{author}{Wyss, D.}, \bibinfo{author}{Wildburger, R.}, \& \bibinfo{author}{M{\"u}ller-Putz, G.~R.} (\bibinfo{year}{2019}).
\newblock \bibinfo{title}{Attempted arm and hand movements can be decoded from low-frequency eeg from persons with spinal cord injury}.
\newblock {\it \bibinfo{journal}{Scientific reports}\/},  {\it \bibinfo{volume}{9}\/}, \bibinfo{pages}{7134}.
%Type = Article
\bibitem[{Ortiz et~al.(2020)Ortiz, Martinez-Murcia, Luque, Gim{\'e}nez, Morales-Ortega \& Ortega}]{ortiz_2020_dyslexia_autoencoder_anomaly}
\bibinfo{author}{Ortiz, A.}, \bibinfo{author}{Martinez-Murcia, F.~J.}, \bibinfo{author}{Luque, J.~L.}, \bibinfo{author}{Gim{\'e}nez, A.}, \bibinfo{author}{Morales-Ortega, R.}, \& \bibinfo{author}{Ortega, J.} (\bibinfo{year}{2020}).
\newblock \bibinfo{title}{Dyslexia diagnosis by eeg temporal and spectral descriptors: An anomaly detection approach}.
\newblock {\it \bibinfo{journal}{International Journal of Neural Systems}\/},  {\it \bibinfo{volume}{30}\/}, \bibinfo{pages}{2050029}.
%Type = Article
\bibitem[{Pang et~al.(2021)Pang, Shen, Cao \& Hengel}]{Guansong_2021_anomaly_detection_review}
\bibinfo{author}{Pang, G.}, \bibinfo{author}{Shen, C.}, \bibinfo{author}{Cao, L.}, \& \bibinfo{author}{Hengel, A. V.~D.} (\bibinfo{year}{2021}).
\newblock \bibinfo{title}{Deep learning for anomaly detection: A review}.
\newblock {\it \bibinfo{journal}{ACM Comput. Surv.}\/},  {\it \bibinfo{volume}{54}\/}. \URLprefix \url{https://doi.org/10.1145/3439950}. \DOIprefix\doi{10.1145/3439950}.
%Type = Article
\bibitem[{Pedregosa et~al.(2011)Pedregosa, Varoquaux, Gramfort, Michel, Thirion, Grisel, Blondel, Prettenhofer, Weiss, Dubourg, Vanderplas, Passos, Cournapeau, Brucher, Perrot \& Duchesnay}]{scikit-learn}
\bibinfo{author}{Pedregosa, F.}, \bibinfo{author}{Varoquaux, G.}, \bibinfo{author}{Gramfort, A.}, \bibinfo{author}{Michel, V.}, \bibinfo{author}{Thirion, B.}, \bibinfo{author}{Grisel, O.}, \bibinfo{author}{Blondel, M.}, \bibinfo{author}{Prettenhofer, P.}, \bibinfo{author}{Weiss, R.}, \bibinfo{author}{Dubourg, V.}, \bibinfo{author}{Vanderplas, J.}, \bibinfo{author}{Passos, A.}, \bibinfo{author}{Cournapeau, D.}, \bibinfo{author}{Brucher, M.}, \bibinfo{author}{Perrot, M.}, \& \bibinfo{author}{Duchesnay, E.} (\bibinfo{year}{2011}).
\newblock \bibinfo{title}{Scikit-learn: Machine learning in {P}ython}.
\newblock {\it \bibinfo{journal}{Journal of Machine Learning Research}\/},  {\it \bibinfo{volume}{12}\/}, \bibinfo{pages}{2825--2830}.
%Type = Article
\bibitem[{Pion-Tonachini et~al.(2019)Pion-Tonachini, Kreutz-Delgado \& Makeig}]{pion2019iclabel}
\bibinfo{author}{Pion-Tonachini, L.}, \bibinfo{author}{Kreutz-Delgado, K.}, \& \bibinfo{author}{Makeig, S.} (\bibinfo{year}{2019}).
\newblock \bibinfo{title}{Iclabel: An automated electroencephalographic independent component classifier, dataset, and website}.
\newblock {\it \bibinfo{journal}{NeuroImage}\/},  {\it \bibinfo{volume}{198}\/}, \bibinfo{pages}{181--197}.
%Type = Article
\bibitem[{Prost et~al.(2022)Prost, Houdard, Papadakis \& Almansa}]{prost_2022_super_resolution_hierarchical_vae}
\bibinfo{author}{Prost, J.}, \bibinfo{author}{Houdard, A.}, \bibinfo{author}{Papadakis, N.}, \& \bibinfo{author}{Almansa, A.} (\bibinfo{year}{2022}).
\newblock \bibinfo{title}{Diverse super-resolution with pretrained deep hiererarchical vaes}.
\newblock {\it \bibinfo{journal}{arXiv preprint arXiv:2205.10347}\/}, .
%Type = Article
\bibitem[{Qiu et~al.(2019)Qiu, Du \& Qian}]{Qiu_2019_time_series_anomaly_detection_classifier}
\bibinfo{author}{Qiu, J.}, \bibinfo{author}{Du, Q.}, \& \bibinfo{author}{Qian, C.} (\bibinfo{year}{2019}).
\newblock \bibinfo{title}{Kpi-tsad: A time-series anomaly detector for kpi monitoring in cloud applications}.
\newblock {\it \bibinfo{journal}{Symmetry}\/},  {\it \bibinfo{volume}{11}\/}. \URLprefix \url{https://www.mdpi.com/2073-8994/11/11/1350}. \DOIprefix\doi{10.3390/sym11111350}.
%Type = Article
\bibitem[{Qiu et~al.(2018)Qiu, Zhou, Yu \& Du}]{yang_2018_DSAE_EEG}
\bibinfo{author}{Qiu, Y.}, \bibinfo{author}{Zhou, W.}, \bibinfo{author}{Yu, N.}, \& \bibinfo{author}{Du, P.} (\bibinfo{year}{2018}).
\newblock \bibinfo{title}{Denoising sparse autoencoder-based ictal eeg classification}.
\newblock {\it \bibinfo{journal}{IEEE Transactions on Neural Systems and Rehabilitation Engineering}\/},  {\it \bibinfo{volume}{26}\/}, \bibinfo{pages}{1717--1726}. \DOIprefix\doi{10.1109/TNSRE.2018.2864306}.
%Type = Inproceedings
\bibitem[{Razavi et~al.(2019)Razavi, van~den Oord \& Vinyals}]{razavi_NEURIPS2019_vqVAE2}
\bibinfo{author}{Razavi, A.}, \bibinfo{author}{van~den Oord, A.}, \& \bibinfo{author}{Vinyals, O.} (\bibinfo{year}{2019}).
\newblock \bibinfo{title}{Generating diverse high-fidelity images with vq-vae-2}.
\newblock In \bibinfo{editor}{H.~Wallach}, \bibinfo{editor}{H.~Larochelle}, \bibinfo{editor}{A.~Beygelzimer}, \bibinfo{editor}{F.~d\textquotesingle Alch\'{e}-Buc}, \bibinfo{editor}{E.~Fox}, \& \bibinfo{editor}{R.~Garnett} (Eds.), {\it \bibinfo{booktitle}{Advances in Neural Information Processing Systems}\/}.
\newblock \bibinfo{publisher}{Curran Associates, Inc.} volume~\bibinfo{volume}{32}.
\newblock \URLprefix \url{https://proceedings.neurips.cc/paper_files/paper/2019/file/5f8e2fa1718d1bbcadf1cd9c7a54fb8c-Paper.pdf}.
%Type = Article
\bibitem[{Riyad et~al.(2021)Riyad, Khalil \& Adib}]{MI_EEGNet_AC_TAB_10}
\bibinfo{author}{Riyad, M.}, \bibinfo{author}{Khalil, M.}, \& \bibinfo{author}{Adib, A.} (\bibinfo{year}{2021}).
\newblock \bibinfo{title}{{MI-EEGNET}: A novel convolutional neural network for motor imagery classification}.
\newblock {\it \bibinfo{journal}{Journal of Neuroscience Methods}\/},  {\it \bibinfo{volume}{353}\/}, \bibinfo{pages}{109037}. \URLprefix \url{https://www.sciencedirect.com/science/article/pii/S016502702030460X}. \DOIprefix\doi{https://doi.org/10.1016/j.jneumeth.2020.109037}.
%Type = Article
\bibitem[{{Sakhavi} et~al.(2018){Sakhavi}, {Guan} \& {Yan}}]{Sakhavi_AC_KS_TAB_4_5}
\bibinfo{author}{{Sakhavi}, S.}, \bibinfo{author}{{Guan}, C.}, \& \bibinfo{author}{{Yan}, S.} (\bibinfo{year}{2018}).
\newblock \bibinfo{title}{Learning temporal information for brain-computer interface using convolutional neural networks}.
\newblock {\it \bibinfo{journal}{IEEE Transactions on Neural Networks and Learning Systems}\/},  {\it \bibinfo{volume}{29}\/}, \bibinfo{pages}{5619--5629}.
%Type = Article
\bibitem[{Sakoe \& Chiba(1978)}]{Sakoe_1978_DynamicPA_DTW_Paper}
\bibinfo{author}{Sakoe, H.}, \& \bibinfo{author}{Chiba, S.} (\bibinfo{year}{1978}).
\newblock \bibinfo{title}{Dynamic programming algorithm optimization for spoken word recognition}.
\newblock {\it \bibinfo{journal}{IEEE Transactions on Acoustics, Speech, and Signal Processing}\/},  {\it \bibinfo{volume}{26}\/}, \bibinfo{pages}{159--165}. \URLprefix \url{https://api.semanticscholar.org/CorpusID:17900407}.
%Type = Inproceedings
\bibitem[{Satopaa et~al.(2011)Satopaa, Albrecht, Irwin \& Raghavan}]{kneed_package}
\bibinfo{author}{Satopaa, V.}, \bibinfo{author}{Albrecht, J.}, \bibinfo{author}{Irwin, D.}, \& \bibinfo{author}{Raghavan, B.} (\bibinfo{year}{2011}).
\newblock \bibinfo{title}{Finding a "kneedle" in a haystack: Detecting knee points in system behavior}.
\newblock In {\it \bibinfo{booktitle}{2011 31st International Conference on Distributed Computing Systems Workshops}\/} (pp. \bibinfo{pages}{166--171}).
\newblock \DOIprefix\doi{10.1109/ICDCSW.2011.20}.
%Type = Article
\bibitem[{Schirrmeister et~al.(2017)Schirrmeister, Springenberg, Fiederer, Glasstetter, Eggensperger, Tangermann, Hutter, Burgard \& Ball}]{Schirrmeister_EEG_CNN}
\bibinfo{author}{Schirrmeister, R.~T.}, \bibinfo{author}{Springenberg, J.~T.}, \bibinfo{author}{Fiederer, L. D.~J.}, \bibinfo{author}{Glasstetter, M.}, \bibinfo{author}{Eggensperger, K.}, \bibinfo{author}{Tangermann, M.}, \bibinfo{author}{Hutter, F.}, \bibinfo{author}{Burgard, W.}, \& \bibinfo{author}{Ball, T.} (\bibinfo{year}{2017}).
\newblock \bibinfo{title}{Deep learning with convolutional neural networks for {EEG} decoding and visualization}.
\newblock {\it \bibinfo{journal}{Human Brain Mapping}\/}, . \DOIprefix\doi{10.1002/hbm.23730}.
%Type = Article
\bibitem[{Straetmans et~al.(2022)Straetmans, Holtze, Debener, Jaeger \& Mirkovic}]{straetmans2022neural}
\bibinfo{author}{Straetmans, L.}, \bibinfo{author}{Holtze, B.}, \bibinfo{author}{Debener, S.}, \bibinfo{author}{Jaeger, M.}, \& \bibinfo{author}{Mirkovic, B.} (\bibinfo{year}{2022}).
\newblock \bibinfo{title}{Neural tracking to go: auditory attention decoding and saliency detection with mobile eeg}.
\newblock {\it \bibinfo{journal}{Journal of neural engineering}\/},  {\it \bibinfo{volume}{18}\/}, \bibinfo{pages}{066054}.
%Type = Article
\bibitem[{Teplan et~al.(2002)}]{teplan2002fundamentals}
\bibinfo{author}{Teplan, M.} et~al. (\bibinfo{year}{2002}).
\newblock \bibinfo{title}{Fundamentals of eeg measurement}.
\newblock {\it \bibinfo{journal}{Measurement science review}\/},  {\it \bibinfo{volume}{2}\/}, \bibinfo{pages}{1--11}.
%Type = Article
\bibitem[{Tran et~al.(2022)Tran, Tran, Le, Huynh, Tran \& Dao}]{Tran_2022_chb_mit_dataset}
\bibinfo{author}{Tran, L.~V.}, \bibinfo{author}{Tran, H.~M.}, \bibinfo{author}{Le, T.~M.}, \bibinfo{author}{Huynh, T. T.~M.}, \bibinfo{author}{Tran, H.~T.}, \& \bibinfo{author}{Dao, S. V.~T.} (\bibinfo{year}{2022}).
\newblock \bibinfo{title}{Application of machine learning in epileptic seizure detection}.
\newblock {\it \bibinfo{journal}{Diagnostics}\/},  {\it \bibinfo{volume}{12}\/}. \URLprefix \url{https://www.mdpi.com/2075-4418/12/11/2879}. \DOIprefix\doi{10.3390/diagnostics12112879}.
%Type = Misc
\bibitem[{Vahdat \& Kautz(2021)}]{vahdat_2021_nvae}
\bibinfo{author}{Vahdat, A.}, \& \bibinfo{author}{Kautz, J.} (\bibinfo{year}{2021}).
\newblock \bibinfo{title}{Nvae: A deep hierarchical variational autoencoder}.
\newblock \href{http://arxiv.org/abs/2007.03898}{\tt arXiv:2007.03898}.
%Type = Article
\bibitem[{Wambura et~al.(2020)Wambura, Huang \& Li}]{wambura_2020_OFAT_NN_timeseries}
\bibinfo{author}{Wambura, S.}, \bibinfo{author}{Huang, J.}, \& \bibinfo{author}{Li, H.} (\bibinfo{year}{2020}).
\newblock \bibinfo{title}{Long-range forecasting in feature-evolving data streams}.
\newblock {\it \bibinfo{journal}{Knowledge-Based Systems}\/},  {\it \bibinfo{volume}{206}\/}, \bibinfo{pages}{106405}.
%Type = Inproceedings
\bibitem[{Wang et~al.(2020)Wang, Wei, Zhang, Huang, Liang, Li \& Zhang}]{Wang_2020_EEGNet_autoencoder_1}
\bibinfo{author}{Wang, J.}, \bibinfo{author}{Wei, M.}, \bibinfo{author}{Zhang, L.}, \bibinfo{author}{Huang, G.}, \bibinfo{author}{Liang, Z.}, \bibinfo{author}{Li, L.}, \& \bibinfo{author}{Zhang, Z.} (\bibinfo{year}{2020}).
\newblock \bibinfo{title}{An autoencoder-based approach to predict subjective pain perception from high-density evoked eeg potentials}.
\newblock In {\it \bibinfo{booktitle}{2020 42nd Annual International Conference of the IEEE Engineering in Medicine and Biology Society (EMBC)}\/} (pp. \bibinfo{pages}{1507--1511}).
\newblock \DOIprefix\doi{10.1109/EMBC44109.2020.9176644}.
%Type = Article
\bibitem[{Welch(1967)}]{welch1967}
\bibinfo{author}{Welch, P.} (\bibinfo{year}{1967}).
\newblock \bibinfo{title}{The use of fast fourier transform for the estimation of power spectra: a method based on time averaging over short, modified periodograms}.
\newblock {\it \bibinfo{journal}{IEEE Transactions on audio and electroacoustics}\/},  {\it \bibinfo{volume}{15}\/}, \bibinfo{pages}{70--73}.
%Type = Article
\bibitem[{Xing et~al.(2020)Xing, Demertzis \& Yang}]{xing_2020_spiking_nn_anomalies}
\bibinfo{author}{Xing, L.}, \bibinfo{author}{Demertzis, K.}, \& \bibinfo{author}{Yang, J.} (\bibinfo{year}{2020}).
\newblock \bibinfo{title}{Identifying data streams anomalies by evolving spiking restricted boltzmann machines}.
\newblock {\it \bibinfo{journal}{Neural Computing and Applications}\/},  {\it \bibinfo{volume}{32}\/}, \bibinfo{pages}{6699--6713}.
%Type = Misc
\bibitem[{Zancanaro et~al.(under review)Zancanaro, Cisotto, Manzoni \& Zoppis}]{Zancanaro_CCIS}
\bibinfo{author}{Zancanaro, A.}, \bibinfo{author}{Cisotto, G.}, \bibinfo{author}{Manzoni, S.~L.}, \& \bibinfo{author}{Zoppis, I.~F.} (\bibinfo{year}{under review}).
\newblock \bibinfo{title}{veegnet: learning latent representations to reconstruct eeg raw data via variational autoencoders}.
%Type = Inproceedings
\bibitem[{Zancanaro et~al.(2021)Zancanaro, Cisotto, Paulo, Pires \& Nunes}]{zancanaro_CIBCB_article}
\bibinfo{author}{Zancanaro, A.}, \bibinfo{author}{Cisotto, G.}, \bibinfo{author}{Paulo, J.~R.}, \bibinfo{author}{Pires, G.}, \& \bibinfo{author}{Nunes, U.~J.} (\bibinfo{year}{2021}).
\newblock \bibinfo{title}{{CNN}-based approaches for cross-subject classification in motor imagery: From the state-of-the-art to {D}ynamic{N}et}.
\newblock In {\it \bibinfo{booktitle}{2021 IEEE Conference on Computational Intelligence in Bioinformatics and Computational Biology (CIBCB)}\/} (pp. \bibinfo{pages}{1--7}).
\newblock \DOIprefix\doi{10.1109/CIBCB49929.2021.9562821}.
%Type = Inproceedings
\bibitem[{Zancanaro et~al.(2023)Zancanaro, Zoppis, Manzoni \& Cisotto}]{Zancanaro_2023_vEEGNet}
\bibinfo{author}{Zancanaro, A.}, \bibinfo{author}{Zoppis, I.}, \bibinfo{author}{Manzoni, S.}, \& \bibinfo{author}{Cisotto, G.} (\bibinfo{year}{2023}).
\newblock \bibinfo{title}{veegnet: A new deep learning model to classify and generate eeg}.
\newblock In {\it \bibinfo{booktitle}{Proceedings of the 9th International Conference on Information and Communication Technologies for Ageing Well and e-Health - Volume 1: ICT4AWE,}\/} (pp. \bibinfo{pages}{245--252}).
\newblock \bibinfo{organization}{INSTICC} \bibinfo{publisher}{SciTePress}.
\newblock \DOIprefix\doi{10.5220/0011990800003476}.
%Type = Article
\bibitem[{Zheng \& Lu(2015)}]{zheng_2015_SEED_dataset}
\bibinfo{author}{Zheng, W.-L.}, \& \bibinfo{author}{Lu, B.-L.} (\bibinfo{year}{2015}).
\newblock \bibinfo{title}{Investigating critical frequency bands and channels for eeg-based emotion recognition with deep neural networks}.
\newblock {\it \bibinfo{journal}{IEEE Transactions on Autonomous Mental Development}\/},  {\it \bibinfo{volume}{7}\/}, \bibinfo{pages}{162--175}. \DOIprefix\doi{10.1109/TAMD.2015.2431497}.
%Type = Inproceedings
\bibitem[{Zhou \& Paffenroth(2017)}]{zhou_2017_robust_autoencoder}
\bibinfo{author}{Zhou, C.}, \& \bibinfo{author}{Paffenroth, R.~C.} (\bibinfo{year}{2017}).
\newblock \bibinfo{title}{Anomaly detection with robust deep autoencoders}.
\newblock In {\it \bibinfo{booktitle}{Proceedings of the 23rd ACM SIGKDD International Conference on Knowledge Discovery and Data Mining}\/} KDD '17 (p. \bibinfo{pages}{665–674}).
\newblock \bibinfo{address}{New York, NY, USA}: \bibinfo{publisher}{Association for Computing Machinery}.
\newblock \URLprefix \url{https://doi.org/10.1145/3097983.3098052}. \DOIprefix\doi{10.1145/3097983.3098052}.

\end{thebibliography}

\end{document}